\documentclass[12pt, aomart]{amsart}

\usepackage{tikz-cd1}
\usepackage{amsmath}
\usepackage{relsize1}
\usepackage{stackengine}
\usepackage{scalerel}

\usepackage[english]{babel}
\usepackage[light,condensed,math]{anttor}
\usepackage[T1]{fontenc}
\usepackage{ulem}
\usepackage{kpfonts}
\usepackage{comment}
\usepackage{array}
\usepackage[T1]{fontenc}
\usepackage{lmodern}
\usepackage[mathscr]{euscript}
\usepackage[T2A]{fontenc}
\usepackage[utf8x]{inputenc}
\usepackage{breqn}
\usepackage{todonotes}
\usepackage{slashed}
\usepackage{youngtab}
\usepackage{amsmath}
\usepackage{amssymb}
\usepackage{amsxtra}
\usepackage{color}
\usepackage[margin=1.2in]{geometry}

\newcommand{\boxit}[1]{\vbox{\hrule\hbox{\vrule\kern8pt
\vbox{\hbox{\kern8pt}\hbox{\vbox{#1}}\hbox{\kern8pt}}
\kern8pt\vrule}\hrule}}
\newcommand{\mathboxit}[1]{\vbox{\hrule\hbox{\vrule\kern8pt\vbox{\kern8pt
\hbox{$\displaystyle #1$}\kern8pt}\kern8pt\vrule}\hrule}}

\newcommand{\picit}[2]{\includegraphics[width=#1cm]{#2.eps}}
\makeatletter

\DeclareFontFamily{OMX}{MnSymbolE}{}
\DeclareSymbolFont{MnLargeSymbols}{OMX}{MnSymbolE}{m}{n}
\SetSymbolFont{MnLargeSymbols}{bold}{OMX}{MnSymbolE}{b}{n}
\DeclareFontShape{OMX}{MnSymbolE}{m}{n}{
    <-6>  MnSymbolE5
   <6-7>  MnSymbolE6
   <7-8>  MnSymbolE7
   <8-9>  MnSymbolE8
   <9-10> MnSymbolE9
  <10-12> MnSymbolE10
  <12->   MnSymbolE12
}{}
\DeclareFontShape{OMX}{MnSymbolE}{b}{n}{
    <-6>  MnSymbolE-Bold5
   <6-7>  MnSymbolE-Bold6
   <7-8>  MnSymbolE-Bold7
   <8-9>  MnSymbolE-Bold8
   <9-10> MnSymbolE-Bold9
  <10-12> MnSymbolE-Bold10
  <12->   MnSymbolE-Bold12
}{}

\let\llangle\@undefined
\let\rrangle\@undefined
\DeclareMathDelimiter{\llangle}{\mathopen}%
                     {MnLargeSymbols}{'164}{MnLargeSymbols}{'164}
\DeclareMathDelimiter{\rrangle}{\mathclose}%
                     {MnLargeSymbols}{'171}{MnLargeSymbols}{'171}
\makeatother

 \newcommand{\beq}{\begin{equation}}
                \newcommand{\bea}{\begin{eqnarray}}
                \newcommand{\eea}{\end{eqnarray}}
                 \newcommand{\eeq}{\end{equation}}  

\newcommand{\vev}[1]{\left\langle\ #1 \ \right\rangle}

\newcommand{\my}[1]{{\mathscr Y} \left( #1 \right)}
\newcommand{\y}{{\mathscr Y}}
\newcommand{\q}{{\mathscr Q}}
\newcommand{\x}{{\mathscr X}}

\newcommand{\mA}{{\mathscr A}}
\newcommand{\mN}{{\mathscr N}}

\newcommand{\iM}{{\mathscr M}}

\newcommand{\qM}{{\mathbb M}}
\newcommand{\qI}{{\tilde I}}
\newcommand{\qJ}{{\tilde J}}
\newcommand{\mM}{{\mathfrak M}}
\newcommand{\mW}{{\mathscr W}}

\newcommand {\BC}   {\mathbb C}

\newcommand {\BH}   {\mathbb H}

\newcommand {\BR}   {\mathbb R}
\newcommand {\BP}   {\mathbb P}
\newcommand {\BQ}   {\mathbb Q}

\newcommand {\bR}   {\mathbf{R}}

\newcommand {\qe} {\mathfrak q}

\newcommand {\ib} {\mathbf{i}}

\newcommand {\jb} {\mathbf{j}}
\newcommand {\Pf} {\mathfrak{P}}
\newcommand {\Xf} {\mathfrak{X}}
\newcommand {\Yf} {\mathfrak{Y}}
\newcommand {\Hf} {\mathsf{H}}

\newcommand{\Hilb} {\mathsf{Hilb}}
\newcommand{\HM} {\mathsf{HM}}
\newcommand{\Ci} {\mathsf{C}_{\ib, \bv}}

\newcommand {\xb} {\mathbf{x}}
\newcommand {\ii} {\mathrm{i}}
\newcommand {\bT}   {\mathbf{T}}
\newcommand {\ba}  {\underline{\ac}}

\newcommand {\bg} {\mathbf{g}}

\newcommand {\bn}{\underline{\mathbf{n}}}
\newcommand {\mm}{\underline{\mathbf{m}}}
\newcommand {\mt} {\tt m}
\newcommand {\Det} {\tt Det}
\newcommand {\dt} {\tt d}
\newcommand {\bc} {\underline{\mathbf{c}}}
\newcommand {\bv} {\underline{\mathbf{v}}}
\newcommand {\bk}{  \mathbf{k}}

\newcommand {\bw} {\underline{\mathbf{w}}}
\newcommand {\vt} {\tt v}
\newcommand {\wt} {\tt w}

\newcommand {\tw} {\text{w}}

\newcommand {\bnu} {\underline{\boldsymbol{\nu}}}
\newcommand {\bmu} {\boldsymbol{\mu}}
\newcommand {\bmt} {\underline{\boldsymbol{\mt}}}
\newcommand {\bzt} {\underline{\boldsymbol{\zeta}}}
\newcommand {\btt} {\underline{\boldsymbol{\tau}}}
\newcommand {\ept} {\underline{\ve}}

\newcommand {\bqt} {\underline{\qe}}

\newcommand {\bkt} {\underline{\bk}}

\newcommand {\bla} {\underline{\boldsymbol{\lambda}}}

\newcommand {\bx}{  \mathbf{x}}
\newcommand {\bX}{ \mathbf{X}}

\newcommand {\bS}{ \mathbf{S}}
\newcommand {\BS}   {\mathbb S}

\newcommand {\BZ}   {\mathbb Z}

\newcommand {\ac} {\mathfrak{a}}

\newcommand {\fe} {\mathfrak{f}}

\newcommand {\CalB} {\mathcal B}
\newcommand {\CalC} {\mathcal C}

\newcommand {\CalD} {\mathcal D}
\newcommand {\CalE} {\mathcal E}
\newcommand {\CalF} {\mathcal F}

\newcommand {\CalH} {\mathcal H}
\newcommand {\CalI} {\mathcal I}

\newcommand {\CalM} {\mathcal M}
\newcommand {\CalN} {\mathcal N}
\newcommand {\CalO} {\mathcal O}

\newcommand {\CalQ} {\mathcal Q}
\newcommand {\CalR} {\mathcal R}
\newcommand {\CalS} {\mathcal S}
\newcommand {\CalT} {\mathcal T}

\newcommand {\CalX} {\mathcal X}
\newcommand {\CalY} {\mathcal Y}
\newcommand {\CalW} {\mathcal W}
\newcommand {\CalZ} {\mathcal Z}

\newcommand {\ee} {\mathfrak E}

\newcommand{\al}{\alpha}
\newcommand{\ve}{\varepsilon}
\newcommand{\ep}{\epsilon}

\renewcommand{\hat}{\widehat}

\newcommand{\Gg}{\mathsf{G}_{\mathbf{g}}}

\newcommand{\Gf}{\mathsf{G}_{\text{\tiny f}}}
\newcommand{\Gr}{\mathsf{G}_{\text{\tiny rot}}}

\newcommand{\gq}{\mathfrak{g}_{\gamma}}

\newcommand{\Ver}{\mathsf{Vert}_{\gamma}}

\newcommand{\Edg}{\mathsf{Edges}_{\gamma}}
\newcommand{\Arr}{\mathsf{Arrows}_{\gamma}}
\newcommand{\Path}{\mathsf{Paths}_{\gamma}}
\newcommand{\Obs}{\mathsf{Obs}}

\newcommand{\diag}{\mathsf{diag}}
\newcommand{\Tr}{\mathsf{Tr}\,}

\newcommand{\subsubsec}[1]{\subsubsection{\uwave{\sl #1}}}
\newcommand{\subsec}[1]{\subsection{\underline{\bf #1}}}
\newcommand{\secc}[1]{\section{$\mathbf{ #1}$}}

\begin{document}
\title[BPS/CFT, Dyson-Schwinger, qq-characters]{BPS/CFT correspondence:\\ \ \\
Non-perturbative\\ Dyson-Schwinger\ equations\\
and\ $qq$-characters}

\author{Nikita Nekrasov}

\address{Simons Center for Geometry and Physics\\
Stony Brook University, Stony Brook NY 11794-3636, USA
\\
E-mail: nikitastring@gmail.com\footnote{on leave of absence from:
IHES, Bures-sur-Yvette, France\\
ITEP and IITP, Moscow, Russia}}


\begin{abstract}
We  study symmetries of quantum field theories involving topologically distinct sectors of the field space. To exhibit these symmetries we define special gauge invariant observables, which we call the $qq$-characters.  In the context of the BPS/CFT correspondence, using  these observables, we derive an infinite set of Dyson-Schwinger-type relations. These relations imply that  the supersymmetric partition functions in the presence of $\Omega$-deformation and defects obey the Ward identities of two dimensional conformal field theory and its $q$-deformations. The details will be discussed in the companion papers.

\end{abstract}

\maketitle

\setcounter{tocdepth}{2} 
\tableofcontents

\vfill\eject
\centerline{}
\vskip 5cm
\centerline{$\mathbf{In\ memory\ of\ Lev\ Borisovich\ Okun\ (1929-2015)}$}
\bigskip
\vfill\eject
\secc{ Introduction}\label{aba:sec1}

\subsec{Dyson-Schwinger equations}

The correlation functions of Euclidean quantum field theory  are defined by the  path integral:
\begin{equation}
\langle {\CalO}_{1}(x_{1}) \ldots {\CalO}_{n} (x_{n} ) \rangle = \frac{1}{Z}
\int_{\Gamma} \ D{\Phi}\ e^{- \frac{1}{\hbar} S[{\Phi}]}\ {\CalO}_{1}(x_{1}) \ldots {\CalO}_{n} (x_{n} )  \, , 
\label{eq:corrf}
\end{equation}
suitably regularized and renormalized. 
The classical theory is governed by the Euler-Lagrange equations, which are derived from the variational principle:
\begin{equation}
{\delta}S [ {\Phi}_{cl} ] = 0
\end{equation}

These equations are modified in the quantum theory: consider an infinitesimal transformation
\begin{equation}
{\Phi} \longrightarrow {\Phi} + {\delta}{\Phi}
\label{eq:chngv}
\end{equation}
Assuming \eqref{eq:chngv} preserves the measure $D{\Phi}$ in \eqref{eq:corrf}
(no anomaly), then 
\begin{multline}
\langle {\CalO}_{1}(x_{1}) \ldots {\CalO}_{n} (x_{n} ) {\delta}S [ {\Phi} ] \rangle =\\
 {\hbar} \sum_{i=1}^{n} 
\langle {\CalO}_{1}(x_{1}) \ldots {\CalO}_{i-1} (x_{i-1}) {\delta}{\CalO}(x_{i}) {\CalO}_{i+1} (x_{i+1}) \ldots {\CalO}_{n} (x_{n}) \rangle
\label{eq:dse}
\end{multline}
In \eqref{eq:chngv} the change of variables can be also interpreted as a small modification of the integration contour $\Gamma$ in \eqref{eq:corrf}, ${\Gamma} \to {\Gamma}' =  {\Gamma} + {\delta}{\Gamma}$, as in the picture below 

\bigskip
\centerline{\picit{5}{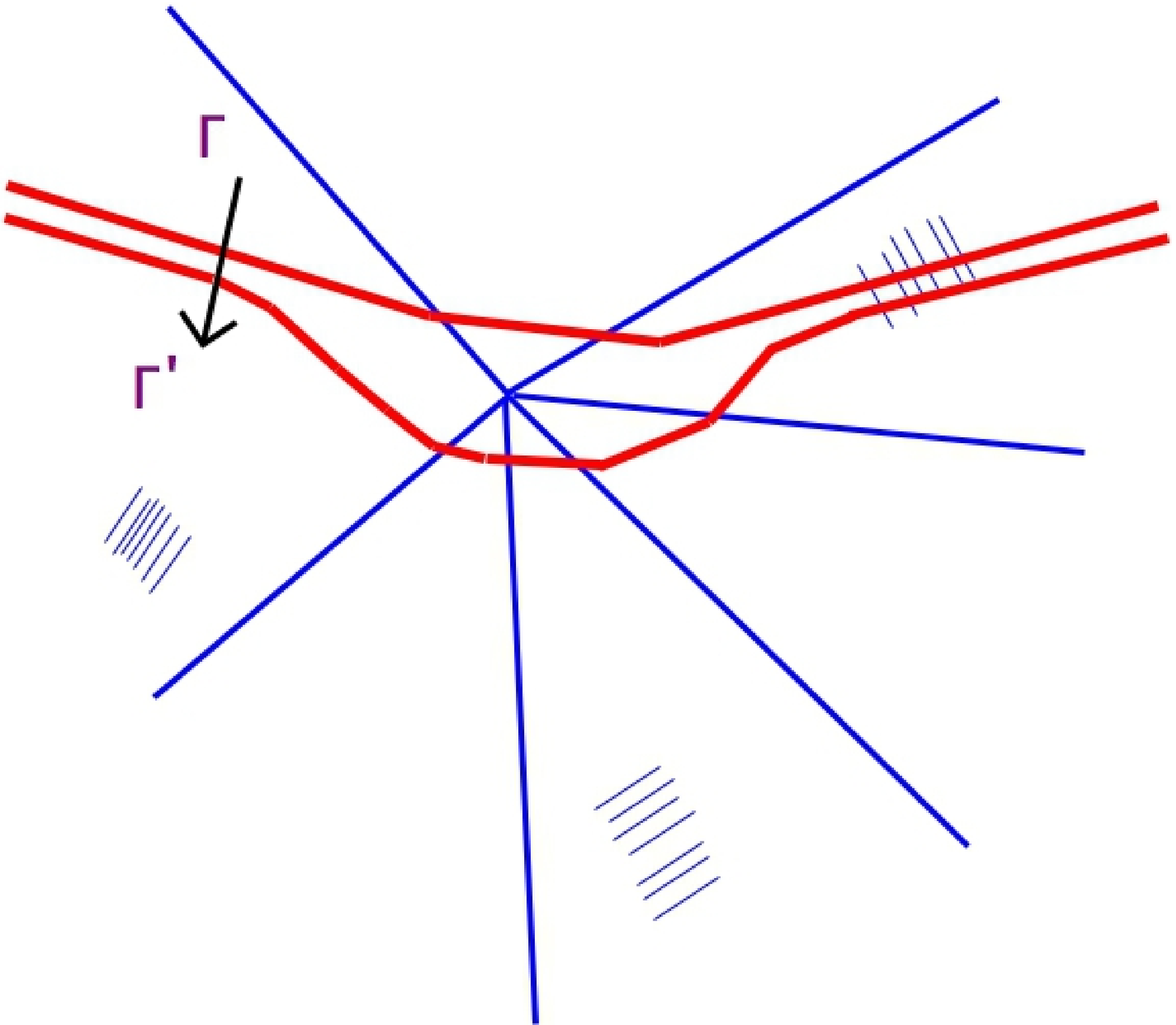}}
\bigskip
\centerline{{\bf Fig.1}}
\bigskip

The small change of contour does not change the integral of a closed form. 

The usefulness of the Dyson-Schwinger equations depends on whether one can find a convenient set of observables ${\CalO}_{i}$ in \eqref{eq:dse} and perhaps also take a limit in order to get a closed system of equations. Formally, the loop equations \cite{MakMig}, \cite{Migdal:1984gj} are an example of such a system. Another, related example, is the matrix model, i.e. the zero dimensional gauge theory. The simplest model is the single matrix integral:
\beq
Z = \int_{{\mathrm Lie} U(N)} \left[ \frac{{\CalD}{\Phi}}{{\rm Vol}U(N)} \right] e^{-\frac{1}{g_{s}} {\Tr}_{N} V_{p+1}({\Phi}) }
\label{eq:matmod}
\eeq
with the polynomial potential
\beq
V_{p+1}(x) = \sum_{k=0}^{p}  \frac{t_{k}}{(k+1)!} \, x^{k+1}
\label{eq:vofx}
\eeq 
The convenient observable is 
\beq
{\bf Y}(x) = g_{s} {\Tr}_{N} \ \frac{1}{x - {\Phi}} - V^{\prime}(x) 
\label{eq:resolve}
\eeq
In the limit $N \to \infty$, $g_{s} \to 0$, with ${\hbar} = g_{s}N$ fixed, the expectation value  $Y(x) = \vev{{\bf Y}(x)}$ obeys:
\beq
Y(x)^2 = V_{p+1}^{\prime}(x)^{2} + f_{p-1}(x)
\label{eq:yyvv}
\eeq
where $f_{p-1}(x)$ is a degree $p-1$ polynomial of $x$:
\beq
f_{p-1} (x) = \vev{{\Tr}_{N}\ \left( \frac{V^{\prime}({\Phi}) - V^{\prime}(x)}{{\Phi}-x} \right)}
\eeq 
whose coefficients encode the expectation values of degree $\leq p-1$ Casimirs of $\Phi$.  
One can reformulate  \eqref{eq:yyvv} somewhat more invariantly by stating that the singularities of ${\bf Y}(x)$ disappear in 
$\vev{{\bf Y}(x)}^{2}$, in the planar limit $N \to \infty$. 
For finite $N, \hbar$ the Dyson-Schwinger equation has the form
\beq
\vev{{\bf Y}(x)^{2}\,  -\  g_{s}\, {\partial}_{x} {\bf Y}(x) } = V_{p+1}^{\prime}(x)^{2} + f_{p-1}(x)
\label{eq:dseq}
\eeq
Although the equation \eqref{eq:dseq} is not a closed system of equations per se, it illustrates a principle, which we shall generalize below: given the basic operator ${\bf Y}(x)$ which, as a function of the auxiliary parameter $x$ has singularities, one constructs an expression, e.g.
\beq
{\bf T}(x) = {\bf Y}(x)^{2} - {\hbar} {\partial}_{x} {\bf Y}(x)
\label{eq:tdseq}
\eeq
whose expectation value has no singularities in $x$ for finite $x$, cf. \eqref{eq:dseq}. We shall be able to generalize this procedure for the supersymmetric gauge theories in various spacetime dimensions. 
\vfill\eject
\subsec{ Non-perturbative Dyson-Schwinger identities}

Let us now study the identities, which can be interpreted as the analogs of \eqref{eq:dse}, \eqref{eq:dseq} corresponding to non-trivial permutations of  homology classes ${\bf\Gamma} = \sum_{a}\,  n_{a} {\Gamma}_{a} \longrightarrow
{\bf\Gamma}' = \sum_{a}\,  n_{a}' {\Gamma}_{a}$, where $({\Gamma}_{a})$ is some basis in the relative homology, cf. \cite{Arnold:1985}
\[ 
H_{\frac 12 \text{dim}} ({\CalF}^{\BC}, \ {\CalF}^{\BC}_{+}) .
\] 
where ${\CalF}^{\BC}$ is the space of  complexified fields, and ${\CalF}^{\BC}_{+} \subset {\CalF}^{\BC}$
is the domain, where Re$S[{\Phi}] \gg 0$.

H.~Nakajima \cite{Nakajima:1994r} discovered that the cohomology of the moduli spaces of instantons
carries representations of the infinite-dimensional algebras (this fact was used in the first strong coupling tests of $S$-duality of maximally supersymmetric gauge theories \cite{Vafa:1994tf}). These algebras naturally occur in physics as symmetries of two dimensional conformal theories. This relation suggests the existence of a novel kind of symmetry in quantum field theory which acts via some sort of permutation of the integration regions in the path integral. The transformations of cohomology classes do not, typically, come from the point symmetries of the underlying space. Indeed, the infinitesimal symmetries, e.g. generated by some vector field $v \in \mathsf{Vect}({\Xf})$ act trivially on the de Rham cohomology $H^{*}({\Xf})$ of $\Xf$, as 
closed differential forms change by the exact forms:
\begin{equation}
{\delta}{\omega} = Lie_{v} {\omega} = d ( \iota_{v}{\omega} ) 
\ \Longrightarrow \ [ {\delta}{\omega} ] = 0 \in H^{*}({\Xf})
\end{equation}
The symmetries of the cohomology spaces come, therefore, from the {\it large} transformations
$f : {\Xf} \to {\Xf}$ or, more generally, the correspondences $L \subset {\Xf} \times {\Xf}$:
\begin{equation}
{\phi}_{L}  = t_{*} \left( {\delta}_{L} \wedge s^{*} \right) :  H^{*}({\Xf}) \to H^{*}({\Xf})
\end{equation}
where ${\delta}_{L}$ is the Poincare dual  to  $L$, the maps $s, t$ are the projections
\beq
\begin{matrix}
&  \ {\Xf} \times {\Xf} &  \\
\quad^{s} \swarrow & \, & \searrow^{t} \\
{\Xf} \quad &  \,  & \quad {\Xf} \end{matrix} \qquad\qquad\qquad\\
\label{eq:corr}
\eeq
and we assume the compactness and smoothness. There exist generalizations which relax these assumptions. 
 
 The physical realization of the symmetries generated by \eqref{eq:corr} is yet to be understood. It was proposed in \cite{Nekrasov:2009zz, Nekrasov:2009ui, Nekrasov:2009uh} that there are symmetries acting {\it between} different quantum field theories, for example changing the gauge groups.  Conjecturally \cite{Nekrasov:2009st}  supersymmetric domain walls in quantum field theory separating different phases of one theory or even connecting, e.g. in a supersymmetric fashion, two different quantum field theories  can be used to generate the generalized symmetries of the sort we discussed earlier. More precisely, one exchanges the spatial and the
 temporal directions, producing the $S$-brane \cite{Gutperle:2002ai} version of the domain wall. 

This paper deals with another type of {\it large} symmetries. They are generated by the transformations  \eqref{eq:chngv} changing the topological sector, i.e. mapping one connected component of the space of fields to another. We shall be concerned with gauge theories, i.e. the Yang-Mills theory on the space-time $\mN$, 
\begin{equation}
Z = \int\, DA \ {\exp} \, \left( - \frac{1}{4g^2} \int_{\mN} {\Tr} F_{A} \wedge \star F_{A} + \frac{{\ii}{\vartheta}}{8{\pi}^{2}} \int_{\mN} {\Tr} F_{A} \wedge  F_{A}  \right)
\label{eq:ympf}
\end{equation}
with the gauge group $\Gg$, 
and its supersymmetric generalizations. The connected components of the space of gauge fields are labeled by the topology types of the principal $\Gg$-bundles, and measured, in particular, by the instanton charge
\beq
n = - \frac{1}{8{\pi}^{2}} \int_{\mN} {\Tr} F_{A} \wedge  F_{A} .
\label{eq:instch}
\eeq
Gauge theory path integral is the sum over $n$ of the path integrals over the space of fields of fixed topology:

\bigskip
\centerline{\picit{10}{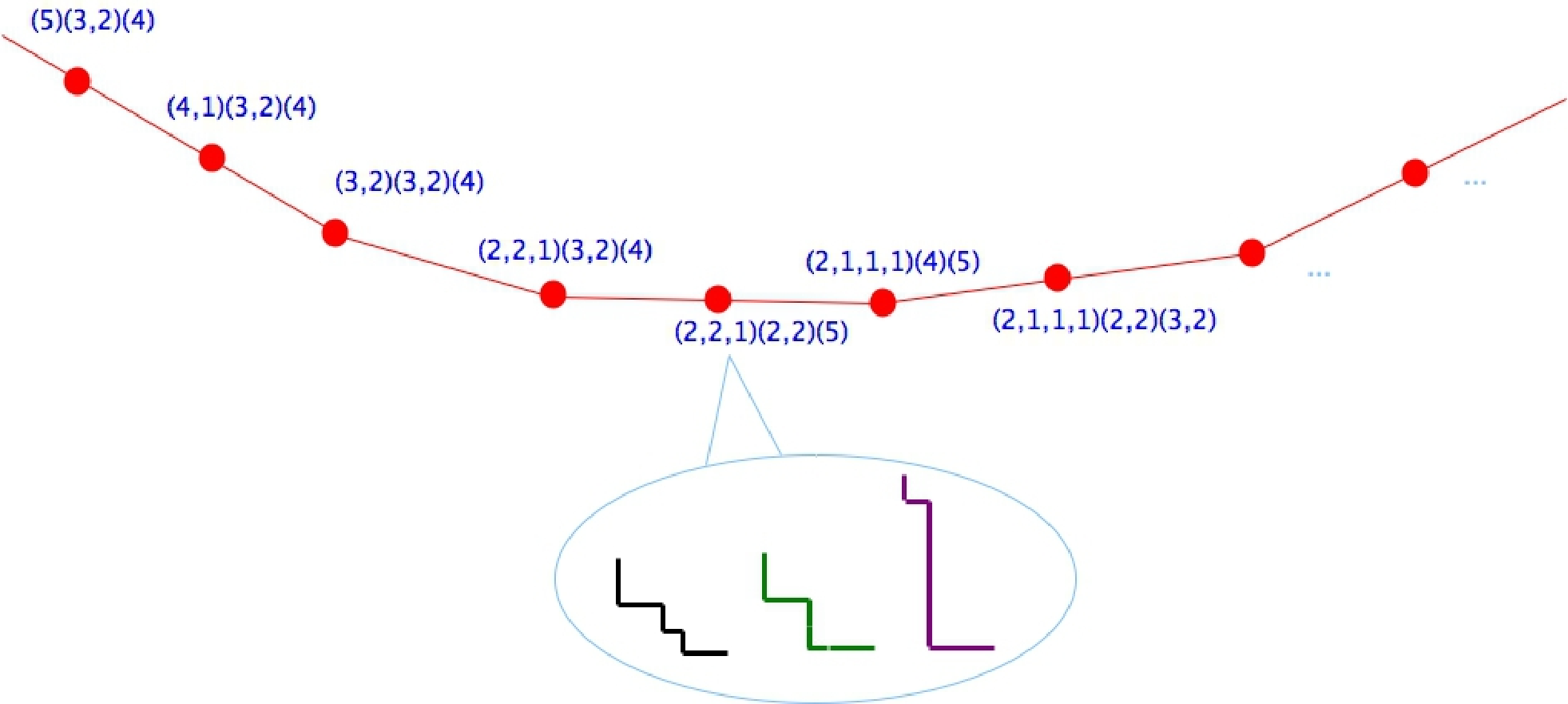}}
\bigskip
\centerline{{\bf Fig.2}}
\centerline{\tiny Path integral in $U(3)$ gauge theory in the sector with $k=14$ instanons.} 
\centerline{\tiny The labels $({\lambda}^{(1)}_{1},{\lambda}^{(1)}_{2},\ldots ) ({\lambda}^{(2)}_{1},{\lambda}^{(2)}_{2},\ldots )({\lambda}^{(3)}_{1},{\lambda}^{(3)}_{2},\ldots )$}
\centerline{\tiny denote various instanton configurations}
\bigskip
The analog of the contour deformation \eqref{eq:chngv} is the discrete deformation, as in the picture:

\bigskip
\centerline{\picit{10}{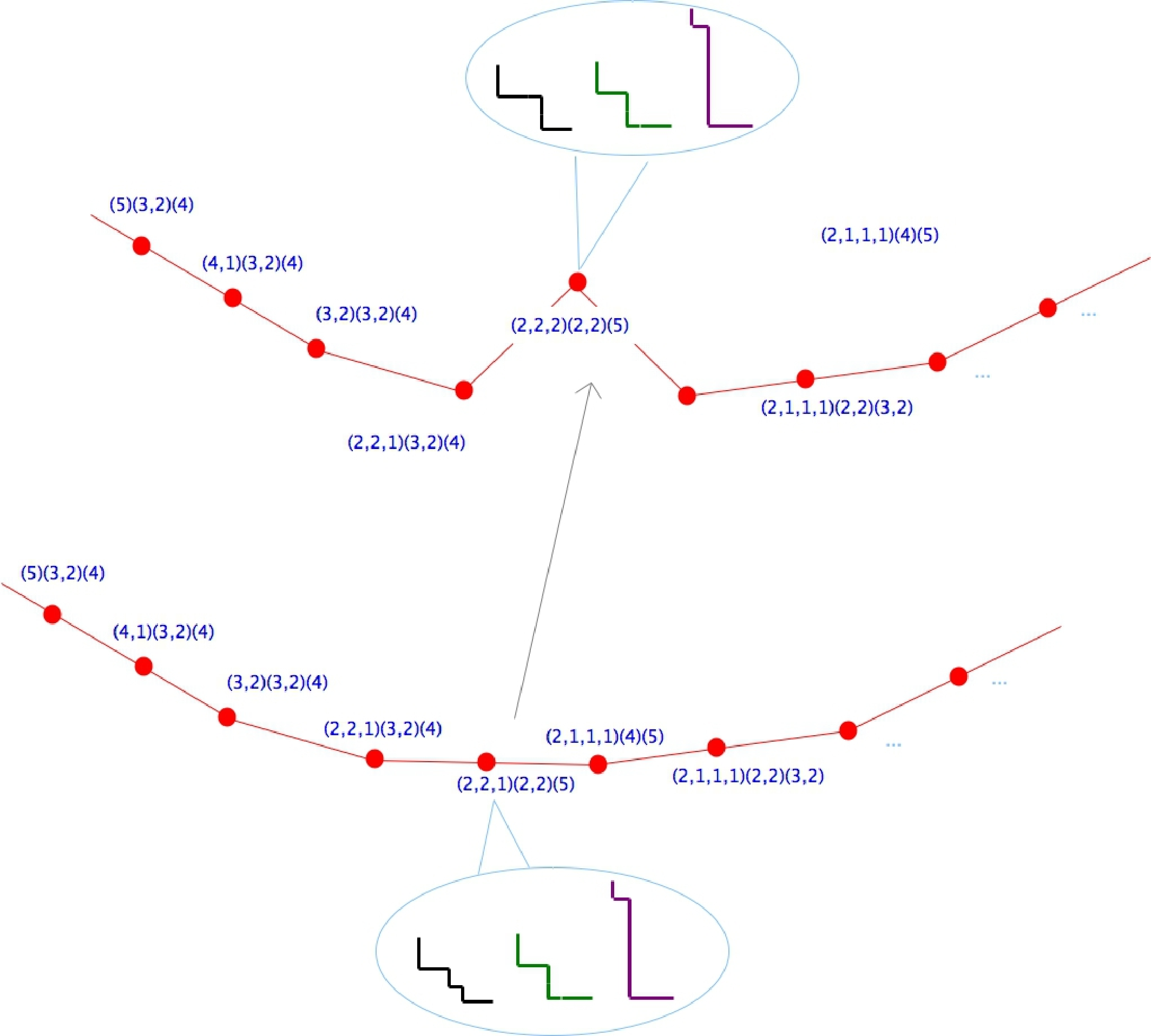}}
\bigskip
\centerline{{\bf Fig.3}}
\centerline{\tiny Path integral in $U(3)$ gauge theory  in the sector with $k=14$ instanons,}
\centerline{\tiny and discrete deformation  to account for $k=15$ instantons}
\bigskip

{}There is no a priori way to deform a connection $A_0$ on a principal bundle $P_0$ to a connection $A_1$ on 
 a principal bundle $P_1$, which is not isomorphic to $P_0$.  However, imagine that we modify 
 $P_0$ in a small neighborhood of a point $x \in \mN$ so that it becomes isomorphic to $P_1$. 
 It means that outside a small disk $D_{x} \subset {\mN}$ there is a gauge transformation, which makes
 $A_0$ deformable to $A_1$.  One can loosely call such a modification {\it adding a point-like instanton at $x$}
 \beq
 A \longrightarrow A + {\delta}_{x}^{(1)} A
 \label{eq:addpi}
 \eeq
 One can imagine a successive application of the modifications ${\delta}_{x_{1}}^{(1)} {\delta}_{x_{2}}^{(1)}$, which add point-like instantons at two distinct points $x_{1} \neq x_{2} \in \mN$, or adding two instantons at the same point, $A \longrightarrow A +  {\delta}_{x}^{(2)}A$, and so on.

The specific realization of such modifications is possible in
the string theory context, where the gauge theory instantons
are the codimension four D-branes dissolved inside another brane \cite{Douglas:1995bn}. The modification changing the instanton number is then a transition where, say, a point-like instanton becomes a D0-brane departed from the D4-brane.

 \subsec{ Organization of the presentation}

We want to study such modifications in the gauge theory language. Specifically, we shall work in 
  the context of ${\CalN}=2$ supersymmetric gauge theories subject to $\Omega$-deformation. We explore these theories  using the special observables ${\x}$ and ${\y}$, which will help us to organize the non-perturbative Dyson-Schwinger identities reflecting the invariance of the path integral with respect to the transformations \eqref{eq:addpi}. We shall see that these identities are organized in a structure, \uwave{the $qq$-characters}, which suggest a deformation of the $q$-deformed Kac-Moody symmetry. The latter is familiar from the study of lattice and massive integrable field theories in two dimensions \cite{Sklyanin:1978pj, Sklyanin:1979, Takhtajan:1979iv, Sklyanin:1980ij, Sklyanin:1982tf, Baxter:1985, Jimbo:1985, Drinfeld:1985, Drinfeld:1986, Faddeev:1987ih, Drinfeld:1987, Smirnov:1991me, Chari:1991, Smirnov:1992vz, Frenkel:1992,  Chari:1994, Knight:1995, Chari:1996, Frenkel:1998, Frenkel:2001,  Chari:2004, Frenkel:2013dh}.

 The $qq$-characters are local observables, the corresponding operators can be inserted at a point in space-time. 
 One can also define and study non-local observables, which are associated to two-dimensional surfaces in space-time. These will be studied elsewhere. 
 
 It is worth revealing at this point that the $qq$-characters (and the analogous surface operators) can be defined most naturally in the context of string theory, where the gauge theory in question arises as a low energy limit of the theory on a stack of $D3$-branes (the `physical' branes) in some supersymmetric background. The $qq$-characters in this realization are the low-energy limits of the partition function of the auxiliary theory, which lives on a stack of $D3$-branes intersecting the physical branes transversely at a point.    
We define the $qq$-character operators in the presence of the surface operators in \cite{Nekrasov:2015is}.

In the companion paper \cite{Nekrasov:2015ii} these constructions
of gauge theories with and without surface defects, as well as the $qq$-character operators are given a unified treatment using what we call the {\it gauge origami}, a generalized gauge theory, which is best thought of a low-energy limit of a theory on a stack of $Dp$-branes in type II string theory, which span the coordinate ${\BC}^{2}$-planes inside ${\BC}^{4}$ times a common flat $2p-4$-dimensional space. 

Orbifolding this construction by discrete symmetries, preserving supersymmetry and the $\Omega$-deformation, leads to more examples of $qq$-observables and defect operators in quiver gauge theories on asymptotically locally Euclidean spaces.

These constructions can be realized mathematically with the help
of novel moduli spaces, which we call the crossed instantons and the 
spiked instantons. The space of \emph{crossed \ instantons} describes
the low-energy modes of open strings connecting $k$ $D(-1)$ instantons
and two stacks of $D3$-branes, spanning two transversely intersecting
copies of ${\BR}^{4}$ inside ${\BR}^{8}$. When one of the two stacks
is empty the moduli space coincides with the ADHM moduli space
of (noncommutative) instantons on ${\BR}^{4}$, together with the 
obstruction bundle, isomorphic, in this case, to the cotangent bundle.
The space of
\emph{spiked\ instantons} is the further generalization, describing
the low-energy modes of the open strings stretched between the $D(-1)$-instantons and six stacks of $D3$-branes spanning the coordinate
complex $2$-planes in ${\BC}^{4}$, a local model of the maximal 
number of complex surfaces intersecting at a point in a Calabi-Yau fourfold.

In the next section we recall the relevant details about the {\uwave{BPS}} side of the BPS/CFT correspondence, the supersymmetric partition functions of ${\CalN}=2$ theories. In this paper we discuss the bulk partition functions, in the companion papers \cite{Nekrasov:2015is}, \cite{Nekrasov:2015ii} we study the theories with defects. We also give a rough definition of the ${\x}$ and ${\y}$ observables, and some physics behind them. 
 In the section $\bf 3$ we review quiver gauge theories with unitary gauge groups, which are superconformal in the ultraviolet. The section $\bf 4$ gives the mathematical expression for the integrals over instanton moduli spaces. The path integrals in the quiver gauge theories under consideration, with and without defects, reduce to those finite dimensional integrals by localization. The section $\bf 5$ defines the $\y$-observables in gauge theory, both in the physical theory and in the mathematical problem of integration over the instanton moduli. The section $\bf 6$ introduces informally the $\x$-observables, the $qq$-characters, and formulates the main theorem. The section $\bf 7$ presents the examples of $qq$-characters. The section $\bf 8$ defines the $qq$-characters rigorously, by explicit formulas.

{}
\subsec{Acknowledgements}
 I have greatly benefited from discussions with H.~Nakajima and  A.~Okounkov. The suggestion of V.~Pestun that the results of \cite{Nekrasov:2012xe}, \cite{Nekrasov:2013xda} must generalize to the case of the general $\Omega$-deformaton 
 was instrumental in pursuing the constructions presented below.  Special thanks are to O.~Tsymbalyuk who read the preliminary versions of this manuscript and suggested lots of improvements. Part of the work was done while the author visited the Euler International Mathematical Institute in Saint-Petersburg and the Imperial College London. 
 
Research was supported in part by the NSF grant PHY 1404446.

The results of the paper, notably the formulae for the $qq$-characters and their consequences, the equations on the gauge theory correlation functions, were reported at \footnote{various conferences in 2013-2015:\\
$\bullet$ ``Facets of Integrability: Random Patterns, Stochastic Processes, Hydrodynamics, Gauge Theories and Condensed Matter Systems'', workshop at the SCGP, Jan 21-27, 2013\\
$\bullet$ Gelfand Centennial Conference:
A View of 21st Century Mathematics, MIT, Sept 2013\\
$\bullet$ ITEP conference in honor of the 100-th anniversary of Isaac Pomeranchuk, June 2013\\
$\bullet$ MaximFest, IHES, June 2013\\
$\bullet$ Strings'2014, Princeton, June 2014\\
$\bullet$ `Frontiers in Field and String Theory', Yerevan Physics Institute, Sept 2014\\
$\bullet$ ``Gauged sigma-models in two dimensions'', workshop at the SCGP, Nov 3-7, 2014 \\
$\bullet$ ``Wall Crossing, Quantum Integrable Systems, and TQFT'', Nov 17-21, 2014\\
$\bullet$ ``Recent Progress in String Theory and Mirror Symmetry'', FRG workshop at Brandeis, Mar 6-7, 2015\\
$\bullet$ ``Resurgence and localization in string theory and quantum field theory'', workshop at the SCGP, Mar 16-20, 2015\\
$\bullet$ ``Algebraic geometry and physics'', workshop at the Euler Mathematics Institute, Saint-Petersburg, May 2015\\
seminars in 2013-2015:}. The preliminary version of this paper was published under the title ``Non-Perturbative Schwinger-Dyson Equations: From BPS/CFT Correspondence to the Novel Symmetries of Quantum Field Theory'' in the proceedings \cite{Gorsky:2014mva} of the ITEP conference (June 2013) in honor of the 100-th anniversary of Isaac Pomeranchuk.  
  It took us a long time to write up all the details of the story. While the paper was being prepared, several publications have appeared with some degree of overlap. The paper  
  \cite{Kanno:2013aha} supports  the validity of our main theorem in the case of the $A$-type quivers. The papers \cite{Tong:2014cha}, \cite{Tong:2014yna} contain some discussion of the $(0,4)$-sigma model on our moduli space ${\mM} (n,{\wt},k)$ (for $\zeta =0$). The paper \cite{Bullimore:2014awa} contains the first few instanton checks of some of the results of our paper (for the ${\hat A}_{0}$ theory). The paper \cite{Gaiotto:2014ina} studies the codimension defects using the superconformal index and sphere partition functions, and the RG flows from vortex constructions, also at the level of the first few instanton checks. The paper \cite{Bourgine:2015szm} discusses the algebra of our $Y$-observables in the $A$-type quiver theories. 

 \vfill\eject
\secc{ The\ BPS/CFT\ correspondence}

 We start by briefly reviewing the BPS/CFT correspondence \cite{Nekrasov:2004sem} between supersymmetric field theories with eight supercharges in four, five, and six dimensions, and conformal and integrable theories in two dimensions. It is based on the observation that
 the supersymmetric partition functions
  \cite{Nekrasov:2002qd} are the remarkable special functions, which generalize all the known special functions given by the periods, matrix integrals, matrix elements of group, Kac-Moody,  and quantum group representations etc. \cite{Nekrasov:2002qd, Losev:2003py, Nekrasov:2003rj}. 
 The particular implementations of this correspondence are well-known under the names of the AGT conjecture \cite{Alday:2009aq, Wyllard:2009hg}, and the Bethe/gauge correspondence \cite{Nekrasov:2009zz, Nekrasov:2009rc} (see \cite{Moore:1997dj, Gerasimov:2006zt, Gerasimov:2007ap} for the prior work). For details the interested reader may consult the references in, e.g. \cite{Nekrasov:2013xda}.

 \subsec{ ${\CalN}=2$ partition functions}
 
 For the definition and some details see \cite{Nekrasov:2002qd, Nekrasov:2003rj}. 
The supersymmetric partition function of ${\CalN}=2$ theory
 \beq
 {\CalZ}( {\ba}; {\bmt}; {\btt}; {\ept}) =  {\CalZ}^{\mathrm{tree}}( {\ba}; {\bmt}; {\btt}; {\ept})\ {\CalZ}^{\mathrm{1-loop}}( {\ba}; {\bmt}; {\ept}) \ {\CalZ}^{\mathrm{inst}}( {\ba}; {\bmt}; {\bqt}; {\ept}) 
 \label{eq:czfun}
 \eeq
depends on the vacuum expectation value
${\ba}$ of the adjoint Higgs field in the vector multiplet, it belongs to the complexified Cartan subalgebra of the gauge group of the theory, the set ${\bmt}$ of complex masses of the matter multiplets, and the set ${\btt}$ of the complexified gauge couplings, 
\[ {\tau} = \frac{\vartheta}{2\pi} + \frac{4\pi \ii}{g^{2}}\, , \]
one per simple gauge group factor (we shall not discuss the issue of $SU(n)$ versus $U(n)$ gauge factors in this paper). 
We denote by $\bqt$ the set of the exponentiated couplings, the instanton factors,
\[ {\qe}   = {\exp} \, 2\pi {\ii}{{\boldsymbol\tau}} \]
The non-perturbative factor ${\CalZ}^{\mathrm{inst}}( {\ba}; {\bmt}; {\bqt}; {\ept})$ in \eqref{eq:czfun} has the $\bqt$-expansion, for small $| {\bqt} |$:
\beq
{\CalZ}^{\mathrm{inst}}( {\ba}; {\bmt}; {\bqt}; {\ept})  = \sum_{\bkt} \ {\bqt}^{\bkt} \ {\CalZ}_{\bkt}   ( {\ba}; {\bmt}; {\ept}) 
\label{eq:czkfun}
\eeq
Finally, ${\ept} = ({\ve}_{1}, {\ve}_{2}) \in {\BC}^{2}$ are the complex parameters of the $\Omega$-deformation of the theory \cite{Nekrasov:2002qd}.

\subsubsec{ Asymptotics of partition functions}

{}The function
\eqref{eq:czfun} contains non-trivial information about the theory. For example, the asymptotics at ${\ept} \to (0,0)$, for {\it generic} ${\ba}$, produces the prepotential \cite{Seiberg:1994rs, Seiberg:1994aj} of the low-energy effective action of the theory:
\begin{equation}
 {\CalZ}( {\ba}; {\bmt}; {\btt}; {\ept}) \sim 
 {\exp} \, \frac{1}{{\ve}_{1}{\ve}_{2}} {\CalF}( {\ba}; {\bmt}; {\btt}) + {\rm less}\ {\rm singular}\ {\rm in}\ {\ve}_{1}, {\ve}_{2}.
 \label{eq:prep}
 \end{equation}
 the low-energy effective action being given by the superspace integral
 \begin{equation}
 S^{\rm eff} = \int_{{\BR}^{4|4}} d^{4}x d^{4}{\vartheta} \ 
 {\CalF}( {\ba} + {\vartheta} {\psi} + {\vartheta}{\vartheta} F^{-} + {\ldots}; {\bmt}; {\btt}) 
 \label{eq:effaci}
 \end{equation}
 The prepotential ${\CalF}( {\ba}; {\bmt}; {\btt})$ as a function of $\ba$ determines the special geometry \cite{Freed:1999}
 of the moduli space ${\CalM}^{\rm vector}$
 of Coulomb vacua:
 \begin{equation}
\boldsymbol{d} \left( \begin{matrix}  
{\ba} \\
  \\
 \frac{{\partial}{\CalF}}{{\partial}{\ba}} \\
\end{matrix} \right)
= {\rm periods} \ {\rm of}\ {\varpi}^{\BC} 
\label{eq:specgeom}
\end{equation}
along the $1$-cycles on the abelian variety $A_{b}$, the fiber $p^{-1}(b)$
of the Lagrangian projection 
\begin{equation}
p: {\Pf} \longrightarrow {\CalM}^{\rm vector}
\label{eq:pcpcm}
\end{equation}
of a complex symplectic manifold $({\Pf},{\varpi}^{\BC})$, the moduli space
of vacua of the same gauge theory, compactified on a circle 
\cite{Seiberg:1996nz}. The manifold ${\Pf}$ is actually the phase space of an algebraic integrable system \cite{Donagi:1995cf}. The first example of this relation, the periodic Toda chain for the $SU(2)$ pure super-Yang-Mills theory, was found in \cite{Gorsky:1995zq}.
The asymptotics of \eqref{eq:czfun} 
at ${\ve}_{2} \to 0$ with ${\ve}_{1} = {\hbar}$  fixed, for {\it generic} $\ba$, gives the effective twisted
superpotential 
\begin{equation}
 {\CalZ}( {\ba}; {\bmt}; {\btt}; ({\hbar}, {\ve}_{2})) \sim 
 {\exp} \ \frac{1}{{\ve}_{2}} {\CalW}( {\ba}; {\bmt}; {\btt}; {\hbar}) + {\rm less}\ {\rm singular}\ {\rm in}\ {\ve}_{2}, \qquad {\ve}_{2} \to 0
 \label{eq:supp}
 \end{equation}
 of a two dimensional effective theory. This function plays an important role in quantization of
 the symplectic manifold $\Pf$ and the Bethe/gauge correspondence \cite{Nekrasov:2009zz, Nekrasov:2009rc, Nekrasov:2010ka, Nekrasov:2011bc, Nekrasov:2013xda}. 

The asymptotics \eqref{eq:supp}, \eqref{eq:prep} are modified in an intricate way when the {\it genericity} assumption on $\ba$ is dropped. The interesting non-generic points are where $\ba$
and $\ve_1$ (with ${\ve}_{2} \to 0$) are in some integral relation. The behavior near such special points and its r\^ole in the Bethe/gauge correspondence will be discussed elsewhere.

\subsec{Defect operators and lower-dimensional theories}

In addition to the $\CalZ$-functions, which are the partition functions of the theory on ${\BR}^{4}$,  in \cite{Nekrasov:2015is} we also consider the partition functions $\bf\Psi$ of  the same gauge theory in the presence of defects preserving some fraction of supersymmetry. These defects could be point-like, or localized along surfaces. We derive the differential equations, which can be used to relate the theory with a surface operator to the theory without one. As a by-product we get the explicit realization of the Bethe/gauge correspondence \cite{Nekrasov:2009zz, Nekrasov:2009rc} with an additional bonus: the gauge theory produces not only the equations, characterizing the spectrum of the quantum integrable system, but also gives an expression for the common eigenfunction of the full set of quantum integrals of motion. In particular, we shall show \cite{Nekrasov:2015iii} in that a class of surface defect operators in the ${\CalN}=2$ theory with $U(n)$ gauge group and $2n$ fundamental hypermultiplets solves the Knizhnik-Zamolodchikov (KZ) equation, which is obeyed by the $4$-point conformal block of the $SU(n)$ Wess-Zumino-Witten theory on the sphere; that a class of surface defect operators 
in the ${\CalN}=2^{*}$ theory with the $U(n)$ gauge group solves the Knizhnik-Zamolodchikov-Bernard (KZB) equation,
which is obeyed by the $1$-point conformal block of the $SU(n)$ Wess-Zumino-Witten theory on the torus. In the ${\ve}_{2} \to 0$ limit these surface operators become the eigenfunctions of Gaudin Hamiltonian and the elliptic Calogero-Moser 
Schr\"odinger equation, respectively, in agreement with the conjectures in \cite{Nekrasov:2009rc}, \cite{Alday:2010vg} and earlier ideas. 

In addition to the codimension two defects in four or five dimensional theories we can also consider lower dimensional theories. For example, gauge theory on the $AdS_3$ space with appropriate boundary conditions can be viewed as the $U(1)$-orbifold of a four dimensional superconformal gauge theory. We study these cases in \cite{Nekrasov:2015ii}.  

\subsec{The $\y$- and $\x$-observables}

The main tools in our analysis are the gauge invariant observables 
${\y}_{\ib}(x)$ and ${\x}_{\ib}(x)$, defined for each simple factor $U(n_{\ib})$ of the gauge group. Here $\ib$ belongs to the set $\Ver$, which in our story is the set of vertices of a quiver. The ${\y}_{\ib}(x)$ are the suitable generalizations of the characteristic polynomials of the adjoint Higgs field,
\beq {\y}_{\ib}(x)  \sim {\Det} ( x - {\Phi}_{\ib} ) \ . \label{eq:charpol}
\eeq 
They are the gauge theory analogues of the matrix model resolvents \eqref{eq:resolve}. As a function of $x$, each operator ${\y}_{\ib}(x)$ has singularities, i.e. the relation \eqref{eq:charpol} is modified. This modification is  due to the mixing between the adjoint scalar and gluinos, e.g.   
\beq
{\Phi}_{\ib} \sim {\Phi}_{\ib}^{\rm cl} \ +\  {\ve}_{\alpha\beta} {\ve}_{j' j''} \,( d_{A_{\ib}}^{*}d_{A_{\ib}} )^{-1} \, [{\psi}^{\alpha j'}_{\ib}, {\psi}^{\beta j''}_{\ib} ] 
\eeq 
which have zero modes in the presence of gauge instantons, leading to the poles in $x$, 
in a way we make much more precise below. The ${\x}_{\ib}(x)$ are composite operators, built out of ${\y}$'s. They are Laurent polynomials or series in ${\y}_{\ib}(x)$'s with shifted arguments and their derivatives. They are the analogues of the matrix model operators \eqref{eq:tdseq}. Their main property is the absence of singularities in $\vev{{\x}_{\ib}(x)}$ for finite $x$, similarly to the matrix model case, cf. \eqref{eq:dseq}. In the weak coupling limit ${\x}_{\ib}(x) \to {\y}_{\ib}(x) \to $ \eqref{eq:charpol}. We define also the observables ${\x}_{{\bw}, {\bnu}}(x)$, labelled by the string ${\bw} = ({\wt}_{\ib})_{{\ib} \in \Ver} \in {\BZ}_{\geq 0}^{\Ver}$ of non-negative integers and the string ${\bnu} = ({\vec\nu}_{\ib} )_{{\ib} \in \Ver}$, ${\vec\nu}_{\ib} \in {\BC}^{{\wt}_{\ib}}$ of complex numbers. Here $\Ver$ is the set of simple gauge group factors. The expectation values $\vev{{\x}_{{\bw}}(x)}$ also have no singularities, while in the limit of zero gauge couplings ${\x}_{{\bw}, {\bnu}}(x)$ approach 
\[ \prod_{\ib} \prod_{f=1}^{{\wt}_{\ib}}\ {\y}_{\ib}( x + {\nu}_{{\ib}, f}) . \] 

We call ${\x}_{{\bw}, {\bnu}}(x)$ the {\it $qq$-characters}. For ${\bw} = ( {\delta}_{{\ib}, {\jb}} )_{{\jb} \in \Ver}$, and ${\bnu} = \underline{0}$,  we call ${\x}_{{\bw}, {\bnu}}(x) =: {\x}_{\ib}(x)$ the {\it fundamental $qq$-characters}. 
The reason for and the meaning of the terms will hopefully become clear
in the coming chapters.

\subsec{ The physics of $\x$-observables}

The  $\x$-observables can be interpreted as the partition functions of the auxiliary gauge theory living on a space, transverse to the space-time of our gauge theory, the ``physical space-time''. The auxiliary theory has massive degrees of freedom coupled to the degrees of freedom of our gauge theory at some point $p \in {\BR}^4$ in the physical space-time, so that integrating them out induces an operator ${\CalO}_{\bw, \bnu, x} (p)$ inserted at $p$. The data $\bw$ is the choice of the auxiliary gauge theory while $\bnu$ and $x$ fix its vacuum. 

The dimensionality of the auxiliary gauge theory is a somewhat subtle issue. Most of the theories we study in the paper, such as the ${\CalN}=2$ quiver theories with affine quivers have the $\x$-observables which come from a four dimensional auxiliary theory. The theories with finite quivers can be viewed as a sector in the auxiliary four dimensional theory corresponding to an affine quiver. In fact, the finite quivers of $A$-type can be realized as a subsector of the ${\hat A}_{\infty}$ theory, which corresponds to the orbifold of ${\BC}^{2}$ by $U(1)$. Gauge theory living on such an orbifold can be viewed either as a three dimensional theory on a manifold with corners (or, conformally, on the $AdS_3$), or, for the purposes of supersymmetric partition functions, as a two dimensional sigma model \cite{Nekrasov:2015ii}.

Here is a sketch of the string theory construction. Consider IIB string theory on the ten-dimensional manifold of the form ${\BR}^2_{\phi} \times {\mN} \times {\mW}/{\Gamma}$, where
${\mN}= {\BR}^{4}$, ${\mW} = {\BR}^{4}$, and $\Gamma$ a finite subgroup of $SU(2)$ (see \cite{Johnson:1996py} for the discussion of IIB string theory on ALE spaces). 

Recall \cite{Douglas:1996sw} that ${\CalN}=2$ quiver gauge theories with affine $A,D,E$ quivers can be realized as the low energy limit of the theory on a stack of $n$ $D3$-branes located at ${\varphi} \times {\mN} \times 0$, with ${\varphi} \in {\BR}^2_{\phi}$ a point, and $0$ the tip of the ${\mW}/{\Gamma}$ singularity, with ${\Gamma}$ being the discrete subgroup of $SU(2)$, McKay dual \cite{McKay:1980} to the corresponding $A, D,E$ simple Lie group.   

{}
\bigskip
\bigskip
\hbox{\vbox{\hbox{The worldvolume of these $D3$ branes}
\hbox{is a copy of ${\mN}$.  Let us now add a stack}
\hbox{of ${\wt}$ $D3$-branes located at 
$x \times {\bf 0} \times {\mW}/{\Gamma}$,}
\hbox{with the worldvolume being a copy of ${\mW}/{\Gamma}$.}
\hbox{Here $x \in {\BR}^2_{\phi} \approx {\BC}$ is a complex number,}
\hbox{and ${\bf 0} \in {\mN}$ is a fixed point.}
\hbox{Here is the picture:}} \vbox{\hbox{\picit{5}{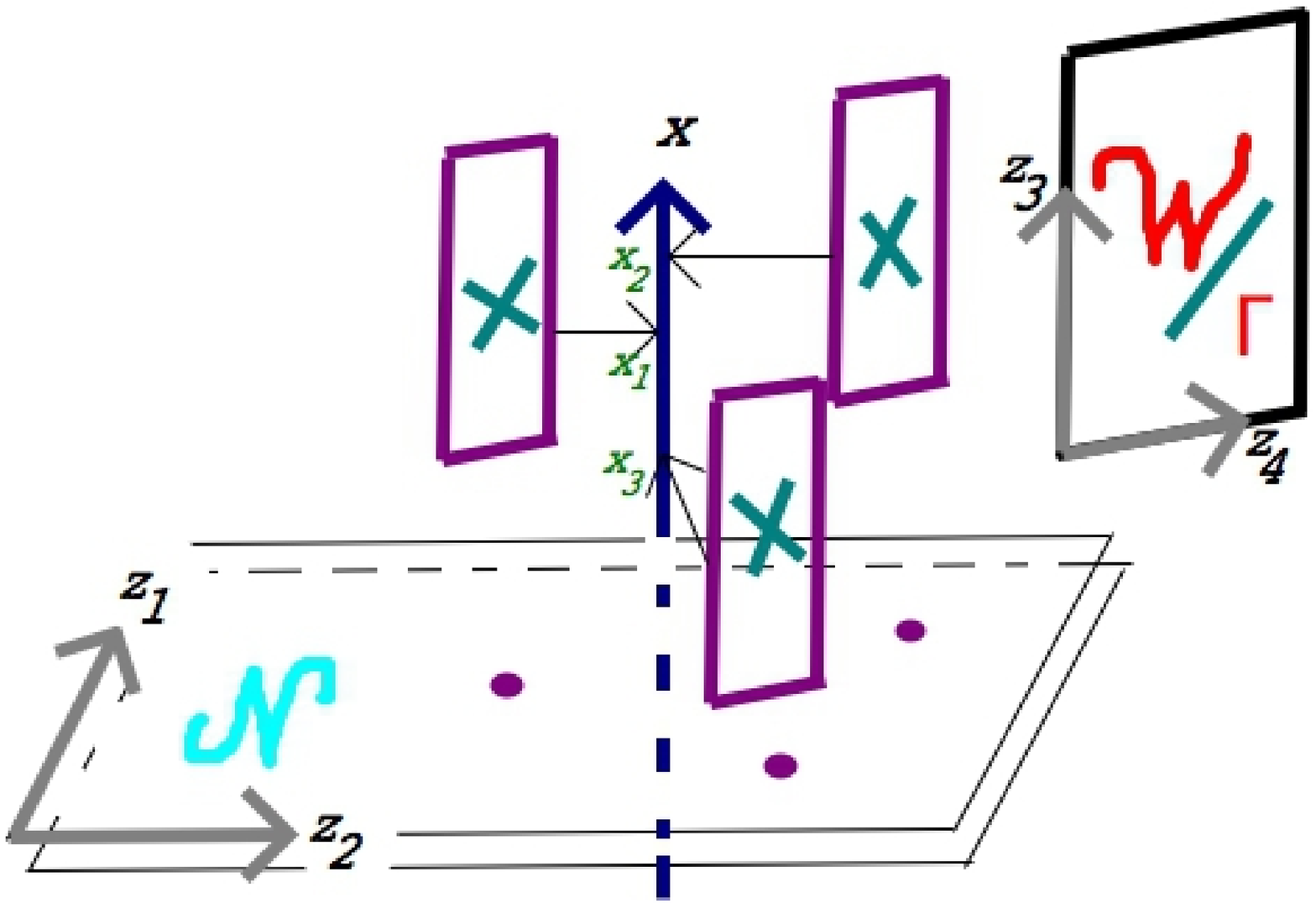}}
\hbox{\qquad\qquad\bf Fig.4}}}
\bigskip
\bigskip

{}The low energy configurations in this system of  two orthogonal stacks  of $D3$ branes are labelled by the separation of branes along ${\BR}^2_{\phi}$ encoded in $\bnu$, and by the choice of flat $U({\wt})$ connection at infinity ${\BS}^{3}/{\Gamma}$ of ${\mW}/{\Gamma}$,  which is equivalent to the choice of
the string $\bw$.  

\bigskip
\bigskip
 \centerline{\picit{5}{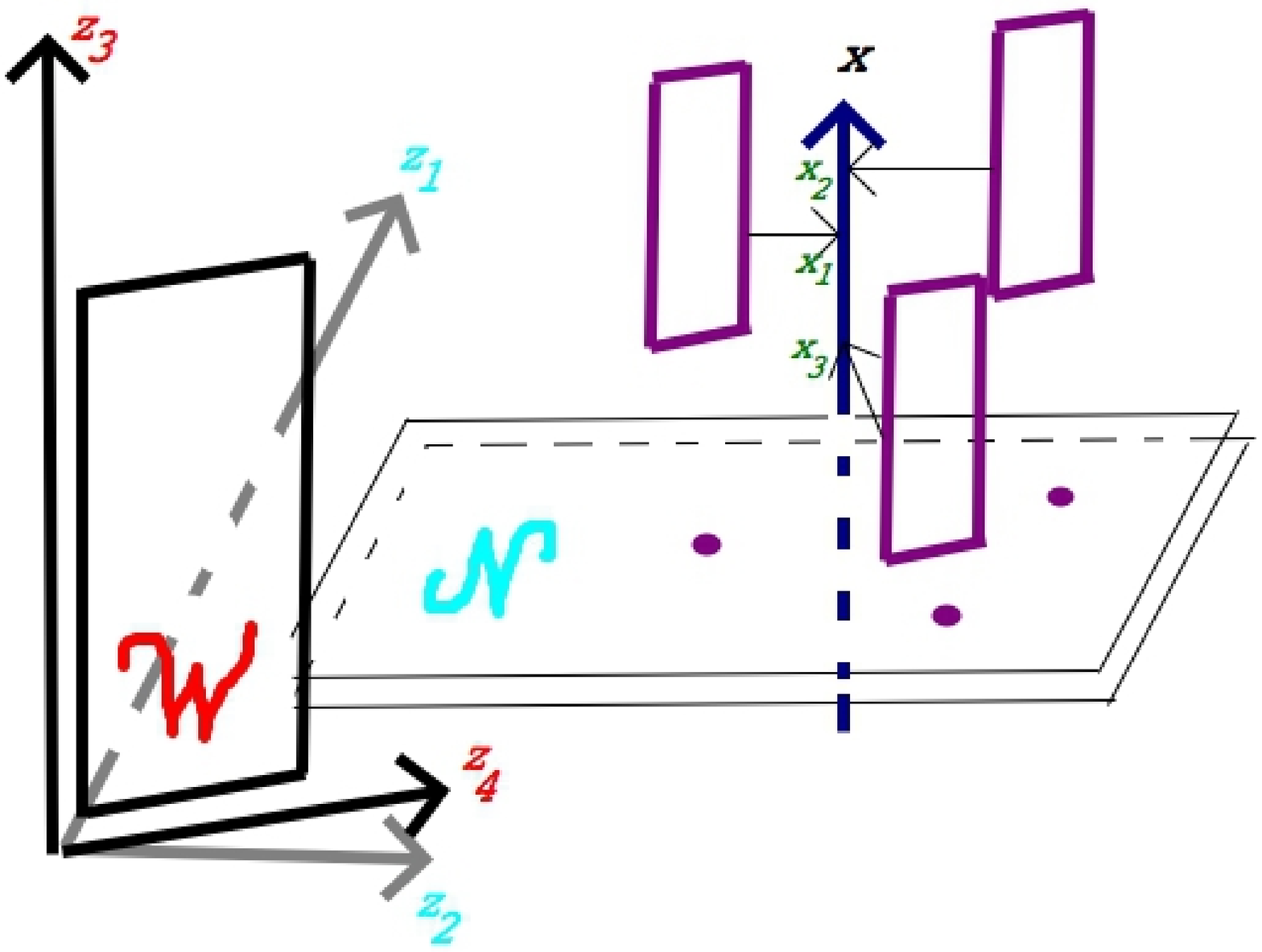}}
\bigskip
\vbox{\centerline{\bf Fig.5}
\centerline{\tiny Gauge theory with ${\x}$-observables in the brane picture}}
\bigskip
\bigskip

The $qq$-character ${\x}_{\bw, {\bnu}}(x)$ is simply the observable in the original theory on the stack of $N$ $D3$-branes living along $\mN$, which is obtained by integrating out the degrees of freedom on the transversal $D3$-branes, in the vacuum corresponding to the particular asymptotic flat connection $\bw$ and the vacuum expectation values $\bnu$ of the scalars in the vector multiplets living on ${\mW}/{\Gamma}$. 

\bigskip
\bigskip
 \centerline{\picit{5}{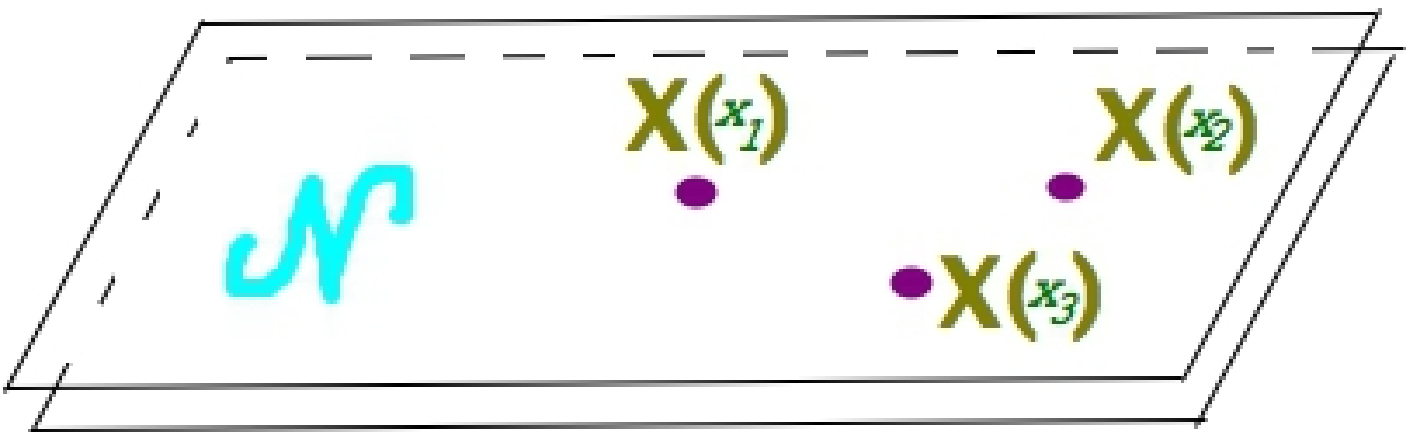}}
\bigskip
\vbox{\centerline{\bf Fig.6}
\centerline{\tiny Gauge theory with the observables ${\x}(x_{1}), {\x}(x_{2}), {\x}(x_{3})$}}
\bigskip
\bigskip

The next piece of our construction is the $\Omega$-deformation using a subgroup of the spin cover of the group $Spin(8)$ of rotations of ${\mW} \times {\mN}$ which commutes with $\Gamma$, preserves the configuration of branes, and some supersymmetry. This subgroup generically has rank two, which enhances to three for $\Gamma$ of $A$ type. The parameters of the $\Omega$-deformation are generically two complex numbers ${\ept} = ({\ve}_{1}, {\ve}_{2})$, and for $\Gamma$ of $A$-type there is an additional parameter $m$. This parameter is the mass of the adjoint hypermultiplet in the ${\hat A}_{0}$-case, and the sum of   masses of all $k+1$ bi-fundamental hypermultiplets in the ${\hat A}_{k}$ case for $k > 0$. 
It is convenient to introduce four $\ve$-parameters, ${\ve}_{a}$, $a = 1, 2, 3,4$, which sum to zero:
\beq
{\ve}_{1} + {\ve}_{2} + {\ve}_{3} + {\ve}_{4} = 0
\label{eq:ve4}
\eeq
so that ${\ve}_{3} = m$, ${\ve}_{4} = - m - {\ve}$, ${\ve} = {\ve}_{1} + {\ve}_{2}$. Together they parametrize the generic
$SU(4)$ $\Omega$-deformation.

The $K$-theoretic and elliptic versions of $qq$-characters correspond to the five- and six-dimensional theories, which are engineered in the analogous fashion, with ${\BR}^2_{\phi}$ replaced by ${\BS}^{1} \times {\BR}^{1}$ and ${\BS}^{1} \times {\BS}^{1}$, respectively.
In the five dimensional case we use IIA string and the D4 branes wrapped on ${\BS}^1$ instead of D3's, in the six dimensional case we are back in the IIB realm with D5 branes wrapped on  ${\BS}^{1} \times {\BS}^{1}$.  

The configuration of $D3$ branes which we described above can be generalized, by considering other orbifolds of the ten dimensional Euclidean space
${\BR}^2_{\phi} \times \mN \times \mW$. For example, the orbifolds ${\BR}^2_{\phi} \times {\mN}/{\Gamma}_{\mN} \times 
{\mW}/{\Gamma}_{\mW}$ with $D3$ branes wrapping $\mN \sqcup \mW$ define the $qq$-characters relevant for the ${\gamma}_{{\Gamma}_{\mW}}$-quiver gauge theory on the ALE space $\widetilde{{\mN}/{\Gamma}_{\mN}}$. The most general orbifold  we could employ is by the subgroup $\Gamma = {\Gamma}_{\mN} \times {\Gamma}_{\Delta} \times {\Gamma}_{\mW}  \subset SU(2)_{\mN, L} \times SU(2)_{\Delta} \times SU(2)_{\mW, L} \subset Spin(4)_{\mN} \times Spin(4)_{\mW} \subset Spin(8)$. We explain the r\^ole of this group and its subgroups in the following chapters. 
Using this construction we also realize various defect operators in various quiver gauge theories on conical spaces.  

The final piece of the construction is turning on the appropriate $B$-field which makes the configurations where the 
$D(-1)$-instantons
are separate from the $D3$-branes non-supersymmetric. 
The $D(-1)$-instantons bound to the two orthogonal stacks of $D3$ branes give rise to what we call the \underline{crossed\ instantons}. We can also study the generalization involving six stacks of $D3$ branes spanning complex
$2$-planes inside ${\BR}^{8} \approx {\BC}^{4}$. The complex coordinate $x$ parametrizes the remaining ${\BR}^{2}
\approx {\BC}^{1}$, which is orthogonal to ${\BC}^{4}$ in the ten dimensional Euclidean space-time of the type IIB
string.

The \uwave{main claim}, i.e. the absence of singularities in $x$ of $\vev{{\x}_{\bw, {\bnu}}(x)}$, is the statement that the combined system of the intersecting stacks of $D3$ branes has no phase transitions and no runaway flat directions at special values of $x$, in the presence of $\Omega$-deformation. 
Mathematically, the argument is the compactness of the moduli space of \emph{crossed} (for two orthogonal stacks
of branes)
and \emph{spiked instantons} (for six stacks of branes), the supersymmetric configurations of the combined system of branes, with the $\Omega$-deformation and appropriate $B$-field turned on. We 
describe the moduli spaces
in \cite{Nekrasov:2015iim}.

One can also apply the orientifold projection (which, unfortunately, would not be consistent with the $B$-field we are using) to arrive at the theory of crossed instantons for the orthogonal and symplectic groups.

\subsec{ Hidden symmetries}

The IIB string theory on ${\BR}^{4}/{\Gamma} \times {\BR}^{1,5}$ contains the non-abelian tensionless strings \cite{Witten:1995zh} with the $A, D, E$ tensor symmetry (it becomes the gauge symmetry of the $A, D, E$ type upon compactification on a circle, i.e. when $\Sigma = {\BS}^{1} \times {\BR}^{1}$). 

In the limit ${\ve}_{1}, {\ve}_{2} \to 0$ our $qq$-characters approach the ordinary characters for the Kac-Moody group built on the quiver (i.e. the affine Lie group ${\hat A}, {\hat D}, {\hat E}$ for affine quivers, and the simple $A, D, E$ groups for the finite quivers), \cite{Nekrasov:2012xe}. 
The non-abelian tensor symmetry seems to be realized on the worldvolume of $D3$ branes by the {\it large field deformations} which lead to the non-perturbative Dyson-Schwinger equations we discuss in this paper. The $qq$-character observables may teach us something important about the nature of the tensor symmetry representation.  
See \cite{Kimura:2015rgi} for the discussion of the $qq$-deformed ${\CalW}$-algebras and their gauge theory
realizations.

\subsec{Some notations.}

\subsubsec{Finite\ sets}  
  
We use the following notations for certain finite sets: 
\beq
\begin{aligned}
& [p]  \equiv \{ 1, 2, \ldots , p \}, \qquad p \in {\BZ}_{+}, \\
& [0,q] \equiv \{ 0, 1, 2, \ldots , q\}, \qquad q \in {\BZ}_{\geq 0} 
\end{aligned}
\eeq
and
\beq
( x_{i} )_{i\in I} \equiv \{ \ x_{i} \ | \ i \in I \ \}
\eeq
Also for the set $(z_{i})_{i \in I}$ of complex numbers indexed by the set $I$ we use the notation
\beq
 z_{I} = \prod_{i \in I}\, z_{i}
\label{eq:zmom} 
\eeq
for their product. This is consistent with the notation \eqref{eq:qsprod}.

\subsubsec{Roots of unity}

\beq
\ii = \sqrt{-1}\ ,
\eeq 
and
\beq
{\varpi}_{p} = {\exp}\, \frac{2\pi\ii}{p}
\label{eq:primrp}
\eeq
so that ${\ii} = {\varpi}_{4}, - {\ii} = {\varpi}_{4}^{3}$.

\subsubsec{ Parameters of $\Omega$-deformations} 

In four dimensions, we have two parameters ${\ept}  = ({\ve}_{1}, {\ve}_{2}) \in {\BC}^{2}$. 
We also use  their sum 
\beq
{\ve} = {\ve}_{1} + {\ve}_{2}\, ,\quad  \label{eq:eee}
\eeq 
their exponents
 \beq
 q_{1} = e^{{\beta}{\ve}_{1}}\, , \ q_{2} = e^{{\beta}{\ve}_{2}}\, , \ q = e^{{\beta}{\ve}} = q_{1} q_{2}\, , 
  \label{eq:qeps}
  \eeq
and the virtual characters
\beq
P = (1-q_{1})(1-q_{2})\, , \qquad  P^{*} = (1-q_{1}^{-1})(1-q_{2}^{-1})
\label{eq:koszul}
\eeq  
The parameter $\beta$ is the circumference of the circle of compactification of a $4+1$ dimensional supersymmetric theory. 
  
 {}In the context of the BPS/CFT correspondence, the parameter 
  \beq
 b^2 = {\ve}_{1}/{\ve}_{2} \label{eq:bpar}
 \eeq 
 is useful. 
 
 {}In eight dimensions, or for the theories in four dimensions with adjoint matter, it will be useful to have four parameters 
 \beq
 {\bar\ve} = ({\ept}, {\tilde\ept}) \equiv ( {\ve}_{1}, {\ve}_{2}, {\ve}_{3}, {\ve}_{4}) \in {\BC}^{4}\ , 
 \label{eq:4ee}
 \eeq
 which sum to zero:
 \beq
{\ve}_{3} + {\ve}_{4} = - {\ve}
\label{eq:ee4}
 \eeq
 We denote ${\bf 4} = \{  1, 2, 3, 4\}$ and
 \beq
 \begin{aligned}
&  q_{a} = e^{\beta\ve_{a}}, \qquad
 P_{a} = ( 1 - q_{a} ),  \qquad a \in {\bf 4}\\
 &  q_{a}^{*} = q_{a}^{-1}, \qquad
 P_{a}^{*} = ( 1 - q_{a}^{-1} ), \qquad  \\
 \label{eq:qapab}
 \end{aligned}
 \eeq
 For any subset $S \subset {\bf 4}$ we define ${\bar S} = {\bf 4} \backslash S$, and:
 \beq
 q_{S} = \prod_{a \in S} q_{a}, \quad q_{S}^{*} = q_{\bar S} , \qquad
 P_{S} = \prod_{a\in S} P_{a}, \quad P_{S}^{*} = (-1)^{|S|} \, q_{\bar S}\, P_{S}\label{eq:qsprod}\eeq
 so that $q_{\emptyset} = q_{\bf 4} = 1$.  
 
 \subsubsec{Chern Characters and Euler Classes}

Let $E \to {\Xf}$ be the rank $m = {\rm rk}E$ complex vector bundle, and $c_{i}(E) \in H^{2i}({\Xf}, {\BZ})$ the corresponding Chern classes. Then $e(E) = c_{\rm top}(E) = c_{m}(E)$ is the Euler class of $E$, and
\beq
{\ep}_{z}(E) = e(E) + z c_{m-1}(E) + z^{2} c_{m-2}(E) + \ldots z^{m} 
\label{eq:epze}
\eeq
is the Chern polynomial. 

\subsubsec{Weights from characters}

For a virtual representation $R$ of a Lie group $\Hf$, 
the {\it weights} $w$ are computed using its character as follows:
\begin{multline}
 \quad R = R^{+} \ominus R^{-}\, , \\
 {\Tr}_{R^{\pm}} ( e^{\theta} )  = \sum_{w \in W(R^{\pm})} e^{w ({\theta})} , \qquad
 {\Tr}_{R} ( e^{\theta} ) = {\Tr}_{R^{+}} ( e^{\theta} ) - {\Tr}_{R^{-}} ( e^{\theta} ) \, , 
\label{eq:weig}
\end{multline}
where $R^{\pm}$ are the vector spaces, the actual representations of $\Hf$, and
$W(R^{\pm})$ are the sets of the corresponding weights (the linear functions on ${\rm Lie}({\Hf})$ which take integer values on the root lattice).

\subsubsec{Chern polynomials from characters}

We denote by ${\ep}_{\theta}(R)$
the following Weyl-invariant rational function on the Cartan subalgebra ${\mathfrak h}_{\BC}$ of ${\rm Lie}({\Hf}_{\BC})$:
\beq
{\ep}_{\theta} (R) = \frac{\prod_{w \in W(R^{-})} w ({\theta})}{\prod_{w \in W(R^{+})} w ({\theta})}, \quad  {\theta} \in {\mathfrak h}_{\BC},
\label{eq:irf}
\eeq
where  the weights $w({\theta})$ are given by \eqref{eq:weig}.  Note that in \eqref{eq:irf} the weights of $R^{+}$ are in the denominator. It follows from the definition \eqref{eq:weig} that 
\beq
{\ep}_{\theta}(R){\ep}_{\theta} (-R)=1\, , 
\label{eq:emerge}
\eeq
 if $R$ does not contain $\pm 1$ as a summand, and, more generally:
\beq
{\ep}_{\theta} (R_{1} \oplus R_{2})  = {\ep}_{\theta} (R_{1}) {\ep}_{\theta} (R_{2})\ . 
\label{eq:addep} \eeq
Also,
 \beq
 {\ep}_{\theta} (R^{*}) = {\ep}_{-\theta}(R) =  (-1)^{{\dt}_{R}} \, {\ep}_{\theta} (R)
 \label{eq:eprs}
 \eeq
where
\beq
{\dt}_{R} = {\dim}_{\BC}R^{+} - {\dim}_{\BC}R^{-}
\eeq
\subsubsec{Chern functions from characters}

In our story we occasionally encounter the generalizations of the formulas like \eqref{eq:irf} where the representations $R^{\pm}$ are infinite dimensional. In order for \eqref{eq:irf} to make sense in this case we use the 
\subsubsec{$\zeta$-function regularization}
The map from the character \eqref{eq:weig} to $\ep_{\theta}(R)$ can be given in the integral form:
\beq
{\ep}_{\theta}(R) = {\exp}\, \frac{d}{ds} \Biggr\vert_{s=0} \frac{{\Lambda}^{s}}{{\Gamma}(s)}
\int_{0}^{\infty} \frac{d\beta}{\beta}{\beta}^{s} {\Tr}_{R} e^{\beta\theta}
\label{eq:conv}
\eeq
where one chooses $s$ and $\theta$ with the real part in the appropriate domain to ensure the convergence of the integral in the right hand side of \eqref{eq:conv}, and then analytically continues. For finite dimensional  $R$ the result does not depend on $\Lambda$. For infinite dimensional $R$ the left hand side of \eqref{eq:conv} really is defined by the right hand side. More precisely, we assume $R$ is graded,   
\beq
R = \bigoplus_{n = 0}^{\infty} \, R_{n}\, , 
\eeq
with the finite dimensional virtual subspaces $R_n = R_{n}^{+} - R_{n}^{-}$, ${\rm dim}R_{n}^{\pm} < \infty$, whose superdimensions grow at most polynomially with $n$.  We define
\beq
{\ep}_{\theta}(R) = {\rm Lim}_{t \to +0} \ {\exp}\, \frac{d}{ds} \Biggr\vert_{s=0} \frac{{\Lambda}^{s}}{{\Gamma}(s)}
\int_{0}^{\infty} \frac{d\beta}{\beta}{\beta}^{s} \sum_{n=0}^{\infty} e^{-{\beta} t (n+1)}\, {\Tr}_{R_{n}} e^{\beta\theta}
\label{eq:convtx}
\eeq
where the integral in the right hand side converges for sufficiently large ${\Re}(t), {\Re}(s)$, defining an analytic function, whose asymptotics near $s, t = 0$ defines the left hand side. 

For example, take $R = {\BC}[z_{1}, z_{2}, z_{3}, \ldots, z_{\delta}]$ with ${\Hf} = \left( {\BC}^{\times}  \right)^{\delta + 1}$ acting via:
\beq
t : f \mapsto f^{t}, \qquad f^{t}(z) = t_{0} f ( t_{1}^{-1}z_{1}, t_{2}^{-1} z_{2}, \ldots , t_{\delta}^{-1} z_{\delta}) 
\eeq
The character ${\Tr}_{R} e^{\beta\theta}$ 
\beq
{\Tr}_{R} e^{\beta\theta} = e^{\beta \theta_0} \prod_{i=1}^{\delta} \frac{1}{1  - e^{\beta\theta_{i}}}
\label{eq:polfun}\eeq
and the refined character (where we define the grading by the polynomial degree)
\beq
\sum_{n=0}^{\infty} e^{-{\beta} t (n+1)}\, {\Tr}_{R_{n}} e^{\beta\theta} = e^{{\beta}( {\theta}_{0} - t )}  \prod_{i=1}^{\delta} \frac{1}{1  - e^{\beta ( {\theta}_{i}- t)}}
\eeq
are easy to compute. 

The integral on the right hand side of \eqref{eq:convtx} absolutely converges for ${\Re} t > {\rm max}_{i=0}^{\delta} {\Re} {\theta}_{i} $ and ${\Re} s > {\delta}$. 
\subsubsec{Asymptotics of the $\zeta$-regularized ${\ep}_{\theta}$'s}

Let $R$ be a virtual representation, $\theta \in {\mathfrak h}_{\BC}$, and  ${\chi}_{R,{\theta}} ({\beta}) = {\Tr}_{R} e^{\beta\theta}$. For $\beta \to 0$ the function ${\chi}_{R,{\theta}} ({\beta})$ has an expansion:
\beq
{\chi}_{R,{\theta}} ({\beta}) = \sum_{n= - {\delta}_{R}}^{+\infty} \,{\beta}^{n}\, {\chi}_{R,{\theta},n}
\label{eq:chirthn}
\eeq
where ${\chi}_{R,{\theta},n}$ is a homogeneous rational function of $\theta$ of degree $n$, obeying:
\beq
{\chi}_{R^{*}, {\theta}, n} = \, {\chi}_{R, - {\theta}, n}\,  = \, (-1)^{n} {\chi}_{R, \theta, n}
\label{eq:refl}
\eeq
We are interested in the large $x$ asymptotics of
\begin{multline}
{\ep}_{-x+ \theta}(R) \equiv {\exp}\ \frac{d}{ds} \Biggr\vert_{s=0} \, \frac{{\Lambda}^{s}}{{\Gamma}(s)}
\int_{0}^{\infty} \frac{d\beta}{\beta}{\beta}^{s} e^{-{\beta}x} {\Tr}_{R} e^{\beta\theta} \sim \\
\sim {\exp} \, \left( \, - \sum_{n=-{\delta}_{R}}^{0} \, {\chi}_{R,{\theta},n}\,  \frac{(-x)^{n}}{(-n)!}  \left( {\log} \left( \frac{x}{\Lambda} \right) - 
\sum_{k=1}^{-n} \frac{1}{k} \right) \, \right)  \\
\times \ {\exp}\, \left( \, 
\sum_{n=1}^{\infty}\,  {\chi}_{R,{\theta},n} \, \frac{(n-1)!}{x^{n}} \, \right)
\label{eq:convx}
\end{multline}
(since the variable $x$ shifts the auxiliary variable $t$ used to regularize an infinite trace, we can safely set $t=0$ in \eqref{eq:convx}). 
Thus, 
\beq
{\ep}_{x + \theta}(R^{*}) = {\ep}_{-x + \theta} (R) \ \times \
{\exp} \, \left( \, - {\pi}{\ii} \sum_{n=-{\delta}_{R}}^{0} \, {\chi}_{R,{\theta},n}\,  \frac{(-x)^{n}}{(-n)!}  \, \right)
\label{eq:signinf}
\eeq

\subsubsec{Flips $\leadsto$}

We shall also use a notation, for a virtual representation $R = R_{1} \oplus R_{2}$,
\beq
R \leadsto R_{1} \oplus R_{2}^{*}
\eeq
and similarly for their characters:
\beq
{\rm Tr}_R \leadsto {\rm Tr}_{R_{1}} + {\rm Tr}_{R_{2}}^{*}
\eeq
where we also use the convention
\begin{equation}
 {\chi}^{*} = \sum_{w \in W(R)} e^{-w({\theta})}, \qquad {\rm for} \qquad {\chi} = \sum_{w \in W(R)} e^{w({\theta})}
 \label{eq:conjconv}
 \end{equation}
Sometimes, when the choice of the element ${\theta} \in {\mathfrak h}_{\BC}$  is understood, we denote the trace ${\rm Tr}_{R} (e^{\theta}) $ in the representation $R$ by the same letter $R$.

For example
\beq
( 1 + q^{-1} ) ( 1- q_{1}) \leadsto 1 - q_{1} + q_{1}q_{2} - q_{2} = P
\label{eq:fromqtoi}
\eeq
The multiplicative measures of the finite dimensional virtual representations $R$, given by the products \eqref{eq:irf} of weights $w({\theta})$ and their K-theoretic analogues, given by the products of $2 {\rm sin} \left( \frac{w({\theta})}{2} \right)$ do not change, up to a sign, under the $\leadsto$ modifications:
\beq
{\ep}_{\theta}(R') =(-1)^{d_{R_{2}}}\ {\ep}_{\theta}(R), \qquad R \leadsto R'
\label{eq:modif}
\eeq 
In the infinite dimensional case the multiplicative anomaly  of the measure \eqref{eq:conv} follows from \eqref{eq:signinf}. 
\subsec{Equivariant virtual Chern polynomials}

Let $R$ be a virtual representation as above, and 
\beq
R = \left( \oplus_{w \in W(R^{+})} R_{w}^{+} \right)  \ominus 
\left( \oplus_{w \in W(R^{-})} R_{w}^{-} \right)
\label{eq:rpm}
\eeq
be the corresponding weight decomposition. Let  $E_{w}$, with the weights $w \in W(R^{+}) \cup W(R^{-})$ be some vector bundles over $\Xf$, and
\beq
{\ee} = \left( \oplus_{w \in W(R^{+})} R_{w}^{+} \otimes E_{w^{+}} \right)  \ominus 
\left( \oplus_{w \in W(R^{-})} R_{w}^{-} \otimes E_{w^{-}} \right)
\eeq
be the associated virtual bundle over $\Xf$.  We denote by (cf. \eqref{eq:epze}, \eqref{eq:irf})
\beq
{\ep}_{\theta}({\ee}) = \frac{\prod_{w \in W(R^{-})} {\ep}_{w ({\theta})}(E_{w^{-}})}{\prod_{w \in W(R^{+})} {\ep}_{w ({\theta})} (E_{w^{+}})}
\eeq
the rational function on the Cartan subalgebra ${\mathfrak h}_{\BC}$ with values in $H^{*}({\Xf}, {\BC})$. 
 
\secc{ Supersymmetric\  gauge\ theories}

In this section we go back  to the gauge theory narrative. Our gauge theories are characterized by a quiver diagram. Let us start by reviewing what we mean by them. 

\subsec{Quivers}

A quiver is an oriented graph ${\gamma}$, with the set $\Ver$ of vertices and the set $\Edg$ of oriented edges. We have two maps $s, t : \Edg \longrightarrow \Ver$, sending each edge $e$ to its source $s(e)$ and the target $t(e)$, respectively.

\noindent
\hbox{\vbox{\hbox{We shall also use an unconventional term \emph{arrow}}
\hbox{which is a pair $(e, {\sigma})$, where $e \in \Edg$, ${\sigma} = \pm 1$.}
\hbox{The  set $\Arr = \Edg \times 2^{\Edg}$ of arrows}
\hbox{is equipped with two maps} 
\hbox{${\bar s}, {\bar t}: {\Arr} \to \Ver$, defined by:}
\hbox{\noindent }
\hbox{$\begin{aligned}
&  {\bar s}(e , {\sigma}) = \left\lbrace \begin{matrix} & s( e), \ & \  {\rm if}\quad {\sigma} = +1 \\
& \ & \\
& t(e) , \ & \  {\rm if}\quad {\sigma} = -1 \end{matrix} \right. \\
& \\
&  {\bar t}(e , {\sigma}) = \left\lbrace \begin{matrix} & t( e), \ & \  {\rm if}\quad {\sigma} = +1 \\
& \ & \\
& s(e) , \ & \  {\rm if}\quad {\sigma} = -1 \end{matrix} \right. \\
\label{eq:stmaps}
\end{aligned}
$}}\vbox{\hbox{ 
\picit{5}{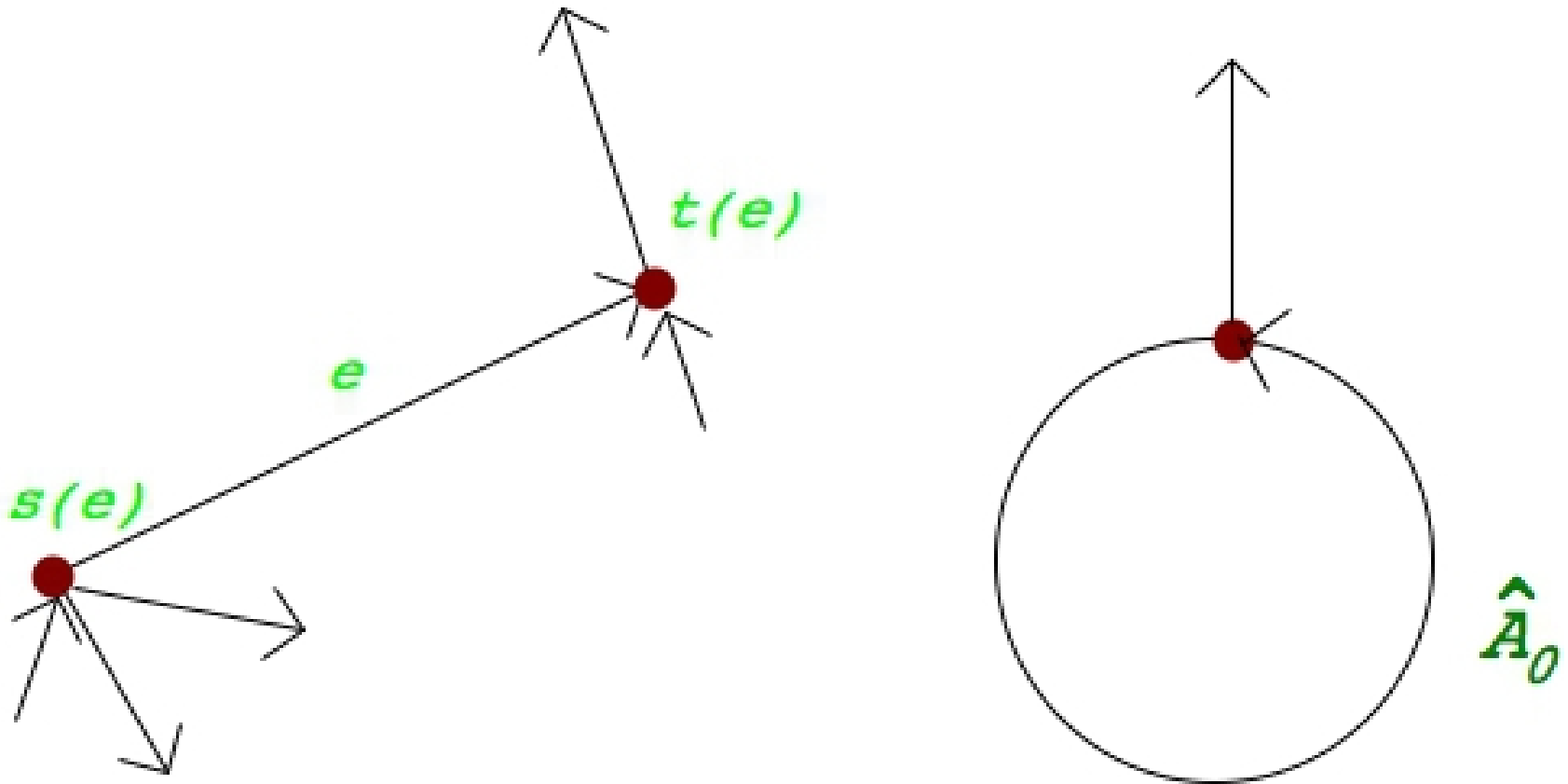}}
\hbox{}
\hbox{\tiny\qquad Note that the source and the target}
\hbox{\tiny\qquad of an edge may coincide,}
\hbox{\tiny\qquad as in the ${\hat A}_{0}$ example above.}}}

\subsec{Quivers with colors}

In addition to the quiver diagram, the gauge theory is
characterized by the vectors $\bn$, $\mm$, sometimes
called the colorings of the quiver:
\begin{equation}
{\bn} = ( n_{\ib} )_{{\ib} \in \Ver} \in {\BZ}_{> 0}^{\Ver}, \qquad {\mm} = (m_{\ib} )_{{\ib} \in \Ver} \in {\BZ}_{ \geq 0}^{\Ver}
\label{eq:dimv}
\end{equation}
to which we associate the vector spaces $N_{\ib} = {\BC}^{n_{\ib}}, M_{\ib} = {\BC}^{m_{\ib}}$. 

\subsec{The symmetry groups}

\subsubsec{The gauge group}
The gauge group $\Gg$ of the theory is the product
\begin{equation}
{\Gg} = \varprod\limits_{{\ib} \in \Ver} \, U(n_{\ib})
\label{eq:gggr}
\end{equation}

\subsubsec{The flavor symmetry}

The theory has the global symmetry which is usually called the flavor symmetry. 
The flavor symmetry group $\Gf$ is a quotient:
\begin{equation}
{\Gf} = \left( \varprod\limits_{{\ib} \in \Ver} \, U( m_{\ib})
 \times U(1)^{\Edg}  \right) \, / \, U(1)^{\Ver}
 \label{eq:gfgr}
 \end{equation} 
 where $U(1)^{\Ver}$ acts on 
 \[ \varprod\limits_{{\ib} \in \Ver} \, U( m_{\ib})
 \times U(1)^{\Edg} \] 
 as follows:
\beq
 ( u_{\ib} )_{\ib \in {\Ver}}:  \left( ( g_{\ib} )_{\ib \in {\Ver}} , ( u_{e} )_{e \in {\Edg}} \right) \mapsto   \left( ( u_{\ib} g_{\ib} )_{\ib \in {\Ver}} , ( u_{s(e)} u_{e} u_{t(e)}^{-1} )_{e \in {\Edg}} \right) 
\label{eq:u1act}
\eeq
This action is equivalently both left and right, therefore $U(1)^{\Ver}$ is a normal subgroup
of $\times_{{\ib} \in \Ver} \, U( m_{\ib})
 \times U(1)^{\Edg}$. In fact, the flavor group $\Gf$ occasionally enhances. For example, the ${\CalN}=4$ theory, viewed as an ${\CalN}=2$ supersymmetric theory, is a particular example of the quiver theory, corresponding to the quiver ${\hat A}_{0}$ with one vertex $v$, and one edge $e$, connecting this vertex to itself $s(e) = t(e) = v$. The flavor symmetry is enhanced from $U(1)$ to $SU(2)$ in this case. This is a subgroup of the $R$-symmetry group $SU(4)$ which commutes with the $SU(2) \times U(1)_{A}$ $R$-symmetry of the particular ${\CalN}=2$ subalgebra of the ${\CalN}=4$ theory.

\subsubsec{Rotational symmetries}

Our four dimensional gauge theories, in the absence of defects to be discussed below, are Poincare invariant. In what follows we shall be breaking the translational invariance by deforming the theory in a rotationally covariant way. The spin cover $Spin(4)_{\mN}$ of the group of rotations is the product
\beq
 Spin(4)_{\mN} = SU(2)_{{\mN}, L} \times SU(2)_{{\mN}, R} \label{eq:spin4}
\eeq
The regularization of the instanton integrals which we employ
in \cite{Nekrasov:2002qd} and here breaks the $Spin(4)_{\mN}$ invariance down to its subgroup $\Gr = SU(2)_{\mN, L} \times U(1)_{\mN, R} \approx U(2) \subset Spin(4)_{\mN}$ which is the group of rotations of the Euclidean space-time ${\mN}  = {\BR}^{4}$, preserving the  identification  of the latter with the complex vector space
${\BC}^{2}$. 

Let $S_{\mN}^{\pm}$ be the defining two dimensional representations (chiral spinors) of $SU(2)_{{\mN}, L}$ and $SU(2)_{{\mN}, R}$, respectively, so that ${\mN}^{\BC} = S_{\mN}^{+} \otimes S^{-}_{\mN}$. Under $\Gr$, $S^{-}_{\mN}$ splits as $L_{\mN} \oplus L^{-1}_{\mN}$. Let us denote the two dimensional representation of ${\Gr}$ by $Q_{\mN} \approx {\BC}^{2}$. Then 
\beq
Q_{\mN} = S_{\mN}^{+} \otimes L_{\mN} \ .
\label{eq:qrep}
\eeq

\subsec{The parameters of Lagrangian}
 
 {}The field content of the theory is the set of ${\CalN}=2$ vector multiplets
$\boldsymbol{\Phi}_{\ib} = ( {\Phi}_{\ib} , \ldots ,  A_{\ib} )$, ${\ib} \in \Ver$, transforming in the adjoint representation of $\Gg$, the set $\boldsymbol{Q}_{\ib} = (Q_{\ib} , \ldots ,  {\tilde Q}_{\ib})$, ${\ib} \in \Ver$ of hypermultiplets transforming in the fundamental representation ${\BC}^{n_{\ib}}$ of $\Gg$, and the antifundamental representation ${\BC}^{m_{\ib}}$ of $\Gf$, and the set $\boldsymbol{Q}_{e}$, $e \in \Edg$ of hypermultiplets transforming in the bi-fundamental representation $\left( \overline{{\BC}^{n_{s(e)}}}, {\BC}^{n_{t(e)}} \right)$ of $\Gg$. 

{}The Lagrangian $\mathbb{L}$ of the theory is parametrized by the 
complexified gauge couplings \[ {\btt} = ({\tau}_{\ib})_{{\ib} \in \Ver} \, , \]
via
\[ \mathbb{L} = \, - \frac{1}{8{\pi}^{2}} \, \sum_{\ib \in \Ver} \ {\ii}\mathrm{Re}{\tau}_{\ib} \int_{\mN} {\Tr}_{N_{\ib}} F_{A_{\ib}} \wedge F_{A_{\ib}} + \]
\[ \qquad\qquad + 
\mathrm{Im}{\tau}_{\ib} \int_{\mN} \, {\Tr}_{N_{\ib}} F_{A_{\ib}} \wedge \star F_{A_{\ib}} + {\Tr}_{N_{\ib}} D_{A_{\ib}} {\Phi}_{\ib} \wedge \star D_{A_{\ib}} {\bar\Phi}_{\ib}  + 
{\Tr}_{N_{\ib}} [ {\Phi}_{\ib}, {\bar\Phi}_{\ib} ]^{2} +  \ldots \]
and the masses
\[ {\bmt} = ({\mt}_{e})_{e \in \Edg} \oplus ({\mt}_{\ib})_{{\ib} \in \Ver}\, , \]
where 
\beq
{\mt}_{e} \in {\BC}, \ {\mt}_{\ib} = {\diag}({\mt}_{{\ib},1}, \ldots , {\mt}_{{\ib}, m_{\ib}})
\in  {\rm End}({\BC}^{m_{\ib}}) \ .
\label{eq:masses}
\eeq 
which enter the superpotential (in the ${\CalN}=1$ language) 
\[ \mathbb{W}   = \sum_{\ib \in \Ver} {\Tr}_{M_{\ib}} \left( {\mt}_{\ib} Q_{\ib} {\tilde Q}_{\ib} \right) + \sum_{e \in \Edg} m_{e} {\Tr}_{N_{s(e)}} \, {\tilde Q}_{e} Q_{e} \, +  \]
\[ \sum_{\ib \in \Ver} {\Tr}_{M_{\ib}} \left( Q_{\ib} {\Phi}_{\ib} {\tilde Q}_{\ib} \right) + \sum_{e \in \Edg}  {\Tr}_{N_{s(e)}} \, \left( {\tilde Q}_{e} {\Phi}_{t(e)} Q_{e} - {\tilde Q}_{e}  Q_{e} {\Phi}_{s(e)}\right) \, , \]
i.e. we view the scalars in the hypermultiplet $\boldsymbol{Q}_{\ib}$ as the linear maps, the matrices:
\[ Q_{\ib} : N_{\ib} \to M_{\ib}, \qquad {\tilde Q}_{\ib}: M_{\ib} \to N_{\ib}, \] 
and
those in $\boldsymbol{Q}_{e}$ as 
\[ Q_{e}: N_{s(e)} \to N_{t(e)},  \qquad {\tilde Q}_{e}: N_{t(e)} \to N_{s(e)}\ . \]  
{}The vacua of the theory are
parametrized by the Coulomb moduli 
\beq
 {{\ba}} = ({\ba}_{\ib})_{{\ib} \in \Ver}\, , \qquad {\ba}_{\ib} = {\diag}({\ac}_{{\ib},1}, \ldots , {\ac}_{{\ib}, n_{\ib}}) \in {\rm End}({\BC}^{n_{\ib}})\ ,
\label{eq:acoul}
\eeq
so that  
\[ \vev{\boldsymbol{\Phi}_{\ib}}_{{\ba}} = {\ba}_{\ib} \ . \] 
It is convenient to package the masses ${\mt}_{\ib}$ and the Coulomb moduli ${\bf a}_{\ib}$ into the polynomials:
\beq
P_{\ib}(x) = \prod_{f=1}^{m_{\ib}} \ (x - {\mt}_{{\ib}, f})\ , \qquad {\mathscr A}_{\ib}(x) = \prod_{{\al}=1}^{n_{\ib}} \ ( x - {\ac}_{{\ib}, {\al}} )
\label{eq:ptpol}
\eeq
We also use the characters
\beq
N_{\ib} = \sum_{{\al}=1}^{n_{\ib}} e^{{\beta}{\ac}_{{\ib}, {\al}}}\, ,  \qquad M_{\ib} = \sum_{f = 1}^{m_{\ib}} e^{{\beta}{\mt}_{{\ib}, f}}
\label{eq:mnchar}
\eeq
which contain the same information about the masses and Coulomb moduli as the polynomials \eqref{eq:ptpol}. 

\subsec{The group $\Hf$}
Define
\beq
{\Hf} = \Gg \times \Gf   \times \Gr\, , 
\label{eq:hgrou}
\eeq
The complexification of the Lie algebra of the maximal torus $T_{\Hf}$ of this group is parameterized by $( {\ba}; {\bmt}; {\ept})$. It is the domain of definition of the supersymmetric partition functions ${\CalZ}_{\bkt}$ in the Eq. \eqref{eq:czkfun}.

\vfill\eject
\subsec{ Perturbative theory}

\subsubsec{Perturbative consistency and asymptotic freedom}

The theory defined by the quiver data is perturbatively asymptoticaly free if the one-loop beta function of all gauge couplings is not positive. For this to be possible we must restrict the gauge group to be the product of special unitary groups
\begin{equation}
{\Gg} \longrightarrow \varprod\limits_{{\ib} \in \Ver} SU(n_{\ib})
\end{equation}
since the abelian factors are not asymptotically free, if there are fields charged  under them. 
For the $SU(n_{\ib})$ gauge coupling the beta function is easy to compute:
\begin{equation}
{\beta}_{\ib} = {\mu}\frac{d}{d{\mu}} {\tau}_{\ib} = - 2 n_{\ib} + m_{\ib} + \sum_{e \in t^{-1}({\ib})} n_{s(e)} + \sum_{e \in s^{-1}({\ib})} n_{t(e)}
\label{eq:bfun}
\end{equation}
The requirement ${\beta}_{\ib} \leq 0$ for all ${\ib} \in \Ver$ implies (see \cite{Howe:1983wj, Katz:1997eq, Lawrence:1998ja, Nekrasov:2012xe} for details) that $\gamma$ is a Dynkin graph
of finite or affine type of a simply-laced finite dimensional
or affine Lie algebra $\gq$. In the latter case ${\mm} = \underline{0}$ (not to be confused with $\bmt \neq 0$). 

\bigskip
\centerline{\picit{10}{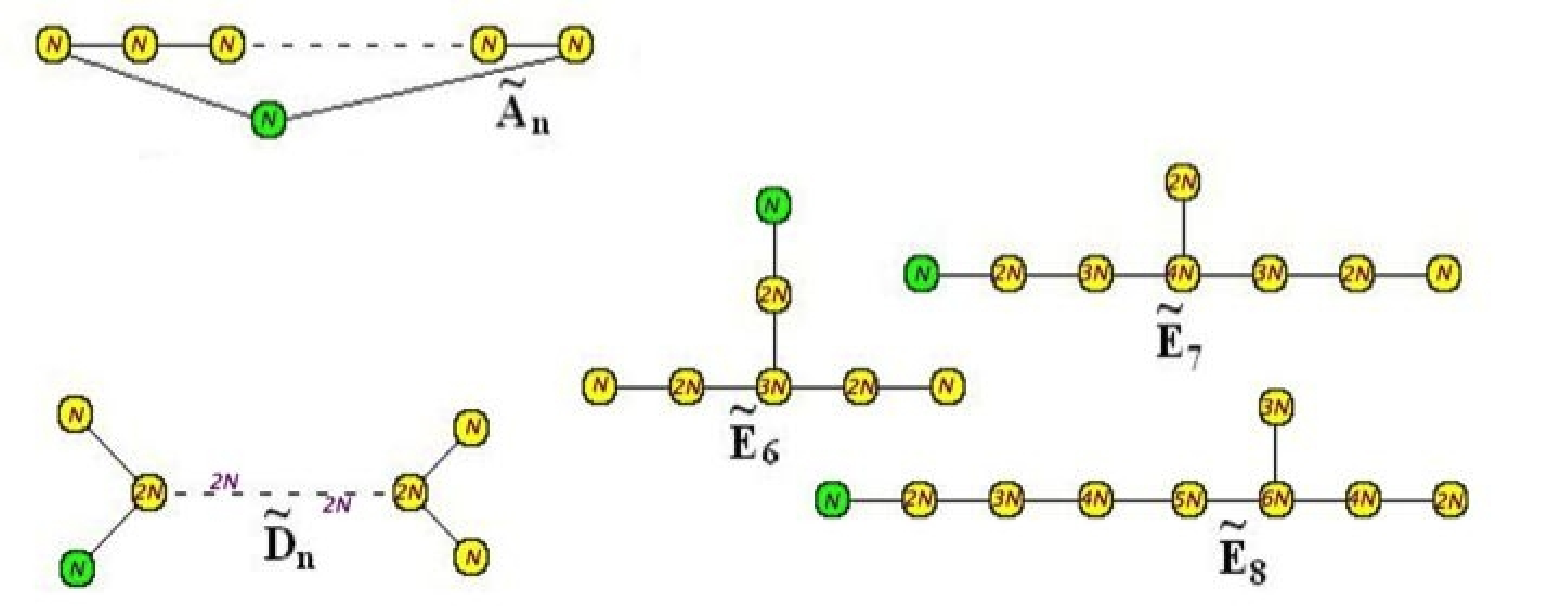}}
\bigskip
\vbox{\centerline{\bf Fig.7}
\centerline{\tiny Affine A,D,E quivers with their ${\bn}$-coloring}
\centerline{\tiny Finite A,D,E quivers are obtained by removing the green node}}
\bigskip

\subsubsec{ Examples.} \begin{enumerate}
\item{}
The $A_{r}$-type quiver $\gamma$, with $r \geq 1$, has:
\beq
 \Ver = [r], \ \Edg = [r-1], \qquad s(e) = e, \ t(e) = e+1, \qquad e = 1, \ldots , r-1 \ .
 \label{eq:finaq}
 \eeq 
 
 \item{}
 The 
 quiver ${\hat A}_{r}$, with $r \geq 0$, has (see the {\bf Fig.6})  
 \beq
 \Ver = \Edg = [r+1],  \qquad s(e) = e, \ t(e) = 1 + \left( e\ {\rm mod} (r+1) \right)\ .
 \label{eq:affaq}
 \eeq 
 
 \item{}
 The $D$ and $E$-type quivers have a single tri-valent vertex, see the {\bf Fig.6}. 

\end{enumerate}

\subsubsec{Perturbative partition function}

The description of the tree level and the perturbative contributions
to the partition function (the latter is given exactly by one loop computation)
can be found in \cite{Nekrasov:2013xda}. 

Here we just quote the results. 
\beq
{\CalZ}^{\mathrm{tree}}_{\gamma}( {\ba}; {\bmt}; {\btt}; {\ept}) = \prod_{{\ib} \in \Ver} \, {\qe}_{\ib}^{\, - \frac{1}{2{\ve}_{1}{\ve}_{2}} \, \sum_{{\alpha}=1}^{n_{\ib}} {\ac}_{{\ib}, {\alpha}}^{2}}\, , 
\label{eq:trlevcz}
\eeq
and
\beq
{\CalZ}^{\mathrm{1-loop}}_{\gamma}( {\ba}; {\bmt}; {\ept})  = {\ep}_{{\ba},{\bmt}, {\ept}} (-{\CalT}^{\rm pert}_{\gamma} )
\label{eq:1loop}
\eeq
where (cf. \eqref{eq:mnchar}): 
\beq
{\CalT}^{\rm pert}_{\gamma} = \frac{1}{(1-e^{-{\beta}{\ve}_{1}})(1-e^{-{\beta}{\ve}_{2}})} \left( \, \sum_{\ib \in \Ver} \left( M_{\ib} - N_{\ib} \right) N_{\ib}^{*} + \sum_{e\in \Edg}
e^{{\beta}{\mt}_{e}} N_{t(e)} N_{s(e)}^{*} \, \right)
\label{eq:chpert}
\eeq 
The character \eqref{eq:chpert} is not a finite sum of exponents as in \eqref{eq:weig}, so the map ${\ep}$ from the sums of exponents to the products of weights is defined by analytic continuation, cf. \eqref{eq:convtx}:
\beq
{\ep}_{{\ba},{\bmt}, {\ept}} (- {\CalT}^{\rm pert}_{\gamma} )
 = \ - \, \frac{d}{ds} \Biggr\vert_{s=0} \, \frac{{\Lambda}^{s}}{{\Gamma}(s)} \, \int_{0}^{\infty} \frac{d{\beta}}{\beta} {\beta}^{s} \ {\CalT}^{\rm pert}_{\gamma} 
 \label{eq:zetfprt}
\eeq
There are subtle points of the regularization of \eqref{eq:zetfprt}  related to boundary conditions in gauge theory. These will be discussed elsewhere. The ultraviolet, $\Lambda \to \infty$ asymptotics of \eqref{eq:zetfprt}, has, a priori, the terms
proportional to ${\Lambda}^{2}$, ${\Lambda}$, ${\Lambda}^{2} {\log}{\Lambda}$, ${\Lambda}{\log} {\Lambda}$, and
${\log}{\Lambda}$. The physically relevant terms are in the last one, they correspond to the one-loop beta-function
of ${\tau}_{\ib}$ if the coefficient of ${\log}{\Lambda}$ contain the terms proportional to 
\beq
ch_{2}(N_{\ib}) \equiv \sum_{{\alpha}=1}^{n_{\ib}} {\ac}_{\ib, \alpha}^{2}
\eeq
Thus, these terms are absent precisely when \eqref{eq:bfun} holds. 

\subsubsec{Beyond asymptotic freedom}

If the asymptotic freedom/conformality conditions are not obeyed, our partition functions are defined as formal power series in the $\qe_{\ib}$ couplings, and some additional couplings, which we call the \emph{higher times}.

\subsubsec{ The extended coupling space}

 Gauge theory can be deformed, in the ultraviolet, by the irrelevant (higher degree) operators, which preserve ${\CalN}=2$ supersymmetry. One adds to the tree level prepotential the terms of the form:
\beq
{\CalF}^{\rm tree} = \sum_{\ib} \sum_{l=0}^{\infty} \frac{1}{(l+2)!} {\tau}_{\ib, l} {\Tr} {\bf\Phi}_{\ib}^{l+2}
\eeq
The parameters ${\tau}_{\ib, l}$ with $l > 0$ are bosonic, in general nilpotent, variables. Actually, for some 
observables one can make sense of the parameters ${\tau}_{\ib, 1}$ in a finite domain near zero \cite{Marshakov:2006ii}. 
One can also add the multi-trace operators $\sim {\Tr} {\bf\Phi}_{\ib}^{l'} {\Tr} {\bf\Phi}_{\jb}^{l''}$ etc. which can be
analyzed with the help of Hubbard-Stratonovich transformation. 

\subsec{Realizations of quiver theories}

\subsubsec{Affine quivers and McKay correspondence}
For affine quivers $\gamma$, the choice of gauge group ${\Gg}$ is characterized by a single integer $N$, for the equation ${\beta}_{\ib} = 0$ for all $\ib \in \Ver$ implies:
\beq
n_{\ib} = N a_{\ib}
\label{eq:nNa}
\eeq
where $a_{\ib} \geq 1$ solves 
\[  2 a_{\ib} =  \sum_{e \in t^{-1}({\ib})}a_{s(e)} +  \sum_{e \in s^{-1} ({\ib})} a_{t(e)}  \]
with the normalization, that for some ${\bf 0} \in \Ver$, $a_{\bf 0} = 1$. It is well-known, that the numbers $a_{\ib} = {\rm dim}{\CalR}_{\ib}$ are the dimensions of the irreducible representations of some finite subgroup $\Gamma \in SU(2)$. 

The $A_{r}$-type subgroup of $SU(2)$ is ${\BZ}_{r+1}$, whose generator ${\Omega}_{A_{r}}$ acts on ${\BC}^{2}$ via (cf. \eqref{eq:primrp}): 
\beq
{\Omega}_{A_{r} }: \left( z_{1}, z_{2} \right) \mapsto \left( {\varpi}_{r+1}\  z_{1} , \quad
{\varpi}_{r+1}^{-1}\  z_{2} \right) , 
\label{eq:amckay}
\eeq
so that ${\Omega}_{A_{r}}^{r+1} = 1$. 
The $D_{r}$-type subgroup of $SU(2)$ ($r \geq 4$) is the product  ${\BZ}_{2(r-2)} \times_{{\BZ}_{2}} {\BZ}_{4}$, whose generators ${\Omega}_{D_{r}}$ and ${\Xi}_{D_{r}}$ act on ${\BC}^{2}$ via:
\beq
{\Omega}_{D_{r}}\, :\, (z_{1}, z_{2}) \mapsto \left( {\varpi}_{2(r-2)} \ z_{1} , \quad
 {\varpi}_{2(r-2)}^{-1} \ z_{2} \right) , \qquad {\Xi}_{D_{r}}\, :\, (z_{1}, z_{2}) \mapsto ( z_{2}, - z_{1} )
\label{eq:dmckay}
\eeq
so that ${\Omega}_{D_{r}}^{r-2} = {\Xi}_{D_{r}}^{2}$, ${\Xi}_{D_{r}}^{4} = 1$. 

\noindent
The $E_{6,7,8}$-type subgroups $SU(2)$ are the binary covers of the symmetry groups of the three platonic solids (and their duals, see \cite{Nekrasov:2012xe} for more details):

\bigskip
\centerline{\picit{10}{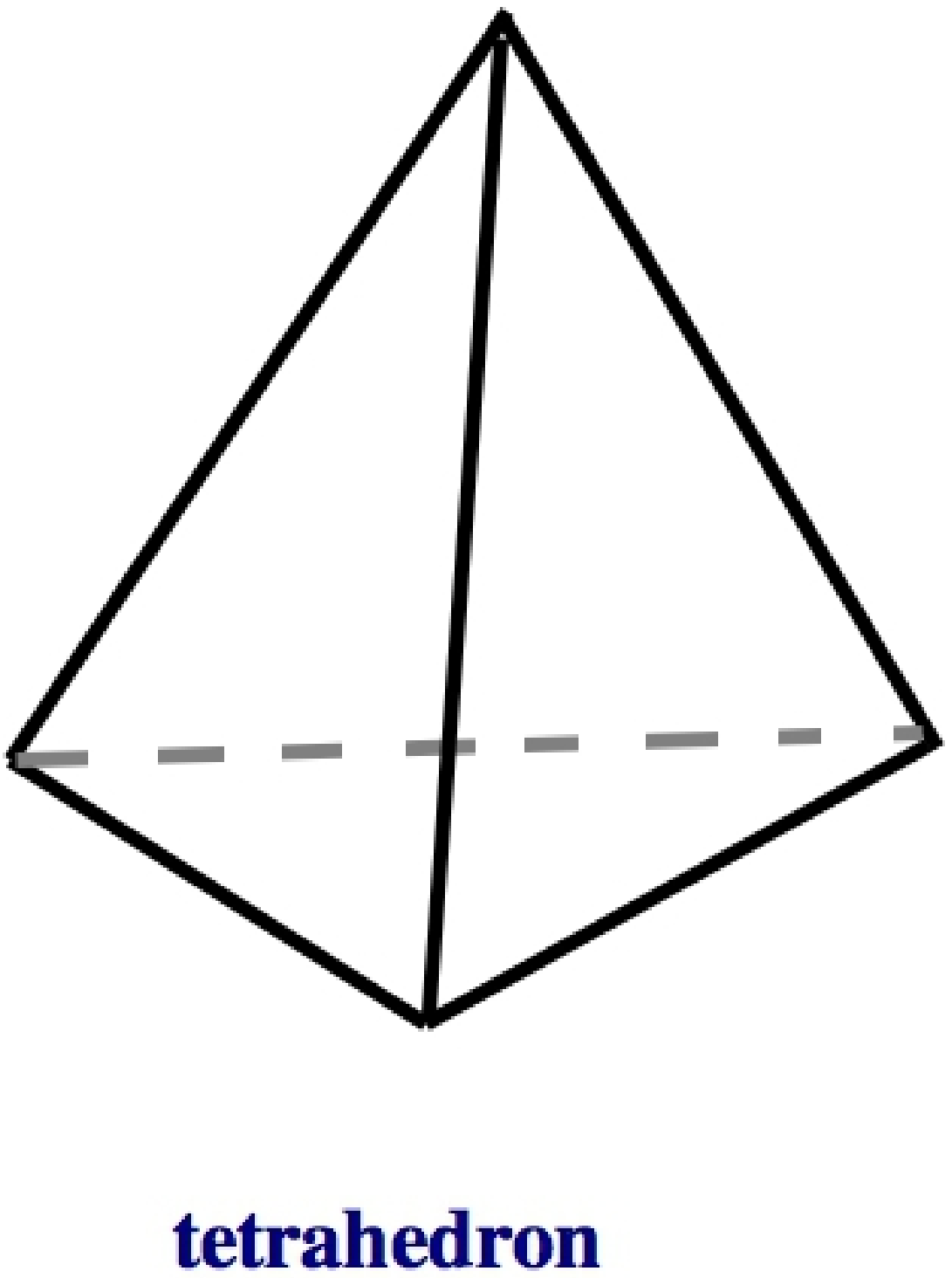}}
\bigskip
\centerline{\bf Fig.8}
\bigskip

The quiver $\gamma$ is  associated to $\Gamma$ as follows: the set $\Ver$ is identified with ${\Gamma}^{\vee}$, the set of irreducible representations of $\Gamma$, ${\bf 0} \in \Ver$ corresponds to the trivial representation ${\CalR}_{\bf 0} = {\BC}^{1}$. The set $\Edg$ of edges is recovered from the tensor products as follows: define the matrix $A : \Ver \times \Ver \to {\BZ}_{\geq 0}$ by
\beq
{\CalR}_{\ib} \otimes S = \bigoplus\limits_{{\jb} \in \Ver} 
{\BC}^{A_{\ib\jb}} \otimes {\CalR}_{\jb}
\label{eq:aijma}
\eeq
where $S \approx {\BC}^{2}$ is the defining two dimensional representation of $SU(2)$. The matrix $A$ is symmetric. There exists another matrix $E: \Ver \times \Ver \to {\BZ}_{\geq 0}$ such that $E + E^{t} = A$. Then 
\beq
\Edg = \bigsqcup\limits_{( {\ib}, {\jb} ) \in \Ver \times \Ver} [E_{\ib, \jb}] \times ( {\ib}, {\jb} ) \, \quad
s \left( k \times ( {\ib}, {\jb} ) \right) = {\ib} , \quad
t \left( k \times ( {\ib}, {\jb} ) \right) = {\jb}
\label{eq:edges}
\eeq
The choice of $E$ given $A$ is the choice of the orientation of edges of $\gamma$.  
Note that this definition associates to ${\Gamma}=1$ the quiver ${\hat A}_{0}$. 

The ${\CalN}=2$ quiver four dimensional gauge theory corresponding to such quiver $\gamma$ can be described most simply by starting with the ${\CalN}=4$ super-Yang-Mills theory with the gauge group $U(N|{\Gamma}|)$, with the fields $A_{\mu} \in {\CalE}\otimes {\CalE}^{*}, 
{\Psi}_{\alpha} \in T \otimes {\CalE}\otimes {\CalE}^{*}, {\bf\Phi} \in {\Lambda}^{2}T \otimes {\CalE}\otimes {\CalE}^{*}$, ${\al} = 1,2$, ${\mu} = 0, 1, 2, 3$, with $\CalE = {\BC}^{N |{\Gamma}|}$ the defining representation of $U(N|{\Gamma}|)$ and $T \approx {\BC}^{4}$ the defining representation representation of the $R$-symmetry group $SU(4)$. Now the space of fields is endowed with the action of $\Gamma$: 
\beq
T = {\BC}^2 \otimes {\CalR}_{\bf 0} \oplus S, \qquad {\CalE} = {\BC}^{N} \otimes {\BC}^{\Gamma}  = \bigoplus_{\ib \in {\Gamma}^{\vee}} {\BC}^{N a_{\ib}} \otimes {\CalR}_{\ib} 
\label{eq:gac}
\eeq
One then defines the new theory by  imposing the 
$\Gamma$-invariance constraint on the fields of the original theory. The ${\CalN}=4$ supersymmetry reduces to ${\CalN}=2$, with $U(Na_{\ib})$-valued vector multiplets ${\bf\Phi}_{\ib}$ labelled by ${\ib} \in \Ver$, and bi-fundamental hypermultiplets
labelled by $e \in \Edg$. The Lagrangian of the original ${\CalN}=4$ theory can be then deformed, preserving the ${\CalN}=2$ supersymmetry.  Since the gauge group $U(N |{\Gamma}|)$ becomes the product 
\beq
U(N |{\Gamma}|) \longrightarrow \varprod\limits_{\ib}  U(Na_{\ib}) \ ,
\eeq
the gauge couplings ${\tau}_{\ib}$ can be chosen independently:
\[ 
{\tau}\ {\Tr}_{N|{\Gamma}|} F^2 \longrightarrow \sum_{{\ib} \in \Ver} {\tau}_{\ib}\ {\Tr}_{Na_{\ib}} F_{\ib}^{2} 
\] 
We have reviewed this well-known construction here because in what follows we shall use its variants on several occasions. 

\subsubsec{Finite quivers}

Some of the finite quiver theories can be obtained as limits of the affine quiver theories. The rest is related to the ones we shall describe below by analytic continuation, sometimes through a strong coupling region.
\begin{enumerate}
\item{}
The $A_{r}$ type theory with $n_{1} = \ldots = n_{r} = N$, 
and $m_{1} = m_{r} = N$, $m_{\ib} = 0$ for $2 \leq {\ib} \leq r-1$, is the limit of the ${\hat   A}_{r+1}$ theory, where one sends ${\qe}_{\bf 0}  \to 0, {\qe}_{r+1} \to 0$. Then ${\ac}_{{\bf 0}, {\al}} - {\mt}_{\bf 0} - {\ve}$, ${\ac}_{r+1, {\al}}+ {\mt}_{r}$ become the masses of the fundamental hypermultiplets, charged under $U(n_{1})$ and $U(n_{r})$, respectively.

\centerline{\picit{6}{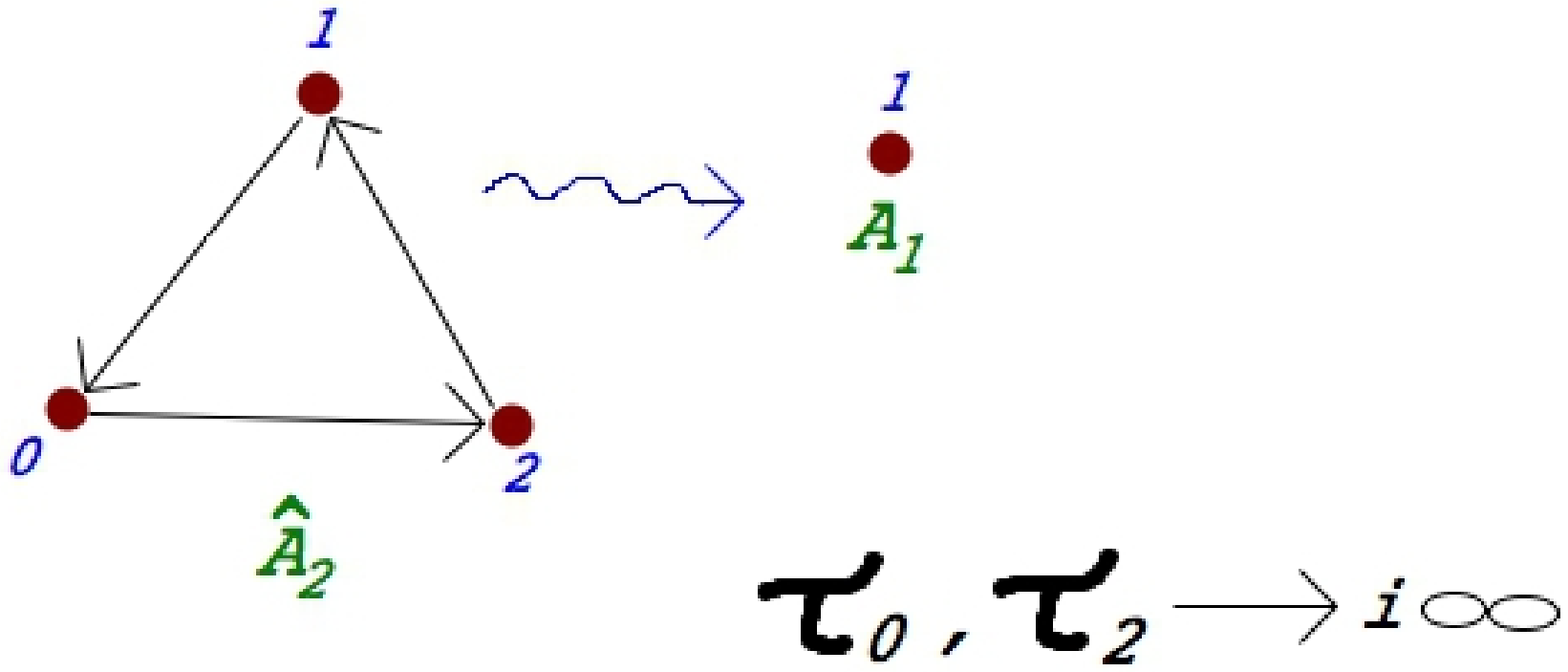}}
 \centerline{\bf Fig.9}
 \centerline{\tiny $A_{1}$ theory as a limit of ${\hat A}_{2}$}

\item{}
A particular $D_{r}$ type theory can be obtained by taking the limit ${\qe}_{\bf 0} \to 0$ limit of ${\hat D}_{r}$ theory. The next-to-last node $\bf 2$ with $n_{\bf 2} = 2N$ has $m_{\bf 2} = N$. 
 \end{enumerate}
 
{}There are other ways of arriving at the quiver ${\CalN}=2$ theories corresponding to finite quivers.

 \secc{ Integration\ over\ instanton\ moduli\ spaces}

In this chapter we recall the mathematical definition of the instanton partition
function
${\CalZ}^{\rm inst}$ of the bulk theory. In \cite{Nekrasov:2015is} we define the defect partition functions ${\bf\Psi}^{\rm inst}$. We give the practical  definition first, without actually describing the relevant instanton moduli spaces. In 
\cite{Nekrasov:2015iim} we describe the moduli spaces ${\iM}_{\gamma}({\bn}, {\bkt})$ whose contributions dominate the gauge theory path integral, explicitly, via modified ADHM construction. More precisely,  the gauge theory path integral localizes to the integral of $1$ over the virtual fundamental cycle of degree (dimension) zero ${\iM}_{\gamma}({\bn}, {\bkt})$ which is represented, in the perfect obstruction theory language
of \cite{obs:1996}  by a smooth (super)-variety ${\iM}_{\gamma}({\bn}, {\bkt})^{c}$  ($c$ stands for coarse) and $\Hf$-equivariant vector bundle  ${\Obs}_{\gamma} \to {\iM}_{\gamma}({\bn}, {\bkt})^{c}$. The $\bkt$-instanton contribution to the gauge theory partition function is the Euler class
\beq
{\CalZ}_{\bkt}^{\rm inst} = \int_{{\iM}_{\gamma}({\bn}, {\bkt})} 1 = \int_{{\iM}_{\gamma}({\bn}, {\bkt})^{c}} {\ep} ( {\Obs}_{\gamma} ) \, , 
\label{eq:czbktg}
\eeq
where we omitted the equivariant parameters.

{}We shall see that for the affine quiver theories (a representative example is the ${\CalN}=2^{*}$ $U(n)$ theory) the underlying variety ${\iM}_{\gamma}({\bn}, {\bkt})^c$ is bosonic, while for the finite quiver theories (a representative example is the $U(n)$ theory with $2n$ fundamental hypermultiplets) ${\iM}_{\gamma}({\bn}, {\bkt})^c$ is a split super-manifold, a vector bundle over an ordinary smooth variety with odd fibers.

{}The same pattern holds for the  theories with defects.

\subsec{Instanton partition function}

\subsubsec{ The bulk partition function ${\CalZ}^{\rm inst}$}
 
 Let ${\bkt} = (k_{\ib})_{{\ib} \in \Ver} \in {\BZ}_{+}^{\Ver}$  be the vector of instanton charges for the gauge group $\Gg$. We denote by ${\iM}_{\gamma}({\bn}, {\bkt})$ the moduli space of framed quiver-graded torsion free sheaves $\boldsymbol{E}_{\gamma} = ( E_{\ib})_{{\ib} \in \Ver} $ on $\BC\BP^2$. More precisely, for each 
${\ib} \in \Ver$, $E_{\ib}$ is a torsion free sheaf on ${\BC\BP}^{2} = {\BC}^{2} \cup {\BC\BP}^{1}_{\infty}$, with the charge ${\rm ch}_{2} (E_{\ib}) = k_{\ib}$, and the framing 
at infinity: 
\beq
E_{\ib} \, \vert_{{\BC\BP}^{1}_{\infty}} \longrightarrow^{\kern -.2in \sim} \quad 
N_{\ib} \ \label{eq:mvsh}
\eeq
Set theoretically, 
\beq
{\iM}_{\gamma}({\bn}, {\bkt})^{c} = \varprod\limits_{\ib \in \Ver}\  {\iM}(n_{\ib}, k_{\ib})
\label{eq:pradhm}
\eeq
is the product of ADHM moduli spaces of $U(n_{\ib})$ instantons of charge $k_{\ib}$. 

Let $\mathscr{E}_{\ib}$ be the universal $\ib$'th sheaf over ${\iM}_{\gamma}({\bn}, {\bkt})^c \times {\BC\BP}^{2}$, and ${\pi}: {\iM}_{\gamma}({\bn}, {\bkt})^c \times {\BC\BP}^{2} \longrightarrow {\iM}_{\gamma}({\bn}, {\bkt})^c $ the projection onto the first factor. 
 Define the {\it obstruction sheaf} ${\Obs}_{\gamma}$ over 
\beq
{\iM}_{\gamma}({\bn})^c = \bigsqcup_{\bkt}\ {\iM}_{\gamma}({\bn}, {\bkt})^c 
\eeq
 by:
 \beq
 {\Obs}_{\gamma} = R{\pi}_{*} \, \bigoplus_{e\in \Edg} {\rm Hom}( \mathscr{E}_{s(e)}, 
 \mathscr{E}_{t(e)}) \, \oplus \, \bigoplus_{{\ib} \in \Ver} {\rm Hom}( \mathscr{E}_{\ib}, M_{\ib})
 \label{eq:obs}
 \eeq
The sheaves above are all ${\Hf}_{\BC}$-equivariant, where
$\Hf$ was defined in \eqref{eq:hgrou}.  

 The complexification of $\Gg$ acts on the isomorphisms ${\CalE}_{i} \vert_{{\BC\BP}^{1}_{\infty}} \longrightarrow^{\kern -.2in \sim} \quad N_{i}$, the complexification
 of $\Gf$ acts on the fibers of \eqref{eq:obs} in the natural way, the complexification  $GL(2, {\BC})$
 of  $\Gr$ acts by the symmetries of ${\BC\BP}^2$, with the fixed point
 $0 \in {\BC}^{2} = {\BC\BP}^2 \backslash {\BC\BP}^1_{\infty}$. Let $T_{\Hf} \subset {\Hf}$, $T_{\Hf}^{\BC}$
 denote the maximal torus of $\Hf$ and its complexification, respectively. 
 
 {}The Coulomb moduli $\ba$ belong to ${\rm Lie}T_{\Gg}^{\BC}$, the masses $\bmt$ belong to ${\rm Lie}T_{\Gf}^{\BC}$. The $\Omega$-deformed theory has two additional complex parameters ${\ept} = ({\ve}_{1}, {\ve}_{2})$ which belong to the Cartan subalgebra of ${\Gr}^{\BC}$, ${\ept} \in {\rm Lie}T_{\Gr}^{\BC} \approx {\BC}^{2}$. 
  
In \cite{Nekrasov:2015iim} we shall discuss the modification of the ADHM construction \cite{Atiyah:1978ri} producing the moduli spaces ${\iM}_{\gamma}({\bn}, {\bkt})$ (cf. \cite{Nakajima:1994}, \cite{Nekrasov:1998ss}) and the obstruction sheaf. More precisely, there is a moduli space  of solutions to a system of matrix equations, determined by the quiver data, which depends on the choice of the Fayet-Illiopoulos (stability) parameters ${\vec\zeta} \in {\BR}^{\Ver}$. It is the choice of these Fayet-Illiopoulos parameters which breaks the rotation symmetry from $Spin(4)$ down to $\Gr$.  When $\vec\zeta$ is in certain chamber ${\CalC} \subset {\BR}^{\Ver}$ the  space of solutions to this system of equations coincides with ${\iM}_{\gamma}({\bn}, {\bkt})$.  The linearization of the equations at the particular solution defines the obstruction sheaf, as the space of solutions to the dual linear system. 
 
 {}The instanton factor in the
  partition function can be shown to reduce to the generating function of the equivariant
  integrals
  \begin{equation}
  {\CalZ}^{\rm inst}_{\gamma} ( {\ba}; {\bmt}; {\bqt}; {\ept} ) = 
  \sum_{\bkt} \ {\bqt}^{\bkt}\  \int_{{\iM}_{\gamma}({\bn}, {\bkt})^c} {\ep}_{{\ba}; {\bmt}; {\ept}}\, ({\Obs}_{\gamma})\, , 
  \label{eq:czobs}
  \end{equation}
with
  \[ {\bqt}^{\bkt}\ =\ \prod_{{\ib} \in \Ver} {\qe}_{\ib}^{k_{\ib}} \]
  Mathematically \eqref{eq:czobs} is just a definition of the left hand side. 
Each term of the $\bqt$-expansion is a rational function on ${\rm Lie}({\Hf}_{\BC})$, of negative degree of homogeneity for the asymptotically free theories, and degree zero (i.e. they are homogeneous functions) for the asymptotically conformal theories.

\subsubsec{ Localization and fixed points}  
 The fixed points ${\iM}({\bn}, {\bkt})^{\Hf}$ of the $T_{\Hf}$-action on ${\iM} ({\bn}, {\bkt})$ are the sheaves
 which split as direct sums of monomial ideals:
 \beq
\boldsymbol{\CalE} \in {\iM}_{\gamma}({\bn}, {\bkt})^{T_{\Hf}} \Leftrightarrow E_{\ib} = \bigoplus\limits_{{\al}=1}^{n_{\ib}}
 \, {\CalI}_{{\ib}, {\al}}\, ,
  \label{eq:moni}
 \eeq
 where ${\CalI}_{{\ib}, {\al}} = I_{{\lambda}^{({\ib}, {\al})}}$, 
{}Thus, the set of fixed points ${\iM}_{\gamma}({\bn}, {\bkt})^{T_{\Hf}}$ 
is in one-to-one correspondence with the set of 
{\it quiver $\bn$-colored partitions}:
\begin{equation}
E_{\bla} \leftrightarrow {\bla} = 
\left\{ {\lambda}^{({\ib}, {\al})} \, \Biggr\vert \ {\ib} \in \Ver, {\al}  \in [n_{\ib}] ,  
\,  {\lambda}^{({\ib}, {\al})} \, \text{\, is\, a \, partition},  \ \sum_{{\al} = 1}^{n_{\ib}} | {\lambda}^{({\ib}, {\al})} | = k_{\ib} \right\} 
\label{eq:fixpar}\end{equation}
These points are also the fixed points of the action of $T_{\Hf}$ on the moduli space of 
$\Gamma$-invariant instantons.
 The fixed point formula expresses the gauge theory path integral as the sum over the set
 of quiver $\bn$-colored partitions. We shall present the explicit formula in the next section. 
 
Now that the path integration is reduced to a finite sum, the non-perturbative field redefinitions involving adding a point-like instanton can be discussed rigorously. 
  
\subsec{ Characters, tangent spaces}

The contribution of a given fixed point to the partition function can be conveniently expressed
using the characters of various vector spaces involved in the local analysis of the path integral measure. 
{}The instanton partition function can be then written as:
 \beq 
 {\CalZ}^{\rm inst}_{\gamma} ( {\ba}; {\bmt};  {\ept} ; {\bqt})  \ = \
\sum_{\bla}\ {\bqt}^{\bla}\  {\bmu}_{\bla}( {\ba}; {\bmt}; {\ept} )  
\label{eq:measz}
\eeq
where 
\beq
{\bmu}_{\bla}( {\ba}; {\bmt}; {\ept} )   = {\ep}_{{\ba}; {\bmt}; {\ept}} \left( - {\CalT}_{\bla} \right) \ , \label{eq:bmubla}
\eeq
and 
\beq
{\bla} = \left( {\lambda}^{({\ib} ,{\al})} \right)_{{\ib} \in \Ver}^{{\al} \in [n_{\ib}]} 
 \  , 
\label{eq:blav}
\eeq
and
\beq
{\bqt}^{\bla} \ = \ \prod_{{\ib} \in \Ver} \,  \prod_{{\al}=1}^{n_{\ib}} 
\
{\qe}_{\ib}^{| {\lambda}^{({\ib}, {\al})} |}  \quad , \label{eq:bqtbla}
 \eeq
and
 \begin{multline}
 {\CalT}_{\bla}  = \left( \sum_{{\ib} \in \Ver} \, \left( N_{\ib}K_{\ib}^{*} + N_{\ib}^{*}K_{\ib} q - P K_{\ib}K_{\ib}^{*}  \right) \right) \\
 - \left( \sum_{{\ib} \in \Ver}  M_{\ib}^{*}K_{\ib} +   \sum_{e \in \Edg} \, e^{{\beta}{\mt}_{e}}\, \left( N_{t(e)} K_{s(e)}^{*} + N_{s(e)}^{*}K_{t(e)} q - P K_{t(e)} K_{s(e)}^{*} \right) \right) \, . \label{eq:ctla}
 \end{multline}
 In writing \eqref{eq:ctla} we adopted a convention where 
 the characters of the vector spaces are denoted by the same letters as the vector spaces themselves.  We are thus using the notations \eqref{eq:mnchar} and
 \beq 
 K_{\ib} = \sum_{{\al}=1}^{n_{\ib}} \left( e^{{\beta}{\ac}_{{\ib}, {\al}}} \sum_{{\square} \in {\lambda}^{({\ib}, {\al})}} e^{{\beta}c_{\square}} \right) 
  \label{eq:chars}
 \eeq
 In \eqref{eq:chars} we use the convention \eqref{eq:conjconv}.

 Note that the Eqs. \eqref{eq:chars} identify $N_{\ib}, K_{\ib}, M_{\ib}$ with representations of $T_{H}$. While $N_{\ib}, M_{\ib}$ are Weyl-invariant, and correspond to representations of $H$, the spaces $K_{\ib}$ do not, in general, carry a representation of $H$. 
 
 \subsec{ Integral representation.} 
The measure \eqref{eq:measz} can be also given an integral representation:
 \beq
 {\CalZ}^{\rm inst}_{\gamma} ( {\ba}; {\bmt}; {\bqt}; {\ept} )  = 
 \sum_{\bkt} \ \frac{{\bqt}^{\bkt}}{{\bkt}!} \
\oint_{\Gamma_{\gamma}} {\displaystyle\prod\limits_{{\ib} \in \Ver} {\Upsilon}_{\ib} \displaystyle\prod\limits_{e \in \Edg} {\Upsilon}_{e}} \ , 
\label{eq:czgammint}
\eeq
where
 \begin{multline}
\qquad\qquad {\Upsilon}_{\ib} = 
{\bigwedge\limits}_{{\al}=1}^{n_{\ib}} \ \frac{\ve}{{\ve}_{1}{\ve}_{2}} \, \frac{\mathrm{d}{\phi}_{{\ib}, {\al}} P_{\ib}({\phi}_{{\ib}, {\al}})}{{\mathscr A}_{\ib}({\phi}_{{\ib}, {\al}}) {\mathscr A}_{\ib}({\phi}_{{\ib}, {\al}}+ {\ve})}
\prod\limits_{{\al}' \neq {\al}''}  \dfrac{1}{{\bS}({\phi}_{{\ib}, {\al}'} - {\phi}_{{\ib}, {\al}''})}  \ ,   \\
 {\Upsilon}_{e} =  \prod\limits_{{\al}'=1}^{n_{s(e)}} {\mathscr A}_{t(e)}({\phi}_{s(e), {\al}'}- {\mt}_{e})  \prod\limits_{{\al}''=1}^{n_{t(e)}} {\mathscr A}_{s(e)}({\phi}_{t(e), {\al}''} + {\ve}+{\mt}_{e}) 
\prod\limits_{{\al}'=1}^{n_{s(e)}}  \prod\limits_{{\al}''=1}^{n_{t(e)}} {\bS}( {\phi}_{t(e), {\al}''} -{\phi}_{s(e), {\al}'} + {\mt}_{e}) 
\end{multline}
where 
\beq
{\bS}(x)  = 1 + \frac{{\ve}_{1}{\ve}_{2}}{x(x+{\ve})}
\eeq
and the choice of the contour 
\beq
\Gamma_{\gamma} \approx \ \varprod_{{\ib} \in {\Ver}} \, {\BR}^{n_{\ib}}
\eeq will be discussed elsewhere. 

For generic values of the parameters ${\ac}_{\ib}$, ${\ept}$ etc. the fixed point formula \eqref{eq:measz} can be used. 
This is equivalent to the statement that \eqref{eq:czgammint} can be evaluated by computing the residues at simple poles. The contributions of the particular toric instanton configuration $\bla$ \eqref{eq:blav} are rational functions with lots of poles. These poles lead to potential divergencies of the instanton partition function. For example, whenever
the ratio \eqref{eq:bpar} is a positive rational number, $b^2 \in {\BQ}_{+}$ some of the individual terms ${\bmu}_{\bla}( {\ba}; {\bmt}; {\ept} )$ blow up. 
However the divergencies cancel between several terms. The contour integral representation is more convenient
in this case, as it remains finite, as long as the contour $\Gamma_{\gamma}$ does not get pinched between two
approaching poles.

Let us briefly explain the reason why these apparent poles occur, and why they potentially cancel between themselves.
A rational relation between the Coulomb parameters and the $\Omega$-deformation parameters means that the symmetry
group used in the equivariant localization is strictly smaller then the maximal torus of $\Hf$. Reduction of the symmetry
group means a potential enhancement of the fixed point locus. For example, instead of a set of isolated points
one may find a copy of ${\BC\BP}^{1}$ or a more complicated positive dimension subvariety. Each component
of the fixed point locus
contributes an integral to the instanton partition function. This contribution is finite if the component is compact. 
In the extreme case ${\ac} = {\ept} = 0$, the symmetry group is trivial. 
The fixed point locus in this case is the whole original moduli space  ${\iM}_{\gamma}({\bn}, {\bkt})$, and the integral 
diverges.  

\subsec{Full partition functions}

The full partition functions  are the products of the instanton partition functions and the tree and one-loop partition functions. They are given by the product of \eqref{eq:trlevcz}, \eqref{eq:1loop} and \eqref{eq:czbktg} leading to the following simple formulas
\beq
{\CalZ}_{\gamma} ( {\ba}, {\bmt}, {\ept} ; {\bqt}) = \sum_{\bla} {\CalQ}({\CalT}_{\bla}) {\ep}_{{\ba}, {\bmt}, {\ept}} ( - {\CalT}[{\bla}])
\eeq
where
\beq
{\CalT}[{\bla}] = \frac{1}{(1- e^{-{\beta}{\ve}_{1}})(1-e^{-{\beta}{\ve}_{2}})}
\left( \, \sum_{\ib \in \Ver} \left( M_{\ib} - S_{\ib}[{\bla}] \right) S_{\ib}^{*}[{\bla}]  + \sum_{e\in \Edg}
e^{{\beta}{\mt}_{e}} S_{t(e)}[{\bla}]  S_{s(e)}^{*}[{\bla}]  \, \right)
\label{eq:chtfull}
\eeq 
and
\beq
{\CalQ}({\CalT}[{\bla}]) = \prod_{\ib \in \Ver} {\qe}_{\ib}^{- \frac{1}{{\ve}_{1}{\ve}_{2}} {\rm ch}_{2}(S_{\ib}[{\bla}])} = \left( \prod_{\ib \in \Ver} {\qe}_{\ib}^{- \frac{1}{2{\ve}_{1}{\ve}_{2}} \sum_{\alpha = 1}^{n_{\ib}} {\ac}_{\ib, \alpha}^2} \right) \ \times\
{\bqt}^{\bla}
\label{eq:instfac}
\eeq
where
\beq
{\rm ch}_{2}(S_{\ib}[{\bla}]) = \frac{1}{2} \sum_{{\al}=1}^{n_{\ib}} {\ac}_{{\ib},{\alpha}}^{2} - {\ve}_{1}{\ve}_{2} k_{\ib} [{\bla}]
\eeq

 \secc{ The\ {\y}\ -observables}

The measures \eqref{eq:measz}  ${\bmu}_{\bla}({\ba}, {\bmt}; {\ept})$  define the complexified statistical models which can be studied without a reference to the original gauge theory. To any function $F= F[{\bla}]$ on the space of quiver $\bn$-colored partitions one associates its normalized expectation value:
\beq
\vev{F}_{\gamma} = \frac{1}{{\CalZ}^{\rm inst}_{\gamma}} \sum_{\bla}\ {\bqt}^{\bla} \ {\bmu}_{\bla}({\ba}, {\bmt}; {\ept}) \, F [{\bla}]
\label{eq:vevf}
\eeq
Sometimes we shall also use the un-normalized expectation value
\beq
\llangle F \rrangle_{\gamma} = {\CalZ}^{\rm tree}_{\gamma}{\CalZ}^{\rm 1-loop}_{\gamma}  \sum_{\bla} \ {\bqt}^{\bla}\ {\bmu}_{\bla}({\ba}, {\bmt}; {\ept}) \ F [{\bla}] = {\CalZ}_{\gamma} \vev{F}_{\gamma}
\label{eq:uvevf}
\eeq
Sometimes, in what follows we shall view such a function $F$ as an {\sl operator},
acting in the infinite-dimensional vector space $\BH$ with the basis $e_{\bla}$ labelled by the quiver $\bn$-colored partitions, 
\beq
F e_{\bla} = F[{\bla}] e_{\bla}
\label{eq:fop}
\eeq
The r\^ole of $\BH$ in gauge theory will be discussed elsewhere. 
 
The functions $F$ which do come from gauge theory will be called {\sl observables}.  An example of observable is the $\ib$'th instanton charge:
\beq
k_{\ib} [ {\bla} ] = \sum_{{\al}=1}^{n_{\ib}}\quad \left\vert \, {\lambda}^{({\ib}, {\al})} \, \right\vert
\label{eq:kib}
\eeq

\subsec{The bulk $\y$-observables}

The important observables are the characteristic polynomials
of the adjoint Higgs fields:
\begin{equation}
{\y}_{\ib}(x) = x^{n_{\ib}} \ {\exp} \, \sum_{l=1}^{\infty} 
-\frac{1}{l x^{l}} \, {\Tr} ({\Phi}_{\ib} \vert_{0} )^{l} 
\label{eq:yibx}
\end{equation} 
Here we denote by ${\Phi}_{\ib} \vert_{0}$ the lowest component of the vector multiplet
${\boldsymbol{\Phi}}_{\ib}$ corresponding to the node $\ib \in \Ver$, evaluated at the specific point $0 \in {\BC}^{2}$ in the Euclidean space-time. This is the fixed point of the rotational symmetry
$Spin(4)_{\mN}$ of which the maximal torus $U(1) \times U(1)$ is generated by the rotations in the two orthogonal two-planes. 

In the ${\CalN}=2$ theory the gauge-invariant polynomials of the scalar components of the vector multiplets, i.e. 
\begin{equation} 
{\CalO}_{l}({\xb}) = {\Tr}\, {\Phi}_{\ib}^{l}({\xb}) \ , \label{eq:trl}
\end{equation} 
for ${\xb} \in {\BR}^{4}$ are invariant under some supersymmetry transformations, which are nilpotent on the physical states.
Moreover, the ${\xb}$-variation of such operators is in itself a supersymmetry variation. Therefore, in the cohomology of such a supercharge, the observable ${\CalO}_{l}({\xb})$ is ${\xb}$-independent. 
The supersymmetry of the $\Omega$-deformed ${\CalN}=2$ gauge theory is such that the operator ${\CalO}_{l}({\xb})$ is invariant only at ${\xb} = 0$, i.e. at the fixed point of the rotations. 
 
 Classically, i.e. for the ordinary matrix-valued function ${\Phi}_{\ib}({\xb})$ the exponential \eqref{eq:yibx} evaluates to the characteristic polynomial of this matrix (cf. \eqref{eq:ptpol}):
 \begin{equation}
 {\y}_{\ib}(x)^{\rm tree} = {\rm Det}_{n_{\ib}} ( x - {\Phi}_{\ib}\vert_{0} ) = {\mA}_{\ib}(x)
 \label{eq:ycl}
 \end{equation}
 
 \subsubsec{$\y$-observables from sheaves}
 
Mathematically ${\y}_{\ib}(x)$ is defined using the 
virtual Chern polynomials of the universal sheaves, localized at the point 
$0 \in {\BC}^{2}$:
\begin{equation}
{\mathscr Y}_{\ib}(x) = c_{x}(R{\pi}_{*}\left[ E_{\ib} \to E_{\ib} \otimes T_{{\BP}^{2}} \to E_{\ib} \otimes {\wedge}^{2}T_{{\BP}^{2}}  \right])
\label{eq:yibxt}
\end{equation} 
Here we used the Koszul resolution of the skyscrape sheaf ${\CalS}_{0}$ supported
at $0 \in {\BC}^{2}$:
\begin{equation}
0 \to {\wedge}^{2}{\CalT}_{{\BP}^{2}}^{*} \to {\CalT}_{{\BP}^{2}}^{*} \to {\CalO}_{{\BP}^{2}} \to {\CalS}_{0}
\label{eq:ress}
\end{equation}
where the second and the third maps are the contraction with the Euler vector field
$z_{1} {\partial}_{z_{1}} + z_{2} {\partial}_{z_{2}}$. 

\subsubsec{$\y$-observables from noncommutative gauge fields}

{}The proper physical definition of the observable \eqref{eq:yibx} is also subtler then the naive expression \eqref{eq:ycl}. In computing the instanton partition function one uses the non-commutative deformation of the gauge theory, in order to make the instanton moduli space smooth with isolated fixed points \cite{Nekrasov:1998ss}. In the noncommutative world,
the notion of a particular point ${\bx}=0$ in the space-time
${\BR}^{4}_{\theta}$ makes no sense. The gauge fields and the adjoint scalar 
${\Phi}_{\ib}$ are the operators in the Hilbert space ${\CalH}$, 
\[ {\CalH}  = \bigoplus_{\ib \in \Ver} N_{\ib} \otimes {\mathfrak H} \]
where ${\mathfrak H}$ is the $2$-oscillator Fock  space
representation of the algebra of functions on ${\BR}^{4}_{\theta}$, generated by ${\hat x}^{\mu}$, ${\mu} = 1, \ldots, 4$, obeying the Heisenberg algebra $[ {\hat x}^{\mu} , {\hat x}^{\nu} ] = {\ii}{\vartheta}^{\mu\nu}$, with constant antisymmetric (non-degenerate) matrix $\theta$:
\[ A_{{\ib}, \mu}(x) \mapsto {\bX}^{\mu}_{\ib} = {\hat x}^{\mu} + {\vartheta}^{\mu\nu} A_{{\ib}, \nu} ({\hat x}) \]
\[ {\Phi}_{\ib} \mapsto {\bf\Phi}_{\ib} = \frac 12 {\mathfrak h}_{\mu\nu} {\bX}^{\mu}_{\ib}{\bX}^{\nu}_{\ib} + {\phi}_{\ib}({\hat x}) \]
where ${\mathfrak h}_{\mu\nu}$ is the symmetric matrix, obeying
\[   {\mathfrak h}_{\mu\nu} {\vartheta}^{\nu\al}  = {\Omega}^{\al}_{\mu}  \]
with $\Omega$ being the matrix of infinitesimal rotation of ${\BR}^{4}$, preserving both the metric and $\vartheta$. In the vacuum 
\[ {\bf\Phi}_{\ib} =  {\diag} ( {\ac}_{{\ib}, 1} , \ldots , {\ac}_{{\ib}, n_{\ib}} ) \otimes {\bf 1}_{\mathfrak H} + {\bf 1}_{N_{\ib}} \otimes
( {\ve}_{1} {\hat n}_{1} + {\ve}_{2} {\hat n}_{2} ) \] 
where ${\hat n}_{\xi} = a^{+}_{\xi} a_{\xi}$ are the oscillator number operators in $\mathfrak H$, ${\xi}=1,2$. 
The observables  like \eqref{eq:trl} are defined, cf. \cite{Nekrasov:2003rj}, as the ratio of infinite dimensional determinants, 
\beq
{\y}_{\ib}(x) = \frac{{\Det}_{\CalH} ( x- {\bf\Phi}_{\ib} )\ {\Det}_{\CalH} ( x- {\bf\Phi}_{\ib} - {\ve}_{1}- {\ve}_{2})}{{\Det}_{\CalH} ( x- {\bf\Phi}_{\ib} - {\ve}_{1}) \ {\Det}_{\CalH} ( x- {\bf\Phi}_{\ib} - {\ve}_{2})}
\label{eq:infdet}
\eeq
or, equivalently,
via a limiting procedure involving the regularized traces
\[ \text{Tr}_{\CalH} \, e^{-t {\bf\Phi}_{\ib}} \]
partly explaining the non-triviality of what follows. Without going into detail, 
let us quote the results which we shall need in this paper. 

\subsubsec{$\y$-observables from Chern classes}

Another, equivalent, definition of ${\y}_{\ib}(x)$ is the following. 
We have the vector bundles $N_{\ib}, K_{\ib}$, ${\ib} \in \Ver$, over ${\iM}_{\gamma}({\bn}, {\bkt})$. Topologically $N_{\ib}$ are trivial bundles, while $K_{\ib}$ are, in general, not. These bundles are $\Hf$-equivariant. 
Then:
\beq
{\y}_{\ib}(x) = {\ep}_{x}(N_{\ib}^{*}) \frac{{\ep}_{x-{\ve}_{1}}(K_{\ib}^{*}){\ep}_{x-{\ve}_{2}}(K_{\ib}^{*})}{{\ep}_{x}(K_{\ib}^{*}){\ep}_{x-{\ve}}(K_{\ib}^{*})}
\label{eq:yobsch}
\eeq

\subsubsec{$\y$-observables on toric instantons}

{}For our calculations, we need the fixed point expression, i.e.
the value ${\y}_{\ib} (x) [{\bla}]$ 
of the observable ${\mathscr Y}_{\ib}(x)$ on the special instanton
configuration ${\CalE}_{\bla}$: 
\beq
\begin{aligned}
{\y}_{\ib}(x)[{\bla}] \ = \ & {\displaystyle\prod\limits_{{\al}=1}^{n_{\ib}}} \left( \left( x - {\ac}_{\ib, \al} \right) \prod_{{\square} \in {\lambda}^{({\ib}, {\al})}} \frac{ \left( x - {\ac}_{\ib, \al}  - c_{\square} - {\ve}_{1} \right) \left( x - {\ac}_{\ib, \al}  - c_{\square} - {\ve}_{2} \right)}{\left( x - {\ac}_{\ib, \al}  - c_{\square}  \right)
\left( x - {\ac}_{\ib, \al}  - c_{\square} - {\ve} \right) } \right) \ = \\
& \qquad\qquad =  \ 
{\displaystyle\prod\limits_{{\al}=1}^{n_{\ib}}} \
 \frac{\textstyle\prod\limits_{{\color{green}{\blacksquare}} \in {\partial}_{+}{\lambda}^{({\ib}, {\al})}} ( x - {\ac}_{{\ib}, {\al}} - c_{\color{green}{\blacksquare}} )}{\textstyle\prod\limits_{{\color{magenta}{\blacksquare}} \in {\partial}_{-}{\lambda}^{({\ib}, {\al})}} ( x - {\ac}_{{\ib}, {\al}} - {\ve} - c_{\color{magenta}{\blacksquare}} )} 
 \label{eq:yiblx}
 \end{aligned}
 \eeq
where for a monomial ideal $I_{\lambda}$, corresponding to the partition $\lambda$ the {\it outer boundary} ${\partial}_{+}{\lambda}$ and the {\it inner boundary} ${\partial}_{-}{\lambda}$ are the monomials corresponding to the generators, and the relations (divided by the factor $z_{1}z_{2}$) of the ideal, cf. $\bf Fig.10$. Explicitly, given the character ${\chi}_{\lambda}$ of the quotient
${\BC}[z_{1}, z_{2}]/I_{\lambda}$ which is the same thing as the character of the partition $\lambda$, 
the contents of the inner and the outer boundaries can be read off the character of the {\it tautological sheaf}:
\begin{equation}
S_{\lambda} = 1 - P {\chi}_{\lambda}  = \sum_{{\square} \in {\partial}_{+}{\lambda}} e^{{\beta} c_{\square}} - q \sum_{{\blacksquare} \in {\partial}_{-}{\lambda}} e^{{\beta} c_{\blacksquare}}
\label{eq:genrel}
\end{equation}
Note that 
\beq
S_{\lambda} = {\Tr}_{S^{+}_{\lambda}} \ {\hat q} - q {\Tr}_{S^{-}_{\lambda}} {\hat q}
\eeq where $S^{\pm}_{\lambda}$ are the fibers over 
${\CalI}_{\lambda} \in {\Hilb}^{\left[ | {\lambda} | \right]} ({\BC}^{2})$ of the vector bundles 
$S^{\pm}$ which we study in more detail in \cite{Nekrasov:2015iim}. It is easy to see from the picture of the Young diagram $\lambda$, to which stratum 
${\HM}_{k,l} \subset {\Hilb}^{\left[ | {\lambda} | \right]} ({\BC}^{2})$ it belongs: 
\beq
{\lambda} \in {\HM}_{k,l}\qquad \Leftrightarrow\qquad k = | {\lambda} |,\qquad  l = 
{\ell}({\lambda}) = \# {\partial}_{-} {\lambda} = \# {\partial}_{+} {\lambda} - 1\ .
\label{eq:klstr}
\eeq  

\centerline{\picit{10}{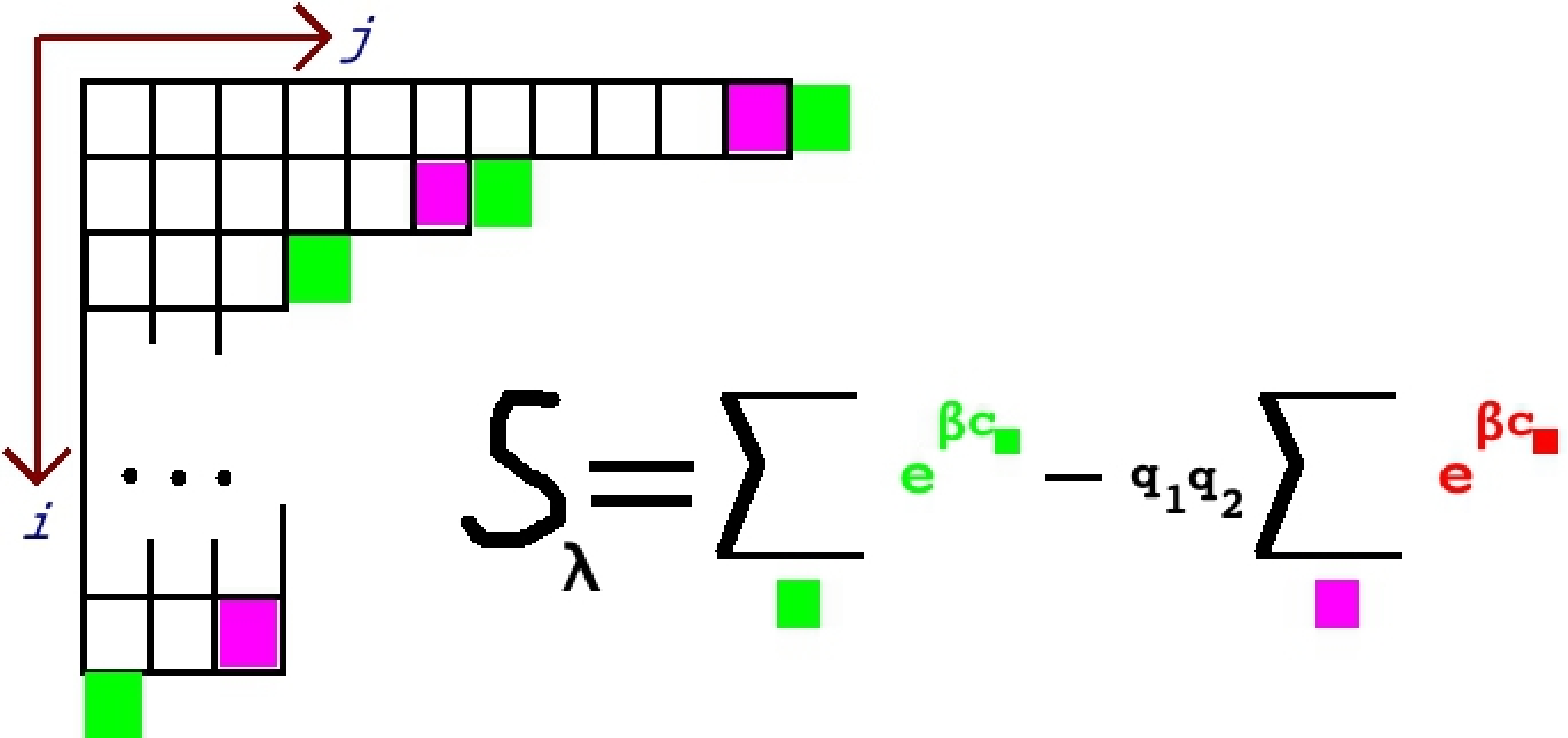}}
\bigskip
\vbox{
\centerline{\bf Fig.10}
\centerline{\sl Generators $\color{green}{\blacksquare}$ and relations $\color{magenta}{\blacksquare}$ of a monomial ideal $I_{\lambda}$}
\centerline{\sl The character of the tautological sheaf $S_{\lambda} = 1 - P {\chi}_{\lambda}$} }
\bigskip
The ${\y}$-observable ${\y}_{\ib}(x)$ is essentially the character of the localized tautological complex ${\mathscr S}_{\ib}$, which is the cohomology (along ${\BC\BP}^{2}$) of the complex 
\begin{equation}
  \mathscr{S}_{\ib} =  \left( \mathscr{E}_{\ib} \to 
  \mathscr{E}_{\ib} \otimes {\CalT}_{{\BP}^{2}}  \to {\mathscr E}_{\ib} \otimes {\wedge}^{2}{\CalT}_{{\BP}^{2}}  \right) [ - 1]
  \label{eq:tauto}
  \end{equation}
  which is the dual Koszul complex tensored with the universal sheaf ${\mathscr E}_{\ib}$. 
The relevant character $S_{\ib}$ is easy to calculate:
\begin{equation}
S_{\ib} = N_{\ib} - P K_{\ib} = \sum_{{\al}=1}^{n_{\ib}}
 e^{{\beta}{\ac}_{{\ib}, {\al}}} S_{\lambda_{{\ib}, {\al}}} 
\label{eq:tausi}
\end{equation}
The previous formulae can be succinctly written as:
\begin{equation}
{\y}_{\ib}(x)[{\bla}] = {\beta}^{-n_{\ib}} \, {\ep} [ e^{{\beta}x} S_{\ib}^{*} ]
\label{eq:yfsi}
\end{equation}
or, in more detail, cf. the notation \eqref{eq:cof}
\beq
{\y}_{\ib}(x)[{\bla}]  = x^{n_{\ib}}\ {\exp}\, \left( - \sum_{l=1}^{\infty} \, \frac{1}{l x^{l}}  [{\beta}^{l}] S_{\ib} \right)
\eeq
 
For large $x$ the observable ${\y}_{\ib}(x)$ can be expanded as:
\beq
{\y}_{\ib}(x) = {\mA}_{\ib}(x) \left( 1 + \frac{{\ve}_{1}{\ve}_{2}}{x^2} k_{\ib} + \ldots \right)
\label{eq:yxki}
\eeq

\subsubsec{ The importance of ${\y}$-observables}

The observables ${\y}_{\ib}(x)$ and the characters $S_{\ib}$ are used in 
the analysis of the non-perturbative Schwinger-Dyson equations. The large field redefinitions \eqref{eq:addpi} we shall employ involve adding a point-like instanton at the $\ib$'th gauge factor, or, conversely, removing a point-like instanton of the $\ib$'th type. This transformation maps one allowed quiver $\bn$-colored partition $\bla$  
to another one $\widetilde{\bla}$, with modified instanton charge
\beq
k_{\bf j} [\widetilde{\bla}] =  k_{\bf j} [{\bla}] \pm {\delta}_{\ib, \bf j} \ . 
\eeq An inspection of the picture ${\bf Fig.}6$ easily shows that the modifications of the indicated type consist of either adding a 
box ${\color{green}{\blacksquare}} \in {\partial}_{+}{\lambda}^{({\ib}, {\al}')}$ for some ${\al}' = 1, \ldots, n_{\ib}$, or removing a 
box ${\color{magenta}{\blacksquare}} \in {\partial}_{-}{\lambda}^{({\ib}, {\al}'')}$ for some ${\al}'' = 1, \ldots, n_{\ib}$. In other words, the allowed modifications of $\bla$ at the $\ib$'th node
correspond to the zeroes and poles of ${\y}_{\ib}(x)[{\bla}]$. 

The measures ${\bmu}_{\bla}( {\ba}; {\bmt}; {\ept} ) $ and
${\bmu}_{\widetilde{\bla}}( {\ba}; {\bmt}; {\ept} )$ are related to each other in a simple manner. Indeed, the character $\CalT_{\bla}$ is quadratic in $K_{\ib}$, more precisely, it is
sesquilinear. The variation ${\CalT}_{\widetilde{\bla}} - {\CalT}_{\bla}$ is, therefore, linear in $K_{\ib}$ and $K_{\ib}^{*}$. In fact, it is linear in $S_{\bf j}$'s
and $S_{\bf j}^{*}$'s. For the modification ${\bla} \to {\widetilde\bla}$ consisting of adding a box ${\color{green}{\blacksquare}} \in {\partial}_{+}{\lambda}^{({\ib}, {\al})}$ for some ${\al} = 1, \ldots, n_{\ib}$:
\begin{multline}
{\CalT}_{\widetilde{\bla}} - {\CalT}_{\bla} =  S_{\ib}[{\bla}] {\xi}^{-1} + S_{\ib}^{*}[{\widetilde\bla}] q {\xi} - M_{\ib}^{*}{\xi} - \\
\sum_{e\in t^{-1}({\ib})} S_{s(e)}^{*}[{\bla}] {\xi} q
e^{\beta{\mt}_{e}} - \sum_{e \in s^{-1}({\ib})} e^{\beta{\mt}_{e}} {\xi}^{-1} S_{t(e)}[{\bla}] 
+ \sum_{e \in s^{-1}({\ib}) \cap t^{-1}({\ib})} e^{\beta{\mt}_{e}} P
\end{multline} 
where
\beq
K_{\jb} [ {\widetilde\bla} ] = K_{\jb} [{\bla}] + {\delta}_{{\ib}, {\jb}} {\xi}, 
\qquad {\xi} = e^{{\beta}({\ac}_{{\ib}, {\al}} + c_{\color{green}{\blacksquare}})}
\eeq
The ratio of the measures can be, therefore, expressed as a product of the values and residues of various functions ${\y}_{\jb}(x)[{\bla}]$ in the variable $x$, for example, as
\begin{multline}
\frac{{\bmu}_{\widetilde{\bla}}( {\ba}; {\bmt}; {\ept} )}{{\bmu}_{\bla}( {\ba}; {\bmt}; {\ept} )}
=  (-1)^{{\kappa}_{\ib}} \ {\qe}_{\ib}  \frac{\ve}{{\ve}_{1}{\ve}_{2}} \, \frac{P_{\ib}(x)}{{\y}_{\ib}(x+{\ve})[{\bla}]{\y}_{\ib}^{\prime}(x)[{\bla}]} \, \times \\
\prod_{e \in t^{-1}({\ib})} {\y}_{s(e)}(x+{\ve}+{\mt}_{e}) [{\bla}]  \prod_{e \in  s^{-1}({\ib})}  {\y}_{t(e)}(x-{\mt}_{e})[{\bla}] \times \\
 \prod_{e \in s^{-1}({\ib}) \cap t^{-1}({\ib})} \frac{({\mt}_{e} + {\ve}_{1})({\mt}_{e} + {\ve}_{2})}{{\mt}_{e} ( {\mt}_{e} + {\ve}) } \\
 x = {\ac}_{{\ib}, {\al}} + c_{\color{green}{\blacksquare}}
 \label{eq:ratio}
\end{multline}
where
\beq
\kappa_{\ib} = n_{\ib}-1 + \sum_{e \in s^{-1}({\ib})} n_{t(e)} 
\label{eq:sign}
\eeq 
Note the identity:
\beq
{\rm res}_{x = {\ac}_{{\ib}, {\al}} + c_{\color{green}{\blacksquare}}} \ {\y}_{\ib}(x+{\ve})[{\widetilde\bla}] = \frac{{\ve}_{1}{\ve}_{2}}{\ve}
\ {\y}_{\ib}({\ac}_{{\ib}, {\al}} + c_{\color{green}{\blacksquare}}+{\ve})[{\bla}] \label{eq:resrel}
\eeq

\subsec{$\q$-observables}

The inspection of the Eq. \eqref{eq:yiblx} shows that
${\y}_{\ib}(x) [ {\bla} ]$ can be represented as a ratio of two entire functions, in two ways:
\beq
{\y}_{\ib}(x) = \frac{{\q}_{\ib}^{(1,2)}(x)}{{\q}_{\ib}^{(1,2)}(x - {\ve}_{2})} = \frac{{\q}_{\ib}^{(2,1)}(x)}{{\q}_{\ib}^{(2,1)}(x - {\ve}_{1})}\ , 
\label{eq:yfrq}
\eeq
where
\beq
{\q}^{(a,b)}_{\ib} (x) [ {\bla} ] \, =\,  \prod_{{\al}=1}^{n_{\ib}} \, \left( \frac{(-{\ve}_{b})^{\left( \frac{x - {\ac}_{\ib, \al}}{{\ve}_{b}} \right)}}{{\Gamma} \left( - \frac{x - {\ac}_{\ib, \al}}{{\ve}_{b}} \right)} \, \prod_{{\square} \in {\lambda}^{({\ib}, {\al})}} \frac{x - {\ac}_{\ib, \al} - c_{\square} - {\ve}_{a}}{x - {\ac}_{\ib, \al} - c_{\square}} \right), \qquad (a,b) = (1,2) \ {\rm or} \ (2,1) 
\label{eq:q12la}
\eeq
The r\^ole of these observables will be revealed in \cite{Nekrasov:2015ii}, \cite{Nekrasov:2015is}. In the limit
${\ve}_{2} \to 0$ with ${\ve}_{1}$-fixed the observables ${\q}_{\ib}^{(2,1)}$ tend to the so-called Baxter operators of the quantum integrable system, which is Bethe/gauge-dual \cite{Nekrasov:2009ui} to the gauge theory under consideration \cite{Nekrasov:2013xda}. 

\secc{Enter\ the\ {\it qq}\ -characters}

Remarkably, the Dyson-Schwinger relations based on \eqref{eq:ratio} can be summarized in the following proposition:

\subsec{ The main theorem}

{\sl For any $\gamma$-graded vector space 
\beq
\boldsymbol{W} = \bigoplus\limits_{{\ib} \in \Ver} W_{\ib}\ , 
\eeq
with the corresponding dimension vector ${\bw} \in {\BZ}_{\geq 0}^{\Ver}$,  $W_{\ib} = {\BC}^{{\wt}_{\ib}}$,  and a choice of $\ell$-weights
${\bnu} = ( {\nu}_{\ib} )_{{\ib} \in \Ver}$, ${\nu}_{\ib} = {\diag} \left( {\nu}_{{\ib}, 1}, \ldots , {\nu}_{{\ib}, {\wt}_{\ib}} \right) \in {\rm End}(W_{\ib})$, there is a Laurent polynomial (Laurent power series for affine $\gamma$) in ${\y}_{\jb}(x + {\xi}_{{\jb}, {\kappa}})$'s , i.e. in ${\y}_{\jb}$'s with possibly shifted arguments, including the nilpotent shifts (i.e. a
finite number of derivatives in $x$ applied to ${\y}_{\jb}$)
\begin{equation}
{\x}_{{\bw}, {\bnu}} \left( {\y} ( x + \ldots ) \right) = \prod_{\ib \in \Ver}\prod_{l=1}^{{\wt}_{\ib}} \ {\y}_{\ib}( x + {\nu}_{{\ib}, l} +{\ve}) \ + \  O ( {\qe} )   
\label{eq:qqdef} 
\end{equation}
such that its expectation value in the 
$\gamma$-quiver gauge theory:}
\beq \vev{{\x}_{{\bw}, {\bnu}} \left( {\y} \right)}_{\gamma} \equiv  
 \frac{1}{{\CalZ}^{\rm inst}_{\gamma}} 
\sum_{\bla} {\x}_{{\bw}, {\bnu}} \left( {\y}[{\bla}] \right)\, {\bqt}^{\bla} \,  {\bmu}_{\bla}( {\ba}; {\bmt}; {\ept} ) = 
T_{{\bw}, {\bnu}} (x), \label{eq:tyeq}
\eeq
{\sl is a polynomial in $x$}. More specifically,  
$T_{{\bw}, {\bnu}} (x)$ is  a  polynomial in $x$ of degree 
\beq
{\rm deg}\, T_{{\bw}, {\bnu}} (x) = {\bw} \cdot {\bn} = \sum_{{\ib} \in \Ver} {\wt}_{\ib} n_{\ib} \ . 
\label{eq:degt}
\eeq  
We call ${\x}_{{\bw}, {\bnu}}(x)$ the Yangian $qq$-character of $Y(\gq)$. For 
${\bw} = ({\delta}_{\jb, \ib})_{\jb \in \Ver}$ and ${\bnu}=0$ the corresponding $qq$-character
will be denoted by ${\chi}_{\ib}(x)$, the $\ib$'th {\it fundamental $qq$-character}. 
 
\noindent {\bf Remark.} The $qq$-characters are the gauge theory generalizations of the matrix model expression ${\bf T}(x)$  \eqref{eq:tdseq}. 

In the limit ${\ve}_{2} \to 0$ ${\x}_{{\bw}, {\bnu}}(x)$ reduces to the Yangian $q$-characters of finite-dimensional representations of the Yangian $Y({\gq})$, constructed for finite $\gamma$ in
\cite{Knight:1995}.  In  \cite{Frenkel:1998} the $q$-characters for the quantum affine algebras $U_{q}({\gq})$ for finite $\gamma$'s
and in \cite{Hernandez:2008} for affine $\gamma$'s are constructed. These correspond to the
$K$-theoretic version of our story in the limit $q_{2} \to 1$, $q_{1} = q$ finite, which was discussed in \cite{Nekrasov:2013xda}. 

The $K$-theoretic version of our story with general $(q_{1}, q_{2})$ produces the $qq$-characters, corresponding to $U_{q}({\gq})$. 
The physical applications of the $qq$-characters are
the five dimensional supersymmetric gauge theories compactified on
a circle \cite{Nekrasov:1996cz}. We shall give the definition and the formulae below, without going into much detail.

\secc{ Examples\ of\ {\it qq}\ -characters}

In this section we prepare the reader by giving a few explicit examples of the $qq$-characters, before unveiling the general
formula in the next section.

\subsec{$A$-type theories: one factor gauge group}

Let us start with a couple of examples for the theories with a single factor gauge group, 
i.e. either the $A_1$ theory or the ${\hat A}_{0}$ theory. 

\subsubsec{ The $A_{1}$ case.} 
The $A_{1}$ theory is the $U(n)$ gauge theory with $N_{f} = 2n$ fundamental hypermultiplets. The theory is characterized by the gauge coupling $\qe$ and $2n$ masses ${\bmt} = ({\mt}_{1}, \ldots , {\mt}_{2n})$, which are encoded in the polynomial 
\[ P(x) = \prod_{{\fe}=1}^{2n}\ ( x - {\mt}_{\fe} ) \]
Since the quiver consists of a single vertex, we omit the subscript
$\ib$ in ${\y}(x)$ and $P(x)$.

The fundamental $A_1$ $qq$-character is equal to 
\beq
{\x}_{1,0} (x) = {\y}(x+{\ve}) + {\qe}\ \frac{P(x)}{{\y}(x)}
\label{eq:fa1ch}
\eeq

The general $A_{1}$ $qq$-character depends on a $\tw$-tuple $\bnu$ of complex numbers, ${\bnu} = ({\nu}_{1}, \ldots , {\nu}_{\tw}) \in {\BC}^{\tw}$.  It 
is given by:
\begin{equation}
{\x}_{{\tw}, {\bnu}}(x) = 
\sum_{[{\tw}] = I \sqcup J}
{\qe}^{|J |} \prod_{i \in I, j \in J} {\bS}({\nu}_{i} - {\nu}_{j}) 
 \prod_{j \in J} \frac{P( x  + {\nu}_{j})}{{\y} (x + {\nu}_{j})}  \prod_{i \in  I} {\y}(x + {\ve} + {\nu}_{i}) 
\label{eq:a1nuch}
\end{equation}
It has potential poles in $\nu$'s, when ${\nu}_{i} = {\nu}_{j}$ or
${\nu}_{i} = {\nu}_{j} + {\ve}$, for $i \neq j$.

The expression \eqref{eq:a1nuch} is actually non-singular
at the diagonals ${\nu}_{i} = {\nu}_{j}$. The limit contains, however, the derivatives $\partial_{x}{\y}$. For example, for ${\wt} =2$, ${\nu}_{1}= {\nu}_{2} = 0$ the $qq$-character is equal to:
\begin{multline}
{\x}_{2,(0,0)} (x) = {\y}(x+{\ve})^{2} \left( 1 - {\qe} \frac{{\ve}_{1}{\ve}_{2}}{\ve} {\partial}_{x} \left( \frac{P(x)}{{\y}(x){\y}(x+{\ve})} \right) \right) + \\
+ 2 {\qe} P(x) \frac{{\y}(x+{\ve})}{{\y}(x)} \left( 1 - \frac{{\ve}_{1}{\ve}_{2}}{{\ve}^{2}} \right) + {\qe}^{2} \frac{P(x)^2}{{\y}(x)^2}
\label{eq:fa1ch2}
\end{multline}
The expression \eqref{eq:a1nuch}
has a first order pole at
the hypersurfaces where ${\nu}_{i} = {\nu}_{j} + {\ve}$ for some pair $i \neq j$. The residue of ${\x}_{{\tw}, {\boldsymbol{\nu}}}$ is equal to the $qq$-character ${\x}_{{\tw}-2, \boldsymbol{\nu} \backslash \{ {\nu}_{i}, {\nu}_{j} \}}$, times the polynomial in $x$ factor
\begin{equation}
\prod_{k \neq i,j} {\bS} ({\nu}_{k} - {\nu}_{j}) P ( x + {\nu}_{k}) \ .
\end{equation} 
The finite part ${\x}_{{\tw}, {\boldsymbol{\nu}}}^{\rm fin}$ of  the expansion of
${\x}_{{\tw}, {\boldsymbol{\nu}}}$ in $\nu_{i}$ near ${\nu}_{i} = {\nu}_{j} +{\ve}$ is the properly defined $qq$-character for the arrangement of weights $\boldsymbol{\nu}$ landing on the hypersurface ${\nu}_{i} = {\nu}_{j}+{\ve}$. It involves the terms with the derivative ${\partial}_{x}{\y}$. For example
\begin{multline}
{\x}_{2, (-{\ve},0)}^{\rm fin} = {\y}(x+{\ve}){\y}(x) + \\
+ {\qe} \left( 1 + \frac{{\ve}_{1}{\ve}_{2}}{2{\ve}^{2}} \right) P(x-{\ve})  \frac{{\y}(x+{\ve})}{{\y}(x- {\ve})} +
{\qe}P(x) \left( 1 -  \frac{{\ve}_{1}{\ve}_{2}}{\ve} \frac{{\partial}_{x}{\y}(x)}{{\y}(x)} \right) + \\
+ {\qe}^{2} \frac{P(x)P(x-{\ve})}{{\y}(x){\y}(x -{\ve})} 
\end{multline}

\subsubsec{ The ${\hat A}_{0}$ theory.}
The ${\hat A}_{0}$ theory (also known as the ${\CalN}=2^{*}$ theory) is characterized by one mass parameter $\mt$, the mass of the adjoint hypermultiplet, and the gauge coupling $\qe$. 

Here we
give the expression for the fundamental character ${\x}_{1}(x) \equiv {\x}_{1,0}(x)$:
\begin{multline}
{\x}_{1}(x) = 
\sum_{\lambda} {\qe}^{|{\lambda}|} \ \prod_{{\square} \in {\lambda}} \,
{\bS} ( {\mt} h_{\square} +{\ve} a_{\square} ) 
\cdot \frac{\prod_{{\square} \in {\partial}_{+}{\lambda}} {\y}(x + {\sigma}_{\square} + {\ve})}{\prod_{{\square} \in {\partial}_{-}{\lambda}} {\y}( x +
 {\sigma}_{\square})} = \\ 
= {\y} ( x + {\ve} )  \sum_{\lambda} {\qe}^{|{\lambda}|} \ \prod_{{\square} \in {\lambda}} 
{\bS} ( {\mt} h_{\square} +{\ve} a_{\square} ) 
\cdot \prod_{{\square} \in {\lambda} } \frac{{\y}(x + {\sigma}_{\square} - {\mt}){\y}(x + {\sigma}_{\square} + {\mt} + {\ve})}{ {\y} (x +
 {\sigma}_{\square}) {\y} (x +
 {\sigma}_{\square} + {\ve})} = \\
 = {\mathscr Y}(x+{\ve}) + \ {\qe} \ {\bS}({\mt}) \, \frac{\my{x-{\mt}} \my{x + {\ve} + {\mt}}}{\my{x}}+ \ldots
 \label{eq:cx10}
\end{multline}
Here 
\begin{equation}
{\sigma}_{\square} = {\mt}(i - j) + {\ve}(1-j)
\label{eq:cont10}
\end{equation}
is the content of ${\square}$ defined relative to the pair of weights $({\mt}, - {\mt} - {\ve})$.  
It is not too difficult to write an expression for the general ${\hat A}_{0}$ $qq$-character
${\x}_{{\wt}, {\boldsymbol{\nu}}}$, 
in terms of an infinite sum over the $\wt$-tuples of partitions,  but we feel it is not very illuminating.

Note that the expression \eqref{eq:cx10} has apparent singularities when ${\mt}$ and $-({\mt}+{\ve})$ are in a positive
congruence, i.e. if 
\beq
{\mt} (p - q) = {\ve} q
\eeq 
for some positive integers $p, q > 0$. In fact, the limit of the expression \eqref{eq:cx10} is finite, but it involves not only the
ratios
of shifted ${\y}$'s, but also its derivatives. The most efficient way to study this asymptotics is to use the geometric
expression to be discussed below. The geometric expression also leads to the contour integral representation
of the $qq$-characters. 
 
\subsec{$A$-type theories: linear quiver theories}

Let us now present the formulas for the general $A_r$ theories, assuming 
\beq
m_{1} = m_{r} = n_{1} = \ldots = n_{r} = N,  \
m_{2} = m_{3} = \ldots = m_{r-1} = 0 \ .
\label{eq:linq}
\eeq 
We  treat the general $A_{r}$ case $m_{i} = 2n_{i} - n_{i-1} - n_{i+1}$, $n_{0} = n_{r+1} = 0$  in the section below. 
We have $r$ observables ${\y}_{\ib}(x)$, and couplings ${\qe}_{\ib}$, for ${\ib} = 1, \ldots , r$. 
Define $r+1$ complex numbers $z_{i}$, $i = 0, 1, \ldots , r$ by \eqref{eq:zcor}: 
\beq
z_{i} = z_{0}\, {\qe}_{1} \ldots {\qe}_{i}, \qquad i = 1, \ldots , r
\label{eq:zfrq}
\eeq
and define $r+1$ functions $\Lambda_{i}(x)$, $i=0, \ldots, r$, by:
\begin{equation}
{\Lambda}_{i}(x) = z_{i} \, \frac{{\y}_{i+1}(x + {\ve})}{{\y}_{i}(x)}, 
\label{eq:laix}
\end{equation}
where we set ${\y}_{0}(x) = P_{1}(x)$, ${\y}_{r+1}(x) = P_{r}(x)$, in other words the masses ${\mt}_{1,f}$, ${\mt}_{r, f}$ of fundamentals are 
denoted as ${\ac}_{0,f}$, ${\ac}_{r+1, f}$, respectively.  We  also choose the normalization $m_{e} = - {\ve}$.

\subsubsec{The height functions}
For a {\sl finite} set $I \subset {\BR}$, we define the {\it height function}:
\beq
h_{I}: I \to [p], \qquad p = | I  |  \equiv {\#}I \, , \qquad 
h_{I}(i) = {\#}\, \{ \, i' \, | \, i' \in I, \ i' < i \, \} 
\label{eq:hf}
\eeq
In other words, for $I = \{ i_{1}, \ldots , i_{p} \}$ with $i_{1} < i_{2} < \ldots < i_{p}$, $h_{i_{b}} = b - 1$, $1 \leq b \leq p$. 
\subsubsec{Pre-character}
{}Define the $l$'th fundamental $qq$ {\sl pre-character} by (cf. \eqref{eq:hf}):
 \beq
 {\chi}_{l}(x) =  \sum_{I \subset [0,r], \, |I| = l} \prod_{i \in I} \ {\Lambda}_{i} \left( x + \left( h_{I}(i) + 1 - l \right) {\ve}\right)
 \label{eq:chilar}
 \eeq
 
 \subsubsec{Fundamental $qq$-character of type $A_r$}
 
 Then  $l$'th {\sl fundamental $qq$-character} is given by the properly normalized
 $\chi_{l}(x)$ :
 \beq
 {\x}_{l}(x) = 
 {\y}_{0}\left(x + \left(1-l \right){\ve} \right) \frac{{\chi}_{l}(x)}{z_{0}z_{1}\ldots z_{l-1}}
 = {\y}_{l}(x+{\ve}) + {\qe}_{l} \frac{{\y}_{l-1}(x){\y}_{l+1}(x+{\ve})}{{\y}_{l}(x)} + \ldots
\label{eq:chilari} 
\eeq

\subsubsec{Comparison to the $q$-characters of E.~Frenkel and N.~Reshetikhin} 
Our formula \eqref{eq:chilar} looks similar to the formula in the section $\bf 11.1$ of \cite{Frenkel:1996} (adapted to the Yangian   $Y({\mathfrak{sl}_{r+1}})$ , of course). The similarity is a little bit misleading, as shows the example of the $D$-type theories. 

\subsubsec{The main property of the $qq$-characters of the $A$-type}

The relations  \eqref{eq:ratio} suffice to prove that the expectation values of the $qq$-characters
of the $A$-type theories do not have singularities as the functions of $x$. It suffices to check the
cancellation of residues between the poles related by adding or removing one square in one
of the Young diagrams in $\bla$. It would take more elaborate arguments to prove 
the analogous claim for all ${\CalN}=2$ theories.

\vfill\eject
\subsec{ The $D$-type theories}

\subsubsec{ The $D_{4}$ theory}

This is the theory with four gauge groups. The quiver is the graph with 
$\Ver = \{ 1, 2, 3, 4 \}$, and $\Edg = \{ 1, 3, 4\}$ with $s(e) = e$, $t(e) = 2$
for all $e \in \Edg$. The graph has the obvious ${\CalS}_{3}$ symmetry, permuting the vertices $1,3,4$. 
We present the formula for the asymptotically conformal theory 
with $n_{1} = n_{3} = n_{4} = N = m_{2}$, $n_{2} = 2N$, $m_{1} = m_{3} = m_{4} = 0$, leaving an obvious extension to the general case to the interested reader as an exercise in the deciphering the general formula of the next section
(essentially one replaces ${\qe}_{i} \mapsto {\qe}_{i} P_{i}(x + a{\ve})$ for
$i = 1,3 ,4$ with some integers $a$). 

In what follows we use the short-hand notation:
\beq
Y_{\ib, a} = {\y}_{\ib}(x + a {\ve}), \ Y_{\ib} = {\y}_{\ib}(x), \ P_{a} = P_{2}(x + a {\ve}), \ P = P_{2}(x)
\eeq
There are four fundamental $qq$-characters, three of which are permuted by the ${\CalS}_{3}$.
The $qq$-character ${\x}_{1,0}$ is given by the sum of $8$ terms (as  would be the case for 
the characters of ${\bf 8}_{\bf v}, {\bf 8}_{\bf s}, {\bf 8}_{\bf c}$  of vector or  spinor representations of $Spin(8)$):
\begin{multline}
{\x}_{1,0} = Y_{1,1} + {\qe}_{1} Y_{1}^{-1}Y_{2}  + {\qe}_{1}{\qe}_{2} P_{-1} Y_{2,-1}^{-1} Y_{3}Y_{4}  + \\
+ {\qe}_{1}{\qe}_{2}{\qe}_{3} P_{-1} Y_{3,-1}^{-1} Y_{4}  + {\qe}_{1}{\qe}_{2}{\qe}_{4} P_{-1} Y_{4,-1}^{-1} Y_{3} + {\qe}_{1}{\qe}_{2}{\qe}_{3} {\qe}_{4} P_{-1} Y_{3,-1}^{-1}Y_{4,-1}^{-1} Y_{2,-1}  + \\
+ {\qe}_{1} {\qe}_{2}^{2} {\qe}_{3}{\qe}_{4} P_{-1}P_{-2} Y_{2,-2}^{-1} Y_{1,-1} + 
{\qe}_{1}^{2} {\qe}_{2}^{2} {\qe}_{3}{\qe}_{4} P_{-1}P_{-2} Y_{1,-2}^{-1} 
\label{eq:1qqD4}
\end{multline}
The formulae for ${\x}_{3,0}, {\x}_{4,0}$ are obtained by the cyclic permutation of the indices $1,3,4$. 
The $qq$-character ${\x}_{2,0}$ reveals a surprising structure,
\beq
{\x}_{2,0} = {\x}_{2}^{+} + {\qe}_{1}{\qe}_{2}^{2}{\qe}_{3}{\qe}_{4} P P_{-1} {\x}_{2}^{-}
\eeq
and contains the derivatives of ${\y}_{\ib}$'s. 
Explicitly,
\begin{multline}
{\x}_{2}^{+} = 
Y_{2,1} + {\qe}_{2}P Y_{2}^{-1} Y_{1,1}Y_{3,1}Y_{4,1} + {\qe}_{1}{\qe}_{2} P Y_{1}^{-1}Y_{3,1}Y_{4,1} + {\qe}_{3}{\qe}_{2} P Y_{3}^{-1}Y_{1,1}Y_{4,1} + {\qe}_{4}{\qe}_{2} P Y_{4}^{-1}Y_{1,1}Y_{3,1} + \\
{\qe}_{1}{\qe}_{2}{\qe}_{3} P Y_{1}^{-1}Y_{3}^{-1}Y_{2}Y_{4,1} +  {\qe}_{1}{\qe}_{2}{\qe}_{4} P Y_{1}^{-1}Y_{4}^{-1}Y_{2}Y_{3,1} +  {\qe}_{2}{\qe}_{3}{\qe}_{4} P Y_{3}^{-1}Y_{4}^{-1}Y_{2}Y_{1,1} + \\
+ {\qe}_{1}{\qe}_{2}^{2}{\qe}_{3} P P_{-1} Y_{2,-1}^{-1} Y_{4}Y_{4,-1}  + {\qe}_{1}{\qe}_{2}^{2}{\qe}_{4} P P_{-1} Y_{2,-1}^{-1} Y_{3}Y_{3,-1} + {\qe}_{3}{\qe}_{2}^{2}{\qe}_{4} P P_{-1} Y_{2,-1}^{-1} Y_{1}Y_{1,-1} + \\
+ {\qe}_{1}{\qe}_{2}{\qe}_{3}{\qe}_{4} P Y_{1}^{-1}Y_{3}^{-1}Y_{4}^{-1} Y_{2}^{2} \\
\label{eq:2qqD4i}
\end{multline}
\beq
{\x}_{2}^{-} = {\x}_{2}^{-,0} + {\x}_{2}^{--}, 
\eeq
\begin{multline}
{\x}_{2}^{-,0} =  \frac{Y_{2}}{Y_{2,-1}} \left( 2 \left( 1 - \frac{{\ve}_{1}{\ve}_{2}}{{\ve}^2} \right)  +
\frac{{\ve}_{1}{\ve}_{2}}{\ve} {\partial}_{x} {\rm log} \left( \frac{Y_{2}Y_{2,-1}}{P_{-1} Y_{1}Y_{3}Y_{4}} \right) \right) +  \\
+ \left( 1 + \frac{{\ve}_{1}{\ve}_{2}}{2{\ve}^{2}} \right) \left(
Y_{1,-1}^{-1}Y_{1,1} + Y_{3,-1}^{-1}Y_{3,1} + Y_{4,-1}^{-1}Y_{4,1} \right)  \\
\label{eq:2qqD4ii}
\end{multline}
\begin{multline}
{\x}_{2}^{--} = {\qe}_{4} Y_{4}^{-1}Y_{4,-1}^{-1}Y_{2} + {\qe}_{3} Y_{3}^{-1}Y_{3,-1}^{-1}Y_{2} + {\qe}_{1} Y_{1}^{-1}Y_{1,-1}^{-1}Y_{2} + 
{\qe}_{2} P_{-1} Y_{2,-1}^{-2} Y_{1}Y_{3}Y_{4} + \\
{\qe}_{2}{\qe}_{4} P_{-1} Y_{2,-1}^{-1}Y_{4,-1}^{-1}Y_{1}Y_{3} + 
{\qe}_{2}{\qe}_{3} P_{-1} Y_{2,-1}^{-1}Y_{3,-1}^{-1}Y_{1}Y_{4} +
{\qe}_{2}{\qe}_{1} P_{-1} Y_{2,-1}^{-1}Y_{1,-1}^{-1}Y_{3}Y_{4} + \\
+ {\qe}_{1}{\qe}_{2}{\qe}_{4} P_{-1} Y_{1,-1}^{-1}Y_{4,-1}^{-1}Y_{3} + 
{\qe}_{2}{\qe}_{3}{\qe}_{4} P_{-1} Y_{3,-1}^{-1}Y_{4,-1}^{-1}Y_{1} +
{\qe}_{1}{\qe}_{2}{\qe}_{3} P_{-1} Y_{1,-1}^{-1}Y_{3,-1}^{-1}Y_{4} + \\
+ {\qe}_{1}{\qe}_{2}{\qe}_{3}{\qe}_{4} P_{-1} Y_{1,-1}^{-1}Y_{3,-1}^{-1}Y_{4,-1}^{-1} Y_{2,-1} + \\
+ {\qe}_{1}{\qe}_{2}^{2} {\qe}_{3}{\qe}_{4} P_{-1}P_{-2} Y_{2,-2}^{-1} \\
\label{eq:2qqD4iii}
\end{multline}

\secc{ The\  {\it qq}-character\ formula}

In order to write the general formula for the $qq$-character we shall use an auxiliary geometric object, the quiver variety ${\qM}({\bw}, {\bv})$, which we presently define.

\subsec{Nakajima quiver variety}

Given a quiver $\gamma$, two dimension vectors 
\beq
{\bv} = (v_{\ib})_{{\ib} \in\Ver} , \, {\bw} = (w_{\ib})_{{\ib} \in \Ver} \, \in {\BZ}_{\geq 0}^{\Ver}
\label{eq:vwvec}
\eeq 
and a choice of {\it stability parameter}
${\bzt} \in {\BR}^{\Ver}$ H.~Nakajima \cite{Nakajima:1994r} defines the quiver variety as the hyperkahler quotient:
\begin{equation}
{\qM}_{\gamma, \bzt} ({\bw}, {\bv}) = {\mu}_{\BC}^{-1}(0) \cap {\mu}^{-1}_{\BR}({\bzt}) / G_{\bv}
\label{eq:mvw}
\end{equation}
where 
\beq
G_{\bv} = \varprod\limits_{{\ib} \in \Ver} \ U(V_{{\ib}})\ , 
\label{eq:qgr}
\eeq
and ${\mu}_{\BC}$, ${\mu}_{\BR}$ are the quadratic maps ${\CalH}_{\gamma} \to {\rm Lie}G_{\bv}^{*} \otimes {\BC}, {\BR}$, respectively, with ${\CalH}_{\gamma}$ the vector space
\begin{equation}
{\CalH}_{\gamma} = T^* \left( \bigoplus_{{\ib} \in \Ver} {\rm Hom}(V_{{\ib}}, W_{{\ib}})
\oplus \bigoplus_{e\in \Edg} {\rm Hom}(V_{s(e)}, V_{t(e)}) \right)
\label{eq:chga}
\end{equation}
of linear operators (matrices) $({\qI}_{\ib}, {\qJ}_{\ib}, B_{e,{\pm}})$:
\begin{equation}
\begin{aligned}
& {\qI}_{\ib}: W_{{\ib}} \to V_{{\ib}}, \quad {\qJ}_{\ib}: V_{{\ib}} \to W_{{\ib}} \\
& B_{e,+}: V_{s(e)} \to V_{t(e)}, \quad B_{e,-} : V_{t(e)} \to V_{s(e)} \end{aligned}
\label{eq:quivdata}
\end{equation}
Explicitly: ${\mu}_{{\BR}, {\BC}} = ( {\mu}^{\BR}_{{\ib}}, {\mu}^{\BC}_{{\ib}} )_{{\ib} \in \Ver}$, with
\begin{equation}
\begin{aligned}
{\mu}^{\BC}_{{\ib}} \ = \ & {\qI}_{\ib} {\qJ}_{\ib} + 
\sum_{e \in s^{-1}({\ib})} B_{e,-} B_{e,+} - \sum_{e \in t^{-1}({\ib})} B_{e,+} B_{e,-} \\
{\mu}^{\BR}_{{\ib}} \ = \ &  {\qI}_{\ib}{\qI}_{\ib}^{\dagger} \, - {\qJ}_{\ib}^{\dagger}{\qJ}_{\ib}  + \sum_{e \in s^{-1}({\ib})} B_{e,-} B_{e,-}^{\dagger} - 
B_{e,+}^{\dagger} B_{e,+}    \ + \\
& \qquad\qquad + \sum_{e \in t^{-1}({\ib})} B_{e,+} B_{e,+}^{\dagger} - B_{e,-}^{\dagger} B_{e,-}  
 \end{aligned}
\end{equation}
 The definition \eqref{eq:mvw} translates to the set of equations:
 \beq
 \begin{aligned}
&  {\mu}^{\BR}_{{\ib}} = {\zeta}_{{\ib}} \, {\bf 1}_{V_{\ib}}\ ,  \\
&  {\mu}^{\BC}_{{\ib}} = 0, \qquad {\ib} \in \Ver
 \label{eq:momm}\end{aligned}
 \end{equation}
 with the identification of solutions related by the $G_{\bv}$ transformations:
 \begin{equation}
 ( B_{e,+} , B_{e,-}, {\qI}_{\ib}, {\qJ}_{\ib} ) \mapsto ( h_{t(e)} B_{e,+} h^{-1}_{s(e)}, h_{s(e)} B_{e,-} h_{t(e)}^{-1} , h_{\ib} {\qI}_{\ib}, {\qJ}_{\ib} h_{\ib}^{-1} ), \qquad h_{\ib} \in U(V_{\ib}) 
 \label{eq:gvac}
 \end{equation}

\subsubsec{Stability parameters}

Solving the real moment map equations (the first line in the Eq. \eqref{eq:momm}) and dividing by $G_{\bv}$ can be replaced by dividing the set of \emph{stable} solutions to the complex moment map equations (the second line in the Eq. \eqref{eq:momm}) by the action of the complexified group 
\beq
G_{\bv}^{\BC} = \times_{\ib}\ GL(V_{\ib}) \ .
\label{eq:qgvc}
\eeq 
The notion of stability depends on the choice of $\bzt$. In this paper we assume ${\zeta}_{\ib} > 0$ for all $\ib \in \Ver$.  The solution $(B_{e, \pm}, {\qI}_{\ib}, 
{\qJ}_{\ib})$ is stable iff any collection $(V_{\ib}^{\prime})_{\ib \in \Ver}$ of subspaces
$V_{\ib}^{\prime} \subset V_{\ib}$, such that
\beq
\begin{aligned}
& \ (1) \qquad  {\qI}_{\ib} (W_{\ib}) \subset V_{\ib}^{\prime}\ {\rm for\ all}\quad \ib \in \Ver
\quad {\rm and}  \\
& \ (2) \qquad {\CalB}_{p} (V_{s(p)}^{\prime}) \subset V_{t(p)}^{\prime},\ {\rm and}\quad
{\tilde\CalB}_{p} (V_{t(p)}^{\prime}) \subset V_{s(p)}^{\prime}\qquad {\rm for\ all}\quad p \in \Path\\
 \label{eq:stabquiv}
 \end{aligned}
 \eeq
 is such that $V_{\ib}^{\prime} = V_{\ib}$ for all $\ib \in \Ver$. 
 Here $p\in \Path$ denotes a sequence $(e_{i}, {\sigma}_{i}) \in \Arr$, $i = 1, \ldots , m$, 
 such that ${\bar s}(e_{1}, {\sigma}_{1})  = s(p)$, ${\bar t}(e_{1}, {\sigma}_{1}) = {\bar s}(e_{2}, 
 {\sigma}_{2})$,  $\ldots $, ${\bar t}(e_{i}, {\sigma}_{i}) = {\bar s}(e_{i+1}, 
 {\sigma}_{i+1})$, $\ldots$, $ {\bar t}(e_{m}, {\sigma}_{m}) = t(p)$.

 {}The proof is simple. Let $P_{\ib}$ be the orthogonal projection
 of $V_{\ib}$ onto the orthogonal complement $(V_{\ib}^{\prime})^{\perp}$ of the ``invariant'' subspace $V_{\ib}^{\prime} \subset V_{\ib}$. We have $P_{\ib} {\qI}_{\ib} = 0$ and
 $P_{t(e)} B_{e,+} (1 - P_{s(e)}) = 0$, $P_{s(e)} B_{e,-} (1 - P_{t(e)}) = 0$. 
 Now compute
 \begin{multline}
0 \leq \sum_{\ib} {\zeta}_{\ib} {\rm dim}\left( V_{\ib}^{\prime} \right)^{\perp} =  \sum_{\ib} {\Tr} P_{\ib} {\mu}_{\ib}^{\BR} P_{\ib}  = 
 - \sum_{\ib} {\Tr} \left( P_{\ib} {\qJ}_{\ib}^{\dagger} {\qJ}_{\ib} P_{\ib} \right) + \\
 \sum_{e} {\Tr} P_{s(e)} B_{e,-} B_{e,-}^{\dagger} P_{s(e)} +
 {\Tr} P_{t(e)} B_{e,+} B_{e,+}^{\dagger} P_{t(e)} - {\Tr} P_{t(e)} B_{e,-}^{\dagger} B_{e,-} P_{t(e)} - {\Tr} P_{s(e)} B_{e,+}^{\dagger} B_{e,+} P_{s(e)} = \\
  - \sum_{\ib} \Vert {\qJ}_{\ib} P_{\ib} \Vert^2 - 
 \sum_{e}  \left(  \Vert (1 - P_{s(e)}) B_{e,-} P_{t(e)} \Vert^2 + \Vert ( 1  - P_{t(e)} ) B_{e,+} P_{s(e)} \Vert^2  \right)  \leq 0
\label{eq:zlz}
\end{multline}
which implies $V_{\ib}^{\prime} = V_{\ib}$ for all $\ib$. The stability condition is equivalent to the condition that
the path operators ${\CalB}_{p}$ and ${\tilde\CalB}_{p}$ for $p \in \Path$ acting on ${\qI}_{\ib'}(W_{\ib'})$ generate $V_{\ib''}$:
\beq
V_{\ib} = \sum_{p \in t^{-1}({\ib})} {\BC} {\CalB}_{p} \, {\qI}_{s(p)} (W_{s(p)}) +  \sum_{p \in s^{-1}({\ib})} {\BC} {\tilde\CalB}_{p} \, {\qI}_{t(p)} (W_{t(p)})
\label{eq:pathgen}
\eeq
Conversely, in order to establish that the stability condition implies that the $G_{\bv}^{\BC}$-orbit  of $(B_{e, \pm}, {\qI}_{\ib}, {\qJ}_{\ib})$ solving the ${\mu}^{\BC}=0$ equations in \eqref{eq:momm} passes through the solution of the $\underline{{\mu}^{\BR}} = {\bzt}$ equations in \eqref{eq:momm}, we use the standard method: 
consider the
Morse-Bott function
\beq
f = \sum_{\ib \in \Ver} \Vert {\mu}_{\ib}^{\BR} - {\zeta}_{\ib} {\bf 1}_{V_{\ib}} \Vert^2
\label{eq:mbf}
\eeq  
The trajectory of the gradient flow 
\beq
\frac{d}{dt} (B_{e, \pm}, {\qI}_{\ib}, {\qJ}_{\ib}) = - ({\nabla}_{B_{e, \pm}^{\dagger}} f, {\nabla}_{{\qI}_{\ib}^{\dagger}} f , {\nabla}_{{\qJ}_{\ib}^{\dagger}} f)
\label{eq:gradf}
\eeq
belongs to the $G_{\bv}^{\BC}$-orbit. Indeed, \eqref{eq:gradf} exponentiates to the transformation:
\beq
 {\exp} \, t {\mu}_{\ib} \in GL(V_{\ib})
 \eeq
The function $f$ decreases along the flow. In the limit $t \to \infty$ the value of $f$ either tends to its absolute minimum, i.e. $f =0$, which is the locus of solutions to the equations \eqref{eq:momm}, or it stops at another critical point with the critical value $f_{*} > 0$. 
Now, the critical points with $f_{*} > 0$ are the configurations $(B_{e, \pm}, {\qI}_{\ib}, {\qJ}_{\ib})$ for which the real moment map $( {\mu}_{\ib} )_{\ib \in \Ver}$ viewed as an element of the Lie algebra of $G_{\bv}^{\BC}$ (more precisely, it is in ${\ii} \, {\rm Lie}G_{\bv} \subset {\rm Lie}G_{\bv}^{\BC}$), is a non-trivial infinitesimal symmetry, i.e.:
\beq
\begin{aligned}
& {\mu}_{\ib} {\qI}_{\ib} = {\zeta}_{\ib} {\qI}_{\ib}, \qquad {\qJ}_{\ib} {\mu}_{\ib} = {\qJ}_{\ib}  {\zeta}_{\ib} \\
& {\mu}_{s(e)} B_{e, -} - B_{e, -} {\mu}_{t(e)}  = ( {\zeta}_{s(e)} - {\zeta}_{t(e)} ) B_{e,-} , \qquad {\mu}_{t(e)} B_{e, +} - B_{e, +} {\mu}_{s(e)} = 
( {\zeta}_{t(e)} - {\zeta}_{s(e)} ) B_{e,+} \\
 \label{eq:fpmm}
 \end{aligned}
 \eeq
Define $V_{\ib}^{\prime} = {\rm ker} \left( {\mu}_{\ib} - {\zeta}_{\ib} {\bf 1}_{V_{\ib}} \right) \subset V_{\ib}$ for all $\ib \in \Ver$. By \eqref{eq:fpmm} these subspaces obey all the conditions of \eqref{eq:stabquiv}, therefore ${\mu}_{\ib} = {\zeta}_{\ib} {\bf 1}_{V_{\ib}}$ for all $\ib \in \Ver$. 

 In what follows we omit the subscripts $\gamma$ and $\bzt$ in the notations for the quiver variety:
 \[  {\qM}_{\gamma, \bzt} ({\bw}, {\bv})\qquad \longrightarrow \qquad {\qM} ({\bw}, {\bv}) \]

\subsubsec{ Symmetries of ${\qM}({\bw}, {\bv})$}
 The group  ${\Hf}_{\bw} = G_{\bw} \times U(1)^{b_{*}({\gamma})}$ acts on ${\qM}({\bw}, {\bv})$ by isometries. Here
 \beq
 G_{\bw} = \varprod\limits_{{\ib} \in \Ver} U(W_{\ib})
 \label{eq:gbwg}
 \eeq
 acts on the ${\qI}, {\qJ}$ maps:
 \beq
 (g_{\ib})_{{\ib} \in \Ver} : ({\qI}_{\ib}, {\qJ}_{\ib})_{{\ib} \in \Ver} \mapsto 
 ({\qI}_{\ib} g_{\ib}, g_{\ib}^{-1}{\qJ}_{\ib})_{{\ib} \in \Ver}
 \label{eq:gw}
 \eeq
The $U(1)^{b_{0}({\gamma})} = U(1)$-factor acts by rotating all of the ${\qI}_{\ib},  B_{e,-}$'s while keeping
 ${\qJ}_{\ib}, B_{e,+}$'s intact (this definition can be extended to the disconnected quivers in a trivial fashion: rotate ${\qI}_{\ib}, B_{e,-}$ belonging to a given connected component):
 \beq
 ({\qI}_{\ib}, {\qJ}_{\ib}, B_{e, +} , B_{e, -} ) \mapsto (u {\qI}_{\ib}, {\qJ}_{\ib}, u B_{e, +} , B_{e, -} )
 \label{eq:urot}
 \eeq
  The group $U(1)^{b_{1}({\gamma})} \approx U(1)^{\Edg}/U(1)^{\Ver}$ acts on the $G_{\bv}$-equivalence classes of the $(B_{e,\pm})$ maps:
  \begin{equation}
 (u_{e})_{e \in \Edg} : (B_{e,+}, \, B_{e,-})_{e \in \Edg} \mapsto (u_{e}B_{e,+}, \, u_{e}^{-1}B_{e,-})_{e \in \Edg} \label{eq:guv}
 \end{equation}
 so that the normal subgroup $U(1)^{\Ver}$ acts by the $G_{\bv}$-transformations
 \begin{equation}
 (u_{\ib})_{{\ib} \in \Ver} : (B_{e,+}, \, B_{e,-})_{e \in \Edg} \mapsto (u_{s(e)}^{-1}u_{t(e)}B_{e,+}, \, u_{s(e)}u_{t(e)}^{-1}B_{e,-})_{e \in \Edg} \label{eq:gv}
 \end{equation}

 \subsubsec{ The canonical complexes and bundles}
 
 For each $\ib \in \Ver$ the vector space $V_{\ib}$ descends to ${\qM}({\bw}, {\bv})$ as a vector bundle. In addition, 
 there are also the canonical complexes of bundles over ${\qM}({\bw}, {\bv})$:
 \begin{equation}
 {\CalC}_{\ib} = \left[ 0 \to V_{\ib} \longrightarrow^{\kern -.2in d_{2}\kern .2in} W_{\ib} \bigoplus_{e \in t^{-1}({\ib})}
 V_{s(e)} \bigoplus_{e \in s^{-1}({\ib})} V_{t(e)} \longrightarrow^{\kern -.2in d_{1}\kern .2in} V_{\ib} \to 0 \right]
 \label{eq:ccom}
 \end{equation}
 where the first and the second maps are given by:
 \beq
 d_{2} = {\qJ}_{\ib} \bigoplus_{e \in t^{-1}({\ib})}
 \left( - B_{e,-} \right) \bigoplus_{e \in s^{-1}({\ib})} B_{e,+} \, , \qquad
 d_{1} =  {\qI}_{\ib} \bigoplus_{e \in t^{-1}({\ib})}
 B_{e,+} \bigoplus_{e \in s^{-1}({\ib})} B_{e,-}  \label{eq:d12}
 \eeq
 The moment map equation \eqref{eq:momm}, ${\mu}^{\BC}_{\ib} = 0$, implies $d_{1} \circ d_{2} = 0$, hence ${\CalC}_{\ib}$ is a complex. 
We set the leftmost term $V_{\ib}$ in \eqref{eq:ccom} to be in degree zero.

\subsec{The bi-observables}
Let 
  \begin{equation}
  {\mathscr G}_{x} =  e^{{\beta} x} \sum_{{\ib} \in \Ver} \left( q S_{\ib}^{*}{\CalC}_{\ib} + M_{\ib}^{*}V_{\ib} \right)
  \label{eq:biobs}
  \end{equation} 
  denote the Chern character of the ${\Hf}_{\bw}$-equivariant complex of vector bundles over
  ${\iM}({\bn},{\bkt}) \times {\qM}({\bw}, {\bv})$:
  \[  {\mathscr G}_{x} = e^{{\beta}x} \,  
  Ch \,\bigoplus_{{\ib} \in \Ver} \left( \, q\, \left[ \, S_{\ib} \to {\CalC}_{\ib}\, \right]  \oplus
  \left[ \, M_{\ib} \to V_{\ib}\, \right] \, \right)  \]
  We identify the equivariant parameters of the $G_{\bw}$ group with ${\bnu}$, 
  the equivariant parameters for the $U(1)$ factor with ${\ve}$, 
  the equivariant parameters for the $U(1)^{\Edg}$-group with 
  ${\mt}_{e} + {\ve}$. 
  Explicitly, the equivariant Chern character of the complex ${\CalC}_{\ib}$ is equal to:
  \begin{equation}
 Ch\,  {\CalC}_{\ib} =    W_{\ib} - V_{\ib}  - q^{-1} V_{\ib} + \sum_{e \in t^{-1}({\ib})} q^{-1} e^{-{\beta}{\mt}_{e}} V_{s(e)} + \sum_{e \in s^{-1}({\ib})} e^{{\beta}{\mt}_{e}}  V_{t(e)} 
  \end{equation}
  where $W_{\ib}$ is a pure $c$-number character (the sum of exponents of $\bnu$-components), while $V_{\jb}$'s are the Chern characters of ${\Hf}_{\bw}$-equivariant bundles, i.e. may have components of positive degrees cohomology 
  classes. 
\subsec{ The formula}
  Finally, we can present the formula for ${\x}_{{\bw},{\bnu}}$. There are  several ways to write it. 
  
  \subsubsec{ Integral over the quiver variety}
  \begin{equation}
  {\x}_{{\bw},{\bnu}} = \sum_{{\bv}}\ {\bqt}^{\bv}\ \int_{{\qM}({\bw}, {\bv})}
  {\ep}_{{\ve}_{2}} ( T{\qM}({\bw}, {\bv}) ) {\ep}_{x} ({\mathscr G} )
  \label{eq:cxwn}
  \end{equation}
\[ {\bqt}^{\bv} = \prod_{{\ib} \in \Ver} {\qe}_{\ib}^{v_{\ib}} \, , \]
 ${\ep}_{x}({\mathscr G})$ is understood as the 
   ${\Hf}_{\bw}$-equivariant cohomology class of ${\qM}({\bw}, {\bv})$: represent ${\CalC}_{\ib}$ as the virtual bundle 
   ${\CalC}_{\ib}^{+}- {\CalC}_{\ib}^{-}$ over ${\qM}({\bw}, {\bv})$, where
   ${\CalC}_{\ib}^{+}$, ${\CalC}_{\ib}^{-}$ are the actual bundles, with the formal Chern roots
   ${\xi}_{\ib, \kappa_{\pm}}^{\pm}$. Then 
   \begin{equation}
   {\ep}_{x}({\mathscr G}) = \prod_{{\ib} \in \Ver} \left(  \ \frac{\prod\limits_{\kappa_{+}} \ {\y}_{\ib} ( x + {\xi}^{+}_{\ib, \kappa_{+}} )}{\prod\limits_{\kappa_{-}} \ {\y}_{\ib} ( x + {\xi}^{-}_{\ib, \kappa_{-}} )} \, \prod_{{\kappa}=1}^{v_{\ib}} P_{\ib}(x + {\eta}_{{\ib}, \kappa}) \ \right)
\label{eq:ifc}
\end{equation}
where ${\eta}_{{\ib}, \kappa}$ are the formal Chern roots of $V_{\ib}$. 
  
For the $A_{1}$, ${\hat A}_{0}$ examples we considered so far the quiver varieties
 ${\qM}({\bw}, {\bv})$ are the cotangent bundle $T^{*}$Gr$({\vt}, {\wt})$ to the Grassmanian of $\vt$-planes in ${\BC}^{\wt}$
  and the Hilbert scheme ${\Hilb}^{[{\vt}]}({\BC}^{2})$ of $\vt$ points on ${\BC}^{2}$, respectively. 
  
\subsubsec{ Contour integral representations} Equivalently, one can write a contour integral representation for \eqref{eq:cxwn} which has the advantage of being explicit, albeit less concise: 
\begin{multline}
{\x}_{{\bw}, {\bnu}}(x) = \sum_{{\bv}}  \prod_{{\ib} \in \Ver}  \frac{1}{{\vt}_{\ib}!} \left( \frac{{\ve} \, {\qe}_{\ib} }{2\pi \sqrt{-1}\, {\ve}_{1}{\ve}_{2}} \right)^{{\vt}_{\ib}} \ \prod_{j =1}^{{\wt}_{\ib}} {\mathscr Y}_{\ib}(x + {\ve}+{\nu}_{{\ib},j})\ \oint_{{\Gamma}_{{\bw}, {\bnu},{\bv}}} {\Upsilon}_{{\bw}, {\bnu},{\bv}}(x)\\
{\Upsilon}_{{\bw}, {\bnu},{\bv}}(x) = \prod_{e\in \Edg} {\Upsilon}_{{\bw}, {\bnu},{\bv}; e}(x) \prod_{{\ib} \in \Ver} {\Upsilon}_{{\bw}, {\bnu},{\bv}; {\ib}}(x)\, , \\
{\Upsilon}_{{\bw}, {\bnu},{\bv}; e}(x) = \prod_{{\kappa}=1}^{{\vt}_{t(e)}} {\mathscr Y}_{s(e)}(x + {\ve}+{\mt}_{e}+ {\phi}^{(t(e))}_{\kappa}) \prod_{{\ell}=1}^{{\vt}_{s(e)}}{\mathscr Y}_{t(e)}(x - {\mt}_{e} + {\phi}^{(s(e))}_{\ell})  \, ,  \\
{\Upsilon}_{{\bw}, {\bnu},{\bv}; {\ib}}(x) =   \prod_{{\kappa}=1}^{{\vt}_{\ib}} 
\left(  \frac{d{\phi}^{({\ib})}_{\kappa}\, P_{\ib}(x+{\phi}^{({\ib})}_{\kappa})}{{\y}_{\ib}(x + {\ve}+{\phi}^{({\ib})}_{\kappa}){\y}_{\ib}(x+{\phi}^{({\ib})}_{\kappa})}\prod_{{\ell} \neq {\kappa}} {\bS} ( {\phi}^{({\ib})}_{\kappa} - {\phi}^{({\ib})}_{\ell} ) \ 
\prod_{j =1}^{{\wt}_{\ib}} {\bS} ( {\phi}^{({\ib})}_{\kappa} - {\nu}_{{\ib},j} )  \right) \, . 
\label{eq:cxwv}
\end{multline} 
The contour ${\Gamma}_{{\bw}, {\bnu},{\bv}}$  is chosen in such a fashion, so as to ignore the poles coming from the zeroes of the ${\y}$-functions in the denominator of \eqref{eq:cxwv} or the poles of ${\y}$-functions in the numerator there. Let us assume that ${\nu}_{{\ib}, j}$ are all real, and that ${\ve}_{1}, {\ve}_{2}$ have positive imaginary part. We also assume that zeroes and poles of ${\y}(x+z)$ in $z$ are far away from the real axis. Then the contour ${\Gamma}_{{\bw}, {\bnu},{\bv}} \approx {\BR}^{|{\bv}|}$, i.e. all ${\phi}^{({\ib})}_{\kappa}$ are real. Now deform ${\nu}_{{\ib}, j}$ and ${\ve}_{1}, {\ve}_{2}$ to whatever values we desire, all the while deforming the contour ${\Gamma}_{{\bw}, {\bnu},{\bv}}$ in a such a way, that the poles of \eqref{eq:cxwv} in ${\phi}^{({\ib})}_{\kappa}$'s do not cross ${\Gamma}_{{\bw}, {\bnu},{\bv}}$. 

The technique to arrive from \eqref{eq:cxwv} to \eqref{eq:cxwn} is well-known, see, e.g. \cite{Moore:1997dj}

\subsec{ Five dimensional theory}

The gauge theories we studied so far in four dimensions canonically lift to five dimensions, with the vector multiplets lifting to vector multiplets. 
The complex scalars ${\ac}_{{\ib}, {\alpha}}$ in the vector multiplet
in four dimensions come from a real scalar in five dimensions
and the fifth component of the gauge field. Now we compactify the
theory on a circle of circumference ${\beta}$, and impose the twisted boundary conditions, rotating the space ${\mN}$ by the angles $( - {\ii}{\beta}{\ve}_{1} , - {\ii}{\beta}{\ve}_{2})$ in the two orthogonal two-planes ${\bR}^{2}$ in ${\mN} = {\bR}^{4}$.  In addition we perform the $SU(2)$ $R$-symmetry rotation 
\[ {\exp} \  \frac{{\ii}{\beta}\ve}{2}  {\sigma}_{3} \] 
and the constant gauge transformation
\[ e^{{\beta}{\ac}_{\ib}} = {\diag}( e^{{\beta}{\ac}_{{\ib}, {\alpha}}} )_{{\alpha}=1, \ldots , {\bv}_{\ib}} \ . \] 

The observables ${\y}_{\ib}(x)$ generalize to:
\begin{equation}
\begin{aligned}
{\CalY}_{\ib}(z) = z^{n_{\ib}} \, {\exp}\, \left( -\sum_{k=1}^{\infty} \frac{1}{k z^{k}} \, {\rm Ch}\, {\psi}^{k}{\CalS}_{\ib}  \right) = \\
= {\rm Det} \left( z - e^{{\beta}{\Phi}_{\ib}\vert_{0}} \right) \\
\end{aligned}
\label{eq:yibz}
\end{equation}
Again, as in the four dimensional theory, the non-perturbative effects make the naive polynomial in the right hand side of \eqref{eq:yibz} a rational function. In particular, on the $U(1) \times U(1)$ invariant instanton configuration $\bla$ the observable ${\CalY}_{\ib}(z)$ evaluates to:
\begin{equation}
{\CalY}_{\ib}(z)[{\bla}] = 
{\displaystyle\prod\limits_{{\al}=1}^{n_{\ib}}} \  
 \frac{{\textstyle\prod\limits_{{\color{green}{\blacksquare}} \in {\partial}_{+}{\lambda}^{({\ib}, {\al})}}} ( z - e^{{\beta}({\ac}_{{\ib}, {\al}} + c_{\color{green}{\blacksquare}})} )}{{\textstyle\prod\limits_{{\color{magenta}{\blacksquare}} \in {\partial}_{-}{\lambda}^{({\ib}, {\al})}}} ( z -  q e^{{\beta}({\ac}_{{\ib}, {\al}} + c_{\color{magenta}{\blacksquare}})} )}  \ .
 \label{eq:yiblxq}
\end{equation}
 The $K$-theoretic version of the $qq$-characters is defined in a similar fashion,  one should use the ${\chi}_{q_{2}^{-1}}$-genus instead of the Chern polynomial
and to use push forwards in equivariant $K$-theory instead of the equivariant integrals. The formula \eqref{eq:cxwn} generalizes to:
\begin{equation}
{\CalX}_{{\bw},{\bnu}}(z) = \sum_{{\bv}, j} {\qe}^{\bv} q_{1}^{-{\hat n}} (-q_{2})^{{\hat n}-j}\int_{{\qM}({\bw}, {\bv})} Td_{T{\qM}({\bw}, {\bv})}
  \, Ch  \left( \bigwedge^{j} T{\qM}({\bw}, {\bv}) \right)  {\Xi}_{z} [  {\mathscr F}_{\bv} ]
  \label{eq:cxwnqq}
  \end{equation}
 where ${\hat n} = {\rm dim}_{\BC} {\qM}({\bw}, {\bv})$
 \begin{equation}
{\Xi}_{z} [  {\mathscr F}_{\bv} ] = \prod_{{\ib} \in \Ver} \left( \prod_{\kappa} P_{\ib}(z e^{{\eta}_{{\ib}, \kappa}}) \frac{\prod\limits_{\kappa_{+}} {\mathscr Y}_{\ib} ( z e^{{\xi}^{+}_{\ib, \kappa_{+}}})}{\prod\limits_{\kappa_{-}} {\mathscr Y}_{\ib} ( z e^{{\xi}^{-}_{\ib, \kappa_{-}}} )} \right)
\label{eq:xifi}
\end{equation}
where
\begin{equation}
\begin{aligned}
Ch ( {\CalC}_{\ib}^{\pm} ) = \sum_{{\kappa}^{\pm}} 
e^{{\xi}^{\pm}_{{\ib}, {\kappa}_{\pm}}} \\
Ch ( V_{\ib} ) = \sum_{\kappa} e^{{\eta}_{{\ib}, {\kappa}}} \\
\label{eq:chrts}
\end{aligned}
\end{equation}

\subsec{The symmetry} ${\ve}_{1} \leftrightarrow {\ve}_{2}$, $q_{1} \leftrightarrow q_{2}$. 

\bigskip

{}The formulas \eqref{eq:cxwn}, \eqref{eq:cxwv}, \eqref{eq:cxwnqq} are not obviously symmetric with respect to the exchange ${\ve}_{1} \leftrightarrow {\ve}_{2}$, $q_{1} \leftrightarrow q_{2}$. However, the symmetry becomes clear once we recall that ${\qM}({\bw}, {\bv})$ is a holomorphic symplectic manifold. Its tangent bundle is isomorphic to the cotangent bundle, the isomorphism being provided by the holomorphic symplectic form ${\omega}^{\BC}$, which descends from the canonical symplectic form on ${\CalH}_{\gamma}$:
\beq
 \sum_{e \in \Edg} 
{\Tr} {\delta}B_{e} \wedge {\delta}{\tilde B}_{e} + 
\sum_{{\ib} \in \Ver} {\Tr} {\delta} {\qI}_{\ib} \wedge {\delta} {\qJ}_{\ib}
\label{eq:holquiv}
\eeq
Since the symplectic form ${\omega}^{\BC}$ is scaled as 
${\omega}^{\BC} \to q^{-1}{\omega}^{\BC}$  by the action of $\Hf_{\bw}$, the equivariant Chern character  
\beq
\sum_{j=0}^{\hat n} (-q_{2})^{-j} Ch \left( \bigwedge^{j} T{\qM}({\bw}, {\bv}) \right) = \prod_{l =1}^{\hat n} ( 1 - e^{{\sl x}_{l}}q_{2}^{-1} ) = (q_{1}/q_{2})^{{\hat n}/2} \prod_{l =1}^{\hat n} ( 1 - e^{-{\sl x}_{l}} q_{2} ) 
\label{eq:chj}
\eeq 
which is equal to $(q_{1}/q_{2})^{{\hat n}/2} \prod_{l =1}^{\hat n} ( 1 - e^{{\sl x}_{l}} q_{1}^{-1} )$ since every equivariant virtual Chern root ${\sl x}_{l}$ is paired with another equivariant virtual Chern root ${\beta}({\ve}_{1} + {\ve}_{2}) - {\sl x}_{l}$. Thus, 
\[ \sum_{j=0}^{\hat n}  q_{1}^{-{\hat n}} (-q_{2})^{{\hat n}-j}
  \, Ch  \left( \bigwedge^{j} T{\qM}({\bw}, {\bv}) \right)  = 
  \sum_{j=0}^{\hat n}  q_{2}^{-{\hat n}} (-q_{1})^{{\hat n}-j}
  \, Ch  \left( \bigwedge^{j} T{\qM}({\bw},{\bv}) \right)  \]

\subsec{Convergence of the integrals}

The integrals \eqref{eq:cxwnqq} may be divergent. Indeed, the quiver varieties
${\qM}({\bw}, {\bv})$ are non-compact. We understand the integrals \eqref{eq:cxwnqq} as the integrals in ${\Hf}_{\bw}$-equivariant cohomology. Practically this means that the differential
form representative for the integrand in \eqref{eq:cxwnqq} contains a factor (cf. \eqref{eq:intone}):
\beq
{\exp} \, D \left( {\bg} ( \cdot , V ({\bar\xi}) ) \right) = e^{- {\bg} ( V ({\xi} ) , V ({\bar\xi}) ) } \times \left( 1 + \ldots \right)
\label{eq:regul} 
\eeq
Here, 
\beq
D = {\rm d} + \iota_{V({\xi})} 
\eeq
is the equivariant de Rham differential, $\xi, {\bar \xi} \in {\rm Lie}({\Hf}_{\bw}^{\BC})$, ${\xi}$ is the collective notation for $({\bnu}, {\ve})$, the equivariant parameters, $V : {\rm Lie}({\Hf}_{\bw}) \to {\rm Vect}({\qM}({\bw}, {\bv}))$ is the infinitesimal action of ${\Hf}_{\bw}$ on  ${\qM}({\bw}, {\bv})$, and
${\bg}$ is any ${\Hf}_{\bw}$-invariant metric on ${\qM}({\bw}, {\bv})$, in which ${\bg} ( V ({\xi} ) , V ({\bar\xi}) )$ grows for generic ${\xi}$ and ${\bar\xi} \approx {\xi}^{*}$ sufficiently fast at ``infinity'' of ${\qM}({\bw}, {\bv})$. 

With this convergence factor understood the integrals over ${\qM}({\bw}, {\bv})$ converge. Moreover, the $D$-exactness of the exponential in \eqref{eq:regul} means that small variations of ${\bar\xi}$ or ${\bg}$ with fixed $\xi$ do not change the integral. The result does, however, depend on $\xi$.
Indeed, for special values of $\xi = ({\bnu}, {\ve})$ it diverges, as we saw in the examples \eqref{eq:a1nuch}. 
Our point is that it converges for all values of $x$. We shall establish this fact in full generality in 
\cite{Nekrasov:2015iim} and  \cite{Nekrasov:2015ii}.

\subsec{Reduction to the fixed loci}

The integrals \eqref{eq:cxwnqq}, \eqref{eq:cxwn} can be computed by localization with respect to the ${\Hf}_{\bw}$-action on ${\qM}({\bw}, {\bv})$. The isolated fixed points contribute rational expressions in ${\y}$'s with shifted arguments, while positive dimension components of the fixed locus contribute terms with derivatives of $\y$'s.

The character of the virtual tangent bundle to the quiver variety
can be transformed to
\beq
T^{\rm virt}{\qM}_{\gamma}({\bw}, {\bv}) \leadsto P \left( \sum_{\ib \in \Ver} ( W_{\ib} - V_{\ib} ) V_{\ib}^{*}  + 
\sum_{e \in \Edg} e^{{\beta}m_{e}} V_{t(e)} V_{s(e)}^{*} \right) 
\label{eq:virttqm}
\eeq
Indeed, the tangent bundle to ${\qM}_{\gamma}({\bw}, {\bv})$ is equal to:
\beq
T{\qM}_{\gamma}({\bw}, {\bv}) = \sum_{\ib \in \Ver} \left( ( W_{\ib} - V_{\ib} ) V_{\ib}^{*} + q^{-1}( W_{\ib} - V_{\ib} )^{*} V_{\ib} \right)  + \sum_{e \in \Edg} \left( e^{{\beta}m_{e}} V_{t(e)} V_{s(e)}^{*}  + q^{-1}e^{-{\beta}m_{e}} V_{s(e)} V_{t(e)}^{*} \right) 
\label{eq:tqmwv} 
\eeq
in the $G_{\bw} \times U(1)^{b_{*}({\gamma})}$-equivariant $K$-theory of ${\qM}_{\gamma}({\bw}, {\bv})$. The virtual tangent bundle is equal to 
\beq
T^{\rm virt}{\qM}_{\gamma}({\bw}, {\bv}) = (1- q_{1}) T{\qM}_{\gamma}({\bw}, {\bv}) \ . 
\label{eq:virtqui}
\eeq
Now, dualize the terms in \eqref{eq:tqmwv} proportional to $q^{-1}$: 
\beq
( 1 - q_{1} ) q^{-1} T \leadsto ( 1 - q_{1}^{-1} ) q T^{*} =  - ( 1 - q_{1} ) q_{2} T^{*}
\eeq
to arrive at \eqref{eq:virttqm}. 

Let ${\bT}_{\gamma, \bw}$ denote the maximal torus in $G_{\bw} \times U(1)^{b_{*}({\gamma})}$. The set of  ${\bT}_{\gamma, \bw}$-fixed points ${\qM}_{\gamma}({\bw}, {\bv})^{{\bT}_{\gamma, \bw}}$ is a union
\beq
{\qM}_{\gamma}({\bw}, {\bv})^{{\bT}_{\gamma, \bw}} = \bigcup_{\bc}\ {\qM}_{\gamma, {\bc}}({\bw}, {\bv})
\label{eq:qmwvc}
\eeq
of connected components. Each component is a product 
\beq
{\qM}_{\gamma, {\bc}}({\bw}, {\bv}) = \varprod_{{\ib} \in \Ver} \varprod\limits_{{\beta}=1}^{{\wt}_{\ib}} \varprod_{n \in L} {\qM}_{\gamma, {\ib}, c_{\ib, \beta}}({\bf e}_{\ib}, {\bv}_{\ib, \beta, n})
\eeq
for some ${\bv}_{\ib, \beta} \in {\BZ}_{\geq 0}^{\Ver}$ such that
\beq
\sum_{{\ib} \in \Ver} \sum\limits_{{\beta}=1}^{{\wt}_{\ib}}{\bv}_{\ib, \beta} = {\bv}
\label{eq:bvfxp}
\eeq
Here we used a notation
\beq
{\qM}_{\gamma, {\ib}, c}({\bf e}_{\ib}, {\bv}) \subset {\qM}_{\gamma} ({\bf e}_{\ib}, {\bv}), \quad c \in {\Ci}
\label{eq:mfunc}
\eeq
for a connected component of the set of ${\bT}_{\gamma, {\bf e}_{\ib}}$-fixed points of Nakajima quiver variety ${\qM}_{\gamma}({\bw}, {\bv})$ with
${\bw} = {\bf e}_{\ib}$:
\beq
{\qM}_{\gamma} ({\bf e}_{\ib}, {\bv})^{{\bT}_{\gamma, {\bf e}_{\ib}}} = \bigsqcup_{c \in {\Ci}} {\qM}_{\gamma, {\ib}, c}({\bf e}_{\ib}, {\bv})
\label{eq:conncom}
\eeq
The vector bundles $W_{\ib}$, $V_{\ib}$, restricted onto each connected component ${\qM}_{\gamma, {\ib}, c}({\bf e}_{\ib}, {\bv})$ of the fixed point set splits, as a sum of ${\bT}_{\gamma, {\bw}}$-equivariant vector bundles:
\beq\begin{aligned}
&  W_{\ib} = \bigoplus_{\beta =1}^{w_{\ib}} \ e^{{\nu}_{\ib, \beta}} \\
& V_{\ib} = \bigoplus_{\beta =1}^{w_{\ib}} \bigoplus_{n} \  e^{{\nu}_{\ib, \beta}} \otimes q^{-n} \, V_{\ib, \beta, n} \\
\label{eq:decovw}
\end{aligned}
\eeq
Here $n$ runs over the lattice of representations of $U(1)^{b_{*}({\gamma})}$, and $q^{n}$ stands for the corresponding character. In all cases except for the affine $A$-type quivers, $q$ is literally $q = q_{1}q_{2}$, and $n$ runs through some lattice $L \subset {\BZ}$. In the ${\hat A}$-case, $q^{-n} = (q_{1}q_{2} e^{\mt})^{-n_{1}} e^{n_{2} \mt}$ for $(n_{1}, n_{2}) \in L \subset {\BZ} \oplus {\BZ}$.

\subsubsec{Example: the $A_1$ case}

Recall the expression \eqref{eq:fa1ch2} for the $A_1$ $qq$-character corresponding to ${\wt}=2$, ${\bnu} = (0,0)$. The corresponding quiver varieties ${\qM}_{A_{1}}(2, {\vt})$,
with ${\vt} = 0, 1, 2$ are the point, $T^{*}{\BC\BP}^{1}$, and another point, respectively. The contribution of $\vt = 1$, i.e. the integral over $T^{*} {\BC\BP}^{1}$ reduces, by the fixed point formula, to the integral over $F = {\BC\BP}^{1}$. 

The vector bundle $V$ reduces to $L^{-1} \approx {\CalO}(-1)$, 
The character-bundle \eqref{eq:virttqm} specifies to:
\[ T^{\rm virt}{\qM} = P ( W - V)V^{*} = P (2 L - 1) \] 
the tangent bundle to $\ell$ is equal to
\[ TF = 2L - 1\] 
(check: $L$ has two sections, while $TF$ has three), 
while the complex $\CalC$ becomes $q( 2 - L^{-1}) - L^{-1}$.  
The contribution of $F$ to the formula \eqref{eq:cxwn} is given by:
\begin{multline} 
{\qe}{\y}(x+ {\ve})^{2} \frac{\ve}{{\ve}_{1}{\ve}_{2}} \int_{\ell} \left( \frac{( {\omega} + {\ve}_{1} ) ( {\omega} + {\ve}_{2} )}{{\omega} + {\ve}} \right)^{2} \frac{P(x - {\omega})}{{\y}(x + {\ve} - {\omega}) {\y}(x - {\omega})}  = \\
- {\qe} {\y}(x+{\ve})^{2}  \frac{{\ve}_{1}{\ve}_{2}}{\ve} {\partial}_{x} \left( \frac{P(x)}{{\y}(x){\y}(x+{\ve})} \right)  + \\
+ 2 {\qe} P(x) \frac{{\y}(x+{\ve})}{{\y}(x)} \left( 1 - \frac{{\ve}_{1}{\ve}_{2}}{{\ve}^{2}} \right) 
\label{eq:fa1reprt}
\end{multline}
where $\omega = c_{1}(L)$, $\int_{F} {\omega} = 1$. It is evident that \eqref{eq:fa1reprt} reproduces the ${\qe}^{1}$ term in \eqref{eq:fa1ch2}. 
\subsubsec{Example: the $D_4$ case}

The $D_{4}$
fundamental character ${\x}_{2}$ provides a representative example. 

Let
${\bw} = (0,1,0,0)$, ${\bv} = (1,2,1,1)$, with ${\Ver} = \{ 1, 2,3,4 \}$. In this subsection
${\qM}_{4} = {\qM}_{D_{4}} ({\bw}, {\bv})$. 

The character \eqref{eq:virttqm} specifies to
\beq
T^{\rm virt}{\qM}_{4} \leadsto P \left( ( 1 - V_{2}) V_{2}^{*} - 3 + V_{2} ( V_{1}^{*}+V_{3}^{*}+V_{4}^{*} ) \right)
\eeq

Now, the three terms in the second line of \eqref{eq:2qqD4ii} are
coming from the isolated fixed points $p_{i}$, $i=1,3,4$ in ${\qM}_{4}$, with 
$W_{2}= 1, V_{2} = 1 + q^{-1}, V_{i} = q^{-1}$, $V_{j} = 1$, $ j \neq i$,$j=1,3,4$. 
This gives $T_{p_{i}}^{\rm virt} {\qM}_{4} \leadsto P q = q  + q^2 - q q_{1} - q q_{2}$, which translates to the factor
$$
\frac{( {\ve}+{\ve}_{1} )({\ve}+ {\ve}_{2})}{{\ve} \cdot 2{\ve}} = 1  + \frac{{\ve}_{1}{\ve}_{2}}{2{\ve}^{2}}
$$
in the second line of \eqref{eq:2qqD4ii}.

The first line in \eqref{eq:2qqD4ii} is the contribution of the non-isolated component of the
fixed point set, the fixed projective line line $F= {\mathbb P}^{1} \subset {\qM}_{4}$, with 
$W_{2}= V_{1} = V_{3} = V_{4} = 1$, $V_{2} = 1 + q^{-1} L$, where $L \approx {\CalO}(-1)$ is a 
non-trivial line bundle over $F$.
 
The corresponding complexes ${\CalC}_{\ib}$ are given by:
\beq
\begin{aligned}
& {\CalC}_{i} =  V_{2} - ( 1 + q^{-1} ) V_{i}, \qquad i = 1, 3, 4 \\
& {\CalC}_{2} =  W_{2} \oplus  \, q^{-1} \left( V_{1} + V_{3} + V_{4} \right)  - \left( 1+ q^{-1} \right)V_{2}    
\end{aligned}
\label{eq:cci2}
\eeq
For $F$ this gives: 
\beq
T^{\rm virt}{\qM}_{4} \vert_{F} = P \left( 2  q^{-1}L - 1 \right) = 
2 (q^{-1} + 1 - q_{1}^{-1} - q_{2}^{-1} ) L - 1 + q_{1} + q_{2} - q
\eeq
The restriction of ${\CalC}_{i}$ onto $F$ is given by:
\beq
\begin{aligned}
& {\CalC}_{i} = L - 1, \qquad i = 1,3,4   \\
& {\CalC}_{2} = 2 - ( 1 +q^{-1} )  L  
\label{eq:ccompl}
\end{aligned}
\eeq
The tangent bundle to $F$ is given by 
\beq
TF = 2 L - 1
\eeq
The corresponding contribution to ${\CalX}_{2,0}$ is the integral over $F$ of the equivariant Euler class of
$T{\qM}_{4} \vert_F$ with the equivariant parameter ${\ve}_{1}$, divided by 
the equivariant Euler class of the virtual normal bundle 
$N^{\rm virt}_{F \subset {\qM}_{4}} = T^{\rm virt}{\qM}_{4} \vert_F - TF$, which is equal to
\beq
N^{\rm virt}_{F \subset {\qM}_{4}} \leadsto (1-q_{1}) T{\qM}_{4} - TF \leadsto 2 ( q^{-1} - q_{1}^{-1} - q_{2}^{-1} ) L + q_{1} + q_{2} - q
\eeq
times the product of $\y$-observables: 
\beq
\begin{aligned}
& \int_{F} \frac{\ve}{{\ve}_{1}{\ve}_{2}} \, \left( \frac{({\omega} + {\ve}_{1}) ({\omega} + {\ve}_{2})}{{\omega} + {\ve}} \right)^{2} \,  \, \frac{P(x)P(x-{\ve}-{\omega}) {\y}_{2}(x)^{2}}{{\y}_{2}( x- {\omega}) {\y}_{2}( x - {\omega} - {\ve})} \, \prod_{i=1,3,4} \frac{{\y}_{i} ( x - {\omega} )}{{\y}_{i}( x)}  = \\
&  \qquad\qquad P P_{-1} \frac{{\y}_{2}}{{\y}_{2,-1}}  \left( 2 \left(1 - \frac{{\ve}_{1}{\ve}_{2}}{{\ve}^2} \right) +  \frac{{\ve}_{1}{\ve}_{2}}{\ve} {\partial}_{x} {\rm log} \left( \frac{{\y}_{2} {\y}_{2,-1}}{ P_{-1} {\y}_{1}{\y}_{3} {\y}_{4}} \right) \right)
\end{aligned}
\label{eq:contrell}
\eeq

\secc{ More\ on\ the\ Physics\ of\ {\it qq}\ -characters}

Let $G$ be a Lie group, and $(R, {\pi})$ its representation, i.e. 
$\pi$ is a group homomorphism ${\pi}: G \to {\rm End}(R)$. The character ${\chi}_{R}(g)$ is a (generalized) function on $G$,  given by the trace of the matrix ${\pi}(g)$ in the representation $R$:
\begin{equation}{\chi}_{R}(g) = {\rm Trace}_{R} \, {\pi}(g)
\label{eq:trr}
\end{equation} By definition $\chi_{R}$ is an adjoint-invariant function, i.e. the function on the space of conjugacy classes:
\begin{equation}
{\chi}_{R}(g) = {\chi}_{R} (h^{-1} g h) , \qquad {\rm for\ any}\ h \in G
\end{equation}
For the compact Lie group $G$ the space of conjugacy classes
$G/Ad(G) = T/W$ is the quotient of the maximal torus $T \subset G$ by the action of discrete group, the Weyl group. 

\subsec{ Characters from supersymmetric quantum mechanics}

A familiar realization of a character ${\chi}_{R}(e^{h})$ in quantum mechanics as the partition functions of a quantum mechanical system with $G$-symmetry, whose space of states is the representation $R$ and the Hamiltonian is a realization ${\pi}(h)$ of an element $h \in t$ of the Lie algebra $t = {\rm Lie}T$ of the maximal torus $T$. 
 
 For example, if $G$ is a compact Lie group, and $R$ is a unitary representation, corresponding to the highest weight $\lambda \in t^{*}$, then the {\it geometric quantization} program associates $R$ to the symplectic manifold ${\Xf} = G/K_{\lambda} \subset {\bg}^{*}$, the coadjoint orbit  of $\lambda$, with the canonical Kirillov-Kostant symplectic form $\omega_{\Xf}/\hbar$. The geometric quantization realizes $R$ as the space of holomorphic sections of the pre-quantization line bundle $L$ over $\Xf$ (which is a K\"ahler manifold), such that $c_{1}(L) = [ \frac{\omega_{\Xf}}{2\pi \hbar} ] \in H^{2}({\Xf}, {\BZ})$. 
\begin{equation}
R = H^{0}({\Xf}, L)
\label{eq:rh0}
\end{equation}
This correspondence extends, with some friction, to a wider class of groups and representations \cite{Kirillov:1999}.

There are various explicit formulas for the character $\chi_{R}$, due to Harish-Chandra, Weyl, Kirillov, and Kac \cite{Pressley:1986}, \cite{Kac:1990}. 
For the dominant weight $\lambda$ the line bundle $L$ has vanishing higher degree cohomology, so that
\begin{equation}
{\rm Trace}_{R} = \sum_{i} (-1)^{i} {\rm Trace}_{H^{i}({\Xf}, L)} 
\label{eq:str}
\end{equation} 
One interpretation of the Kac-Weyl character formula is the equivariant Riemann-Roch-Grothendieck formula applied to \eqref{eq:str}. 

Physically, one takes $({\Xf}, {\omega}_{\Xf})$ as the phase space of the mechanical system. For the Hamiltonian one takes the function $h$ defined as: for $x \in {\Xf}$,  $h(x) = \langle i(x), t \rangle$, where
 $i : {\Xf} \to {\mathfrak g}^{*}$ is the embedding, and $t$ is some fixed element of $t \subset {\mathfrak g}$. Then the character can be realized as the path integral \cite{alekseev:1988, alekseev:1989}:
 \begin{equation}
 {\chi}_{R}(g) = \int\,  D{\Xf}\ {\exp} \  \frac{\ii}{\hbar}  \oint  \left( d^{-1}{\omega}_{\Xf} -  h(x(t)) dt \right)
\label{eq:chipi}
\end{equation} 
where the integral is taken over the space $L{\Xf}$ of parametrized loops 
$x: S^{1} \to {\Xf}$. The loop space $L{\Xf}$ is acted upon by the torus
$T \times U(1)$, where $T$ acts pointwise on $\Xf$, and $U(1)$ acts by the loop rotations: $e^{2\pi {\ii}s} \cdot x (t) = x(t+s)$.

The integral \eqref{eq:chipi} can be evaluated exactly by the infinite-dimensional version of the Duistermaat-Heckman formula.
The loop space $L\Xf$ is viewed as the symplectic manifold with the symplectic form being the integral (``a point-wise sum") 
\begin{equation}
{\Omega} =  \frac 12\ \int_{S^{1}} dt \ {\omega}_{\mu\nu} {\psi}^{\mu} {\psi}^{\nu} 
\label{eq:omg}
\end{equation}
Then the action $\oint  \left( d^{-1}{\omega}_{\Xf} -  h(x(t)) dt \right)
$ is interpreted as the Hamiltonian, generating a one-parametric subgroup in $T \times U(1)$. 
 
The character formula can be also interpreted with the help of the 
supersymmetric quantum mechanics on $\Xf$,  \cite{AlvarezGaume:1983at, Hietamaki:1991qp}. Instead of $X$ one takes the supermanifold ${\Yf} = {\Pi}T{\Xf} \otimes T^{*}{\Xf}$ (the total space of the sum of the cotangent bundle and the tangent bundle with fermionic fibers over $\Xf$). $\Yf$ is endowed with the even symplectic form (as opposed to the BV formalism, where the symplectic form is odd):
\begin{equation}
{\omega}_{\Yf} =  dp_{\mu} \wedge dx^{\mu} + 
g_{\mu\nu}   d{\psi}^{\mu} \wedge d{\psi}^{\nu}
\label{eq:omy}
\end{equation}
where $g_{\mu\nu}$ is a metric on $\Xf$. 
We study the quantum mechanics on $\Yf$ with the Hamiltonian 
\begin{equation}
H_{\Yf} = \frac 12 g^{\mu\nu} \left( p_{\mu} - {\Gamma}_{\mu\tilde\nu}^{\tilde\kappa} g_{\tilde\kappa\tilde\lambda} {\psi}^{\tilde\nu} {\psi}^{\tilde\lambda} \right) \left( p_{\nu} - {\Gamma}_{\nu\hat\nu}^{\hat\kappa} g_{\hat\kappa\hat\lambda} {\psi}^{\hat\nu} {\psi}^{\hat\lambda} \right) 
\label{eq:hamy}
\end{equation}
The supersymmetry is generated by the odd function
\begin{equation}
{\CalQ} = {\psi}^{\mu} \left( p_{\mu} - {\Gamma}_{\mu\tilde\nu}^{\tilde\kappa} g_{\tilde\kappa\tilde\lambda} {\psi}^{\tilde\nu} {\psi}^{\tilde\lambda} \right)
\label{eq:cqch}
\end{equation}
If the target space $\Xf$ itself is a moduli space of solutions to some partial differential equations involving gauge fields on a $d$-dimensional space $B^d$, e.g. vortices in $d=2$, monopoles in $d=3$, or instantons in $d=4$, then the quantum mechanics on $\Xf$ is a low energy approximation to the $d+1$-dimensional gauge theory. The character \eqref{eq:chipi} would be then given by the path integral in the theory on ${\BS}^{1} \times B^{d}$. The parameters $t \in {\bf g}$ in the limit of shrinking ${\BS}^{1}$ would be interpreted as the (twisted) mass parameters for the flavor symmetry $G$ acting on the moduli space $\Xf$.  
If $d=3$, then using the three dimensional mirror symmetry we can exchange these parameters for the Fayet-Illiopoulos parameters for the dual target space ${\Xf}^{\vee}$. Then, the weight subspaces would identify with the contribution of components of fixed topology. This is almost as good as the statement of our theorem.

\subsec{ Gauge theory realization of the $qq$-characters.}
 
 In this section we expand on the
interpretation of $qq$-characters we sketched in the section ${\bf 2.4}$ 
using the intersecting branes, or its string dual. 

We shall mainly consider the case of affine 
$\gamma$. The theories corresponding to the finite dimensional $A,D,E$-type quivers can be viewed as limits of the affine ones, by sending some of the gauge couplings to zero. For example, the $A_1$ theory is a limit
of ${\hat A}_{2}$ theory with two out of three gauge couplings sent to zero. 

As we recalled above the ${\CalN}=2$ quiver gauge theories with affine $A,D,E$ quivers can be realized as the low energy limit of the theory on a stack of $n$ $D3$-branes placed at the tip of the ${\BC}^{2}/{\Gamma}$ singularity, with ${\Gamma} \subset SU(2)$ the McKay dual finite group.

For definiteness, let ${\mN} \approx {\BR}^{4}$ denote the worldvolume of the stack of the `physical' $D3$ branes. The six dimensional transversal slice splits as a product ${\mW}/{\Gamma} \times {\BR}^2_{\phi}$. Here ${\mW} \approx {\BR}^{4}$ is the Euclidean cover of the singularity. The fluctuations along ${\BR}^{2}_{\phi}$ are represented, in the $D3$ theory, by the adjoint complex scalar in the vector multiplet.   

We want to  subject the theory to the  $\Omega$-deformation. To this purpose we choose a point ${\bf 0} \in {\mN}$, and use the symmetry of rotations about $\bf 0$. 

The superconformal vacuum of the theory on $D3$-branes corresponds to the branes located at the origin in $\mW$, fixed by the $\Gamma$ action and at some point $p$ in $\Sigma$. Let us identify $\Sigma \approx {\BC}$ and $p$ with $0 \in {\BC}$. 
 
Let us now add a stack of ${\wt}$ $D3$-branes located at $ 0 \times x \in {\mN} \times \Sigma$, with the worldvolume being a copy of ${\mW}/{\Gamma}$. Here $x \in {\BC}$ is a complex number. 

The low energy configurations in the combined system of $n + {\wt}$ $D3$ branes split into two orthogonal stacks are labelled by some continuous and discrete parameters, such as the separation of branes along $\Sigma$ and the choice of flat $U({\wt})$ connection at infinity ${\BS}^{3}/{\Gamma}$ of ${\mW}/{\Gamma}$, i.e. a homomorphism ${\rho}: {\Gamma} \to U({\wt})$. 

When $\Gamma = 1$ is trivial, the supersymmetry of the combined system of branes is consistent with $\Omega$-deformation, which uses the subgroup $SU(2)_{{\mN}, L} \times SU(2)_{\Delta} \times SU(2)_{{\mW}, L}$ of the group 
\[ Spin(4)_{\mN} \times Spin(4)_{\mW} = SU(2)_{{\mN},L} \times SU(2)_{{\mN},R} \times SU(2)_{{\mW},L} \times SU(2)_{{\mW},R} \] of rotations of the two orthogonal ${\BR}^{4}$'s.

The open string Hilbert space splits as a sum of the spaces corresponding to the strings stretched between different types of D-branes. We have the $(-1)-(-1)$ strings connecting the D(-1)-instantons, we have the $(-1)-3$ strings connecting the D(-1)-instantons to the stack of $n$ D3-branes, we have the $(-1)-3'$ strings connecting the D(-1)-instantons to the stack of $\wt$ D3 branes. There are also the $3-3'$ open strings. In \cite{Nekrasov:2015iim} we define the moduli space using the low-energy
modes of these open strings. 

The $qq$-character ${\x}_{\bw, {\bnu}}(x)$ is simply the observable in the original theory on the stack of $n$ $D3$-branes living along $\mN$, which is obtained by integrating out the degrees of freedom on the transversal $\wt$ $D3$-branes, in the vacuum corresponding to the particular $\bw$ and the vacuum expectation values $\bnu$ of the scalars in the vector multiplets living on ${\mW}/{\Gamma}$.

The integral \eqref{eq:cxwnqq} can be interpreted (cf. \cite{Nekrasov:1996cz}) as the partition function of the supersymmetric quantum mechanics on the moduli space of Yang-Mills instantons on the ALE gravitational instantons $\widetilde{{\BR}^{4}/{\Gamma}_{\gamma}}$, constructed in \cite{Kronheimer:1990nk}.  Here ${\Gamma}_{\gamma} \subset SU(2)$ is the MacKay dual to $\gamma$ discrete group, whose representation theory is encoded in the quiver $\gamma$: $\Ver = {\Gamma}_{\gamma}^{\vee}$. The gauge group $U({\wt})$, with 
 \[ {\wt} = \sum_{\ib \in {\Gamma}_{\gamma}^{\vee}} {\wt}_{\ib} {\rm dim}{\CalR}_{\ib} \]
 is broken at infinity, by the choice of flat connection $ {\Gamma}_{\gamma} \to U({\wt})$, to a subgroup 
 \[ U({\wt}) \longrightarrow \varprod\limits_{\ib \in \Ver} U({\wt}_{\ib}) \]
 The dimensions ${\bv}$ encode the magnetic fluxes through the exceptional two-spheres in the resolved orbifold ${\widetilde{{\BR}^{4}/{\Gamma}_{\gamma}}}$, and the ordinary instanton charge
 \[  v_{\rm tot} = \sum_{\ib \in {\Gamma}_{\gamma}^{\vee}} {\vt}_{\ib} {\rm dim}{\CalR}_{\ib} \]
 In the supersymmetric quantum mechanics the sum over ${\bv}$ is not natural, as it adds the partition functions of different Hilbert spaces. However, in the $4+1$ dimensional gauge theory on ${\widetilde{{\BR}^{4}/{\Gamma}_{\gamma}}} \times {\BS}^{1}$ the sum over $\bv$ is just the sum over various topological sectors which is enforced anyway by the cluster decomposition. A more careful look at \eqref{eq:cxwnqq} reveals that we are dealing with the maximally supersymmetric Yang-Mills theory 
 subject to the $\Omega$-deformation (more on it below) and coupled 
to a point-like source localized along the circle ${\BS}^{1} \times {\tilde 0}$, where ${\tilde 0}$ is the exceptional variety, the joint of the exceptional spheres, which arose in the resolution of singularities.  
 
The coupling ${\qe}^{\bv}$ comes naturally from the usual Chern-Simons couplings on the $D4$-branes
\beq
\int C_{1} \wedge {\Tr} ( F - B_{NS} ) \wedge (F - B_{NS}) + \int C_{3} \wedge {\Tr} (F - B_{NS})
\label{eq:cscd}
\eeq 
In the IIB picture these would translate to the couplings to 
\[ \int_{{\BS}^{2}_{i}} B_{NS} + {\tau} B_{RR}  = {\tau}_{i} \]
as described, e.g. in \cite{Lawrence:1998ja}.

Here is the general construction (the reader is invited to consult \cite{Nekrasov:2005wp, Nekrasov:2002kc} for details). Consider the maximal supersymmetric super-Yang-Mills theory in eight dimensions, on a noncommutative ${\BR}^{8} \approx {\BC}^{4}$. One can view this theory as a particular background in the IKKT matrix model \cite{IKKT} with an dimensional theory with an infinite dimensional gauge group, the group of unitary operators in the Hilbert space. This theory can also be lifted to $8+1$ and $9+1$ dimensions, (also known as the  Matrix theory \cite{BFSS} and the matrix string theory \cite{Verlinde:1995}, respectively).  

Recall that the gauge fields on the noncommutative Euclidean space ${\BR}^{n}_{\theta}$ with the
coordinates ${\hat x}^{\mu}$, ${\mu} = 1 , \ldots , n$, obeying
\[  [{\hat x}^{\mu}, {\hat x}^{\nu} ] = {\ii} {\theta}^{\mu\nu} {\cdot} 1\]
 can be described, more conveniently, as operators
\[ {\hat X}^{\mu} = {\hat x}^{\mu} + {\theta}^{{\mu}{\nu}} A_{{\nu}} ({\hat x} ) \]
In the vacuum, $A_{\nu} = 0$ and ${\hat X}^{\mu} = {\hat x}^{\mu}$. The equations of motion of Yang-Mills
theory on ${\BR}^{n}_{\theta}$ translate to the relations on the commutators of ${\hat X}^{\mu}$'s:
\beq G_{\mu' \mu''} [{\hat X}^{\mu'} , [ {\hat X}^{\mu''}, {\hat X}^{\mu} ]] = 0 \label{eq:ncym}
\eeq
In the form \eqref{eq:ncym}  the equations of motion do not distinguish between the gauge fields and 
adjoint scalars, and are equally applicable both to the $n$-dimensional theory and to its dimensional reductions
to lower dimensions. Everything is hidden in the nature of the operators ${\hat X}^{\mu}$.

Let us now take $n=10$, choose an identification ${\BR}^{10} \approx {\BC}^{4} \times {\BC}$,
assume the metric  to be Euclidean, $G_{\mu\nu} = {\delta}_{\mu\nu}$, and the Poisson tensor 
${\theta}^{\mu\nu}$ to be of the $(1,1)$-type. We assume it vanishes on the last ${\BC}$ factor. 
Define $Z^{i} = {\hat X}^{2i-1} + {\ii} {\hat X}^{2i}$, $i = 1, 2, 3,4$, ${\bf\Phi} = {\hat X}^{9} + {\ii} {\hat X}^{10}$. 
We are interested in the supersymmetric field configurations, the generalized instantons. In the present case the relevant equations are (cf. \cite{Moore:1998et}):
\beq
[ Z^{i}, Z^{j} ] + {\ve}^{ijkl} [Z^{k}, Z^{l}]^{\dagger} = 0, \qquad i,j = 1, 2, 3, 4
\label{eq:zz}
\eeq
\beq
\sum_{i=1}^{4} [Z^{i}, Z^{i\dagger} ] \ = \ -2{\theta} \cdot {\bf 1}_{\CalH}
\label{eq:stab}
\eeq
and
\beq
[ {\bf\Phi} , Z^{i} ] = [ {\bf\Phi}, Z^{i\dagger} ] = 0
\label{eq:phiz}
\eeq
The equations \eqref{eq:zz}, \eqref{eq:stab}, \eqref{eq:phiz} imply \eqref{eq:ncym}. The $\Omega$-deformation modifies
the equations \eqref{eq:phiz} to:
\beq
[ {\bf\Phi}, Z^{i} ] + {\ve}_{i} Z^{i} = 0, \qquad [ {\bf\Phi}, Z^{i\dagger} ] - {\ve}_{i} Z^{i\dagger} = 0
\label{eq:phizom}
\eeq
The equation \eqref{eq:zz} involves the $(4,0)$-form $\frac{1}{4!} {\ve}_{ijkl} dz^i \wedge dz^j \wedge dz^k \wedge dz^l$
which is only invariant under the $SU(4)$ rotations, forcing the constraint  \eqref{eq:ve4}. 

Let us denote by ${\mathfrak H}$ a copy of the two-oscillator Fock space:
\beq
{\mathfrak H} = \bigoplus_{n_{1}, n_{2} =0}^{\infty} {\BC} \vert {\vec n} \rangle, 
\label{eq:hfsp}
\eeq
acted upon by the creation and the annihilation operators:
\beq
{\bf A}_{i}^{\dagger} \vert {\vec n} \rangle = \sqrt{n_{i} +1 } \vert {\vec n} + {\bf e}_{i} \rangle, \qquad
{\bf A}_{i}  \vert {\vec n} \rangle = \sqrt{n_{i}} \vert {\vec n} - {\bf e}_{i} \rangle, \qquad i = 1,2
\label{eq:cran}
\eeq
with ${\vec n} = n_{1} {\bf e}_{1} + n_{2} {\bf e}_{2}$, ${\bf e}_{1} = (1,0)$, ${\bf e}_{2} = (0,1)$. The Hilbert space $\mathfrak H$ 
is the irreducible representation of the $2$-oscillators Heisenberg algebra 
 \[ [ {\bf A}_{i} , {\bf A}_{j}^{\dagger} ] = {\delta}_{ij}, \ [ {\bf A}_{1}, {\bf A}_{2} ] = 0, \ i,j = 1,2\]
A simple solution to the Eqs. \eqref{eq:zz}, \eqref{eq:stab}, \eqref{eq:phiz}  describing a stack of $n$ parallel $D3$-branes stretched in the ${\BR}^{4}_{1234}$ direction is given by: identify ${\CalH} = {\mathfrak H} \otimes N$, with $N$ the fintie dimensional complex vector space of dimension $n$:
\beq
Z^{1} = \sqrt{\theta} {\bf A}_{1}^{\dagger} \otimes {\bf 1}_{N}, \qquad Z^{2} = \sqrt{\theta} {\bf A}_{2}^{\dagger} \otimes {\bf 1}_{N}
\label{eq:zzd3}
\eeq
while for $i = 3,4$, $Z^{i} = {\bf 1}_{\mathfrak H} \otimes {\rm diag} ( z_{(1)}^{i}, \ldots , z_{(n)}^{i} )$, where $z_{(a)}^{i} , {\ac}_{a} \in {\BC}$, $a = 1, \ldots , n$, $i = 3,4$.  The scalar $\bf\Phi$ is equal to ${\bf 1}_{\mathfrak H} \otimes {\rm diag} ( {\ac}_{1}, \ldots , {\ac}_{n} )$ in the absence of $\Omega$-deformation, and to
\beq
{\bf\Phi} = \left( {\ve}_{1} {\bf A}_{1}^{\dagger}{\bf A}_{1} + {\ve}_{2} {\bf A}_{2}^{\dagger} {\bf A}_{2} \right) \otimes {\bf 1}_{N}+ {\bf 1}_{\mathfrak H} \otimes  {\rm diag} ( {\ac}_{1}, \ldots , {\ac}_{n} )
\label{eq:phiom}
\eeq
when the $\Omega$-deformation is turned on.

The solution which preserves less supersymmetry has ${\CalH} = H_{12} \oplus H_{34}$, $H_{12} = {\mathfrak H} \otimes N$, $H_{34} = {\mathfrak H} \otimes W$, with two vector spaces $N$ and $W$, of dimensions $n$ and $w$, respectively
and:
\beq
\begin{aligned}
& Z^{i} =  \sqrt{\theta}\, P_{12} {\bf A}_{i}^{\dagger} \otimes {\bf 1}_{N}  P_{12}\, , \qquad i = 1,2 \\
& Z^{i} = \sqrt{\theta}\, P_{34}  {\bf A}_{i-2}^{\dagger} \otimes {\bf 1}_{W}  P_{34}\, , \qquad i = 3,4 \\
\end{aligned}
\label{eq:cross}\eeq 
 where $P_{ij}: {\CalH} \to H_{ij}$ is the orthogonal projection, $P_{ij}^{2} = P_{ij} = P_{ij}^{\dagger}$, $P_{12}P_{34} = P_{34} P_{12} = 0$, ${\bf 1}_{\CalH} = P_{12} + P_{34}$. The scalar $\bf\Phi$ is given by
\[ {\bf \Phi} = \, P_{12}  {\bf 1}_{\mathfrak H} \otimes {\rm diag} ( {\ac}_{1}, \ldots, {\ac}_{n} ) P_{12} + 
P_{34} {\bf 1}_{\mathfrak H} \otimes {\rm diag} ( {\nu}_{1}, \ldots, {\nu}_{w} ) P_{34} \]
without $\Omega$-deformation, and by 
\beq
\begin{aligned}
& {\bf \Phi} = \, P_{12}  \left( \left( {\ve}_{1} {\bf A}_{1}^{\dagger}{\bf A}_{1} + {\ve}_{2} {\bf A}_{2}^{\dagger} {\bf A}_{2} \right) \otimes {\bf 1}_{N}+ {\bf 1}_{\mathfrak H} \otimes  {\rm diag} \left( {\ac}_{1}, \ldots , {\ac}_{n} \right) \right) P_{12} + \\
& \qquad\qquad  P_{34}\left( \left( {\ve}_{3} {\bf A}_{1}^{\dagger}{\bf A}_{1} + {\ve}_{4} {\bf A}_{2}^{\dagger} {\bf A}_{2} \right) \otimes {\bf 1}_{W}+ {\bf 1}_{\mathfrak H}  \otimes {\rm diag} \left( {\nu}_{1}, \ldots, {\nu}_{w} \right) \right) P_{34} \\
\end{aligned}
\label{eq:phiom4}
\eeq
 with the $\Omega$-deformation corresponding to the generic $SU(4)$ rotation. 

We are mostly interested in the solutions, which asymptotically tend to the $4+4$-dimensional background, corresponding to the 
 intersecting branes solution \label{eq:cross} (the asymptotics does not allow shifting $Z^i$ by a constant). Let us describe the so-called $1$-instanton solutions in the ``abelian'' case $n = w = 1$. Let $e_{N}$ and $e_{W}$ denote the orthonormal bases in $N$ and $W$, respectively. The space of solutions has three components:
 ${\color{violet}{{\CalM}_{N}}} \cup {\color{orange}{{\CalM}_{NW}}} \cup {\color{green}{{\CalM}_{W}}}$. 
 
 \medskip
 \noindent
\hbox{ \vbox{\hbox{The components ${\CalM}_{N}, {\CalM}_{W}$ are both isomorphic}
 \hbox{to ${\BC}^{2}$, while the component ${\CalM}_{NW} \approx {\BC\BP}^{1}$}
 \hbox{is compact. Moreover these components}
 \hbox{intersect, at two points:}
 \hbox{${\color{violet}{{\CalM}_{N}}} \cap {\color{orange}{{\CalM}_{NW}}} = p_{N}$, ${\color{green}{{\CalM}_{W}}} \cap {\color{orange}{{\CalM}_{NW}}} = p_{W}$}
 \hbox{(it is tempting to call $p_{N}$ the North pole,}
 \hbox{and $p_{W}$ the Western pole,}
 \hbox{unfortunately we couldn't place}
 \hbox{the latter on the map, even on the Google map).}}  \qquad {\hfill\picit{5}{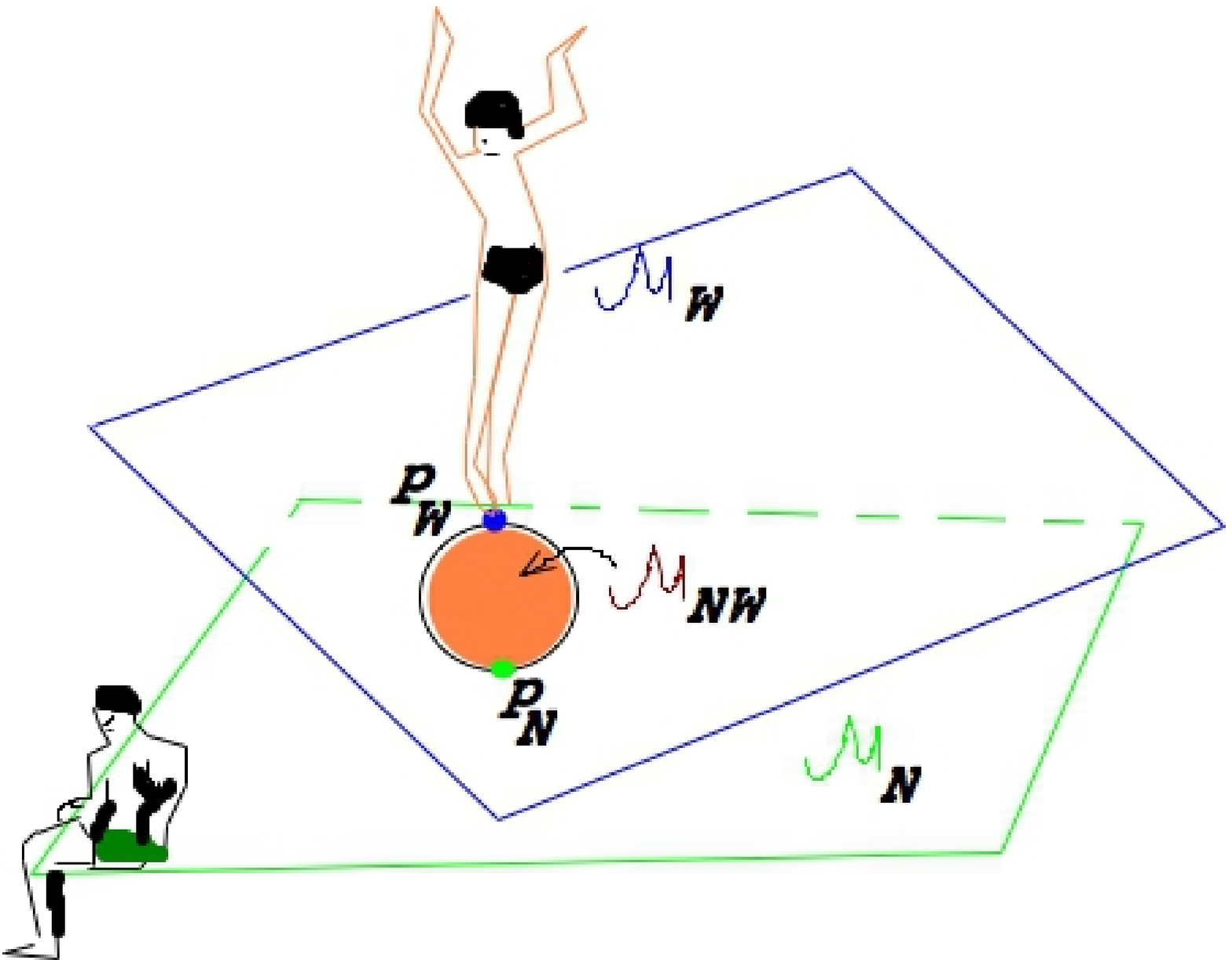}}}
 \bigskip
 
 Explicitly, ${\CalM}_{N}$ parametrizes the solutions where the pair $(Z^{1}, Z^{2})$ in \eqref{eq:cross} is replaced by the one-instanton solution of \cite{Nekrasov:1998ss} (see also \cite{Nekrasov:2000zz} for more details), while $(Z^{3}, Z^{4}, {\bf\Phi})$ are intact. Likewise, 
 ${\CalM}_{W}$ parametrizes the solutions where the pair $(Z^{3}, Z^{4})$ in \eqref{eq:cross} is replaced by the one-instanton solution. Recall \cite{Furuuchi:1999kv}, \cite{Nekrasov:2000zz}, \cite{Nekrasov:2000ih} that these solutions make use of Murray-von Neumann partial isometries $S: {\CalH} \to {\CalH}$, which obey:
\beq
S S^{\dagger} = {\bf 1}_{\CalH}, \qquad S^{\dagger} S = {\bf 1}_{\CalH} - P_{K}
\label{eq:mvn}
\eeq
with $P_{K}$ an orthogonal projection onto a finite-dimensional subspace $K \subset {\mathfrak H}$. For the one-instanton
solutions in the components ${\CalM}_{N,W}$ the subspace $K$ is one-dimensional, $K = {\BC} e^{u{\bf A}_{1}^{\dagger} + v {\bf A}_{2}^{\dagger}} \vert 0, 0 \rangle \otimes e_{N,W}$, 
with $(u,v) \in {\CalM}_{N,W} \approx {\BC}^{2}$ being the instanton modulus. 
The solutions corresponding to the component ${\CalM}_{NW}$
have $K =  {\BC}e^{\perp}$, where
\beq
e^{\perp} = {\bar\alpha} \vert 0,0\rangle \otimes e_{N} + {\bar\beta} \vert 0,0 \rangle \otimes e_{W},
\eeq
with  some ${\alpha}, {\beta}  \in {\BC}$,  
\beq
| {\alpha} |^{2} + | {\beta} |^{2}  = 1
\eeq
Define 
\beq
e = {\beta} \vert 0,0\rangle \otimes e_{N} - {\alpha}  \vert 0,0 \rangle \otimes e_{W}, \qquad\qquad e^{\dagger} K = 0
\eeq
and 
\beq
{\mathfrak H}' = \bigoplus_{n_{1}+n_{2} > 0} {\BC} \, \vert {\vec n} \rangle \subset {\mathfrak H}
\eeq
and
\beq
{\CalH}' = {\BC} e \oplus \left( {\mathfrak H}' \otimes N \right) \oplus \left( {\mathfrak H}' \otimes W \right)
\eeq
the orthogonal complement to $K$. Define:
\beq
\begin{aligned}
& {\tilde Z}^{i} \, \vert {\vec n} \rangle \otimes e_{N} = \sqrt{\theta}\, g_{n+1}^{-1} g_{n} \sqrt{n_{i}+1}\, \vert {\vec n} + {\bf e}_{i} \rangle \otimes e_{N}, \quad i = 1, 2 \\
& \qquad\qquad {\tilde Z}^{i}\, \vert {\vec n} \rangle \otimes e_{W} = \sqrt{\theta}\,{\tilde g}_{n+1}^{-1} {\tilde g}_{n} \sqrt{n_{i-2}+1} \, \vert {\vec n} + {\bf e}_{i-2} \rangle \otimes e_{W}, \quad i = 3, 4 \\
&  \qquad\qquad\qquad {\tilde Z}^{1,2}\, \vert {\vec n} \rangle \otimes e_{W} = 0, \quad {\tilde Z}^{3,4} \, \vert {\vec n} \rangle \otimes e_{N} = 0 \\
& \\
& \qquad\qquad\qquad\qquad n = n_{1} + n_{2} > 0, \qquad g_{n}, {\tilde g}_{n} \in {\BC} \\
& \\
& {\tilde Z}^{1} e =\, \sqrt{\theta}\, {\gamma}_{12} \, \vert 1,0 \rangle \otimes e_{N}, \qquad {\tilde Z}^{2} e =\, \sqrt{\theta}\, {\gamma}_{12} \, \vert 0,1 \rangle \otimes e_{N}, \\
& \\
& {\tilde Z}^{3} e =\, \sqrt{\theta}\, {\gamma}_{34} \, \vert 1,0 \rangle \otimes e_{W}, \qquad {\tilde Z}^{4} e =\, \sqrt{\theta}\, {\gamma}_{34} \, \vert 0,1 \rangle \otimes e_{W}, \\
& \\
& \qquad\qquad {\tilde Z}^{i \dagger} e = 0, \qquad i = 1, 2, 3, 4\\
\end{aligned}
\label{eq:1instsol}
\eeq
 with
 \beq
 {\gamma}_{12}, {\gamma}_{34} \in {\BC}, \qquad |{\gamma}_{12}|^{2} + | {\gamma}_{34} |^{2} = 1, \qquad ( {\gamma}_{12} : {\gamma}_{34} ) \in {\CalM}_{NW}
 \eeq
 Now define:
 \beq
 Z^{i} = S Z^{i} S^{\dagger}
 \eeq
 where $S^{\dagger}$ maps ${\CalH}$ onto ${\CalH}'$ isometrically (use the Hilbert hotel construction) obeying \eqref{eq:mvn}. 
 
 The diagonal matrices $g_{n}, {\tilde g}_{n}$ are fixed, up to the unitary gauge transformations, by the equation \eqref{eq:stab}, 
 \beq
 | g_{n} |^{2} = \frac{(n+1)! n!}{(n+{\kappa}_{12})! (n+1- {\kappa}_{12})!}, \quad  | {\tilde g}_{n} |^{2} = \frac{(n+1)! n!}{(n+{\kappa}_{34})! (n+1- {\kappa}_{34})!}
 \label{eq:1inst}
 \eeq
 where
 \beq
 {\kappa}_{ij} ( 1 - {\kappa}_{ij} ) = 2 ( |{\gamma}_{ij}|^{2} - 1 )
 \eeq 
 In the limiting cases $( {\gamma}_{12} : {\gamma}_{34} ) \to (1: 0) = p_{W}$ or $( {\gamma}_{12} : {\gamma}_{34} ) \to (0: 1) = p_{N}$ the solution \eqref{eq:1instsol} approaches the direct sum of the vacuum solution for $H_{34}$  and one-instanton solution on $H_{12}$ or the direct sum of the one-instanton solution for $H_{34}$ and the vacuum solution on $H_{12}$, respectively.  
 
 In \cite{Nekrasov:2015ii} we shall consider more general intersecting brane solutions. Let
 ${\bf 6} = \left( \begin{matrix} {\bf 4} \\ {\tiny 2} \end{matrix} \right)$, the set of $2$-element subsets of
 $\bf 4$. Fix $6$ vector spaces $N_{A}$. We take the Hilbert space to be the sum  
 \beq
 {\CalH} = \bigoplus_{A \in {\bf 6}} \, H_{A}, \qquad H_{A} = {\mathfrak H} \otimes N_{A}
 \eeq
  Define
 \beq
 Z_{0}^{a} = \sum_{A, a \in A} {\bf A}_{h_{A}(a)+1}^{\dagger} \otimes {\bf 1}_{N_{A}}
 \eeq 
 so that $Z_{0}^{a} \vert_{H_{B}} = 0$ whenever $a \notin B$. This is the reference solution for the generalized instantons in the theory we call the {\emph{gauge origami}} in \cite{Nekrasov:2015ii}, \cite{Nekrasov:2015iim}. 
 
\subsec{Other realizations of ${\x}$-observables}

A natural question is what is the meaning of the ${\x}_{\ib}(x)$ observables on the {\tt CFT} side of the BPS/CFT-correspondence?

It might seem natural, e.g. in the AGT setup \cite{Alday:2009aq, Alday:2009fs} to assign the ${\x}_{\ib}(x)$-observables to the non-intersecting loops on the curve $C$ on which one compactifies the $A_{1}$ $(0,2)$-theory, which define the $\al$-coordinates in the system of Darboux coordinates on the moduli space of $SL_{2}$ local systems \cite{Nekrasov:2011bc}. 

One systematic way to derive such a representation would be to start with the type IIB ten-dimensional background
whose geometry is a rank $4$ complex vector bundle $E$ over a flat two-torus with an $SU(4)$-flat connection. One can add up to six stacks of $D5$ branes, wrapping the base torus and one of the complex rank two sub-bundles of $E$, invariant under the action of the product of the maximal torus $T \subset SU(4)$ and the two-torus translating the base.  
This symmetry can be used to $T$-dualize the configuration of branes leading to various equivalent realizations. 
This direction will be explored elsewhere. 

However, one may try to address the question directly within the realm of the two-dimensional conformal field theory. We
know the ${\CalN}=2^*$ $SU(n)$ theory corresponds to the $A_{n-1}$ Toda theory \cite{Wyllard:2009hg} on a two-torus
${\BC}^{\times}/{\qe}^{\BZ}$ with an insertion of a special vertex operator. The coupling constant $b^2$ of the Toda theory is
determined by the ratio ${\ve}_{2}/{\ve}_{1}$ of two equivariant parameters. The $qq$-character ${\CalX}_{{\bw}, {\bnu}}$
of the ${\hat A}_{0}$-theory
is generated by the auxiliary ${\CalN}=2^*$ theory on the transverse ${\BR}^{4}$ with the equivariant parameters
${\ve}_{3}, {\ve}_{4}$. It corresponds to its own $A_{w-1}$ Toda theory with the coupling constant 
${\tilde b}^{2} = {\ve}_{4}/{\ve}_{3}$. It would be interesting to work out the coupling between these theories generating
the $x$-dependent contributions to the $qq$-character. 

 \secc{ The \ first\ applications}

\subsubsec{Expansion coefficients}

{}For the formal Laurent series $f(z^{-1})$ near $z={\infty}$ we denote by $[z^n]f(z)$ the $z^n$ coefficient:
\beq
[z^{n}]f(z) \equiv {\rm Coeff}_{z^{n}}f(z) = \frac{1}{2\pi {\ii}} \oint_{\infty} f(z) \frac{dz}{z^{n+1}}\, ,  
\label{eq:cof}
\eeq
the  latter equality holding for actual functions $f(z)$.

\subsec{ Effective prepotentials and superpotentials}

One obvious application of our formalism is the solution of the low-energy theories. Indeed, in the limit ${\ve}_{2} \to 0$ the integrals  \eqref{eq:cxwn} simplify (the Chern polynomial of the tangent bundle drops out). In addition, the sum over 
all quiver $\bn$-colored partitions \eqref{eq:tyeq} is dominated  
by a single {\it limit shape} \cite{Nekrasov:2013xda} which maps \eqref{eq:cxwn} to a system
of difference equations for the ${\mathscr Y}_{\ib}$-functions. These equations were studied in 
\cite{Nekrasov:2013xda}. In this case it suffices to study the equations for the  dimension vectors ${\bw}$ corresponding to the fundamental weights of $\gq$. 

\subsec{Instanton fusion}

In the quantum case where ${\ve}_{1}, {\ve}_{2}$ are finite we need all $\bw$. The equations \eqref{eq:cxwn} can  be viewed as a system of Hirota difference equations, which should fix $\CalZ$ uniquely. This direction is currently investigated. Note that for the finite
$A$-type quivers with the special choice of ${\bn}$ the related equations were found in \cite{Kanno:2013aha} by somewhat different methods, although we couldn't match them with our equations for specific $\bw$. It would be very interesting to relate the algebraic structure found in \cite{Kanno:2013aha} to the one we exhibited here.

\subsec{ Undressing the $U(1)$ legs}

Another application of our formalism is the reduction formula, which allows to relate
the partition functions of the gauge theories with $U(1)$ gauge factors
to the partition functions of the gauge theories with these $U(1)$'s being treated as global symmetries. We assume the asymptotic freedom condition ${\beta}_{\ib} \leq 0$ \eqref{eq:bfun} is obeyed. 

Let ${\ib} \in \Ver$ be the node with $n_{\ib} =1$. Shift the argument of ${\mathscr Y}_{\ib}(x)$ so as to set ${\ac}_{\ib} = 0$. Then, we have the expansion (cf. \eqref{eq:yxki}):
\begin{equation}
{\mathscr Y}_{\ib}(x) = x  + \frac{{\ve}_{1}{\ve}_{2}}{x} \,  k_{\ib}  + \ldots , \qquad x \to \infty
\label{eq:asu1}
\end{equation}
There are two possibilities: either the node $\ib$ is connected to itself by an edge $e \in s^{-1}({\ib}) \cap t^{-1}({\ib})$, or $s^{-1}({\ib}) \cap t^{-1}({\ib})$ is empty. 

\subsubsec{ The ${\hat A}_{0}$ theory}

In the first case the theory is the ${\CalN}=2^{*}$ theory. In the $U(1)$ case it is characterized by the mass $\mt$ of the adjoint hypermultiplet, the complexified coupling $\qe$ and the $\Omega$-background parameters $({\ve}_{1}, {\ve}_{2})$. 
The partition function ${\CalZ}_{{\hat A}_{0}}$ is a homogeneous function of ${\mt}, {\ve}_{1}, {\ve}_{2}$, symmetric
in ${\ve}_{1}, {\ve}_{2}$ and invariant under ${\mt} \to - {\mt} - {\ve}$:
\beq
{\CalZ}({\qe}; {\mt}: {\ve}_{1}:{\ve}_{2}) = \sum_{\lambda} {\qe}^{|{\lambda}|} \prod_{{\square} \in {\lambda}} 
\left( 1 + \frac{{\mt}({\mt}+{\ve})}{c_{\square}^{\vee}({\ve} - c_{\square}^{\vee})} \right) \ , 
\label{eq:blam}
\eeq
where $c^{\vee}_{\square} = {\ve}_{1} (l_{\square}+1) - {\ve}_{2} a_{\square}$.
Since for ${\lambda} \neq \emptyset$ there always exists a locally most south-east box $\square$ for which $l_{\square} = a_{\square}  = 0$, the partition function ${\CalZ}({\qe}; {\mt}: {\ve}_{1}:{\ve}_{2}) = 1$ for ${\mt}= - {\ve}_{1}$ or ${\mt} = - {\ve}_{2}$:
\beq
{\CalZ}({\qe}; {\mt}: {\ve}_{1}:{\ve}_{2}) = 1 + \frac{({\mt}+{\ve}_{1}) ({\mt}+{\ve}_{2})}{{\ve}_{1}{\ve}_{2}} {\tilde Z} ({\qe}; {\mt}: {\ve}_{1}:{\ve}_{2})
\label{eq:vani}
\eeq
The normalization
\begin{equation}
{\CalZ}({\qe}; 0: {\ve}_{1}:{\ve}_{2}) = {\CalZ}({\qe}; - {\ve}: {\ve}_{1}:{\ve}_{2}) = 
{\phi}({\qe})^{-1}
\label{eq:phiq}
\end{equation}
follows trivially from \eqref{eq:blam}.
Let us expand the  character ${\x}_{1,0}(x)$ \eqref{eq:cx10} in $x$ near $x = \infty$:
\begin{multline}
 {\x}_{1,0} (x)  = \\
\sum_{\lambda} {\qe}^{|{\lambda}|} \prod_{{\square} \in \lambda}
{\bS} ( {\mt}h_{\square} + {\ve} a_{\square})  \left( x+{\ve}  +  
\frac{{\ve}_{1}{\ve}_{2}}{x}k + \ldots \right) \left(  
1 - \frac{{\mt}({\mt}+{\ve})}{x^2}|{\lambda}| + \ldots \right)
\end{multline}
Recall that the formula above gives the $x$-expansion of an observable.
It has the form ${\x}_{1,0} (x) =  {\x}_{1,0}^{(0)} (x) +  
{\x}_{1,0}^{(1)} (x) k + \ldots$, where $k$ is our familiar observable  
\eqref{eq:kib}.

{}Thus
\beq
[x^{-1}]  {\x}_{1,0} (x) =  {\ve}_{1}{\ve}_{2}  {\CalZ} ( {\qe}; {\ve}_{1}:  - {\mt} - {\ve}: {\mt}) k - {\mt}({\mt}+{\ve}) {\qe} \frac{d}{d{\qe}} {\CalZ} ( {\qe}; {\ve}_{1}: - {\mt} - {\ve}: {\mt})
\eeq
and the consequence of our equations \eqref{eq:tyeq} reads
\begin{multline}
0 = \vev{\vev{[x^{-1}] \ {\x}_{1,0} (x)}}_{{\qe}; {\mt}, {\ve}_{1}, {\ve}_{2}} = \\
{\ve}_{1}{\ve}_{2}  {\CalZ} ( {\qe}; {\ve}_{1}: - {\mt} - {\ve}: {\mt}) {\qe} \frac{d}{d{\qe}} 
{\CalZ} ( {\qe}; {\mt}:  {\ve}_{1}: {\ve}_{2}) - \\
{\mt}({\mt}+{\ve}) {\CalZ} ( {\qe}; {\mt}:  {\ve}_{1}: {\ve}_{2}) 
{\qe} \frac{d}{d{\qe}} {\CalZ} ( {\qe}; {\ve}_{1}: - {\mt} - {\ve}: {\mt})
\label{eq:zdz}
\end{multline}
Introduce:
\begin{equation}
{\Phi}({\qe}; {\mt}:{\ve}_{1}:{\ve}_{2}) = \frac{{\ve}_{1}{\ve}_{2}}{({\mt}+{\ve}_{1})({\mt}+{\ve}_{2})} \text{log} {\CalZ} ( {\qe}; {\mt}:  {\ve}_{1}: {\ve}_{2})
\end{equation}
For fixed $\qe$ it is a priori a meromorphic function on ${\BC\BP}^{2}$, with possible
singularities at ${\ve}_{2}/{\ve}_{1} \in {\BQ}_{\geq 0}$
(but not at ${\mt} = - {\ve}_{1}$, ${\mt} = - {\ve}_{2}$, cf. \eqref{eq:vani}). For $\qe = 0$, $\Phi = 0$. 
Then \eqref{eq:zdz} implies:
\begin{equation}
{\Phi}({\qe};\, {\mt}:{\ve}_{1}:{\ve}_{2}) = {\Phi}({\qe};\, {\ve}_{1}: -{\mt}-{\ve}: {\mt})
\label{eq:symm}
\end{equation}
which shows that $\Phi$ has no singularities in $({\mt} : {\ve}_{1} : {\ve}_{2})$ for fixed $\qe$, i.e. it is a constant. The normalization \eqref{eq:phiq} then implies that
${\Phi}({\qe}; {\mt}:{\ve}_{1}:{\ve}_{2})  = \text{log}{\phi}({\qe})$, i.e.
\begin{equation}
{\CalZ} ( {\qe}; {\mt}:  {\ve}_{1}: {\ve}_{2}) = {\phi}({\qe})^{-\frac{({\mt}+{\ve}_{1})({\mt}+{\ve}_{2})}{{\ve}_{1}{\ve}_{2}}}
\label{eq:adjm}
\end{equation}

\subsubsec{ Other theories}
In this case the node $\ib$ is such that $s^{-1}({\ib}) \cap t^{-1}({\ib})$ is empty. The fundamental $qq$-character ${\x}_{\ib}(x)$ has the following structure:
\begin{equation}
{\x}_{\ib}(x) = {\y}_{\ib}(x+{\ve}) + {\qe}_{\ib} {\Gamma}_{2}(x) {\y}_{\ib}^{-1}(x) + 
{\qe}_{\ib} {\Gamma}_{1}(x) 
\label{eq:chii}
\end{equation}
where ${\Gamma}_{1}(x), {\Gamma}_{2}(x)$ are built out of ${\y}_{\jb}, {\qe}_{\jb}$ with ${\jb} \neq {\ib}$.  For large $x$ the functions $\Gamma_{a}(x)$ behave as $x^{a} ( 1 + O(1/x))$, for $a = 1, 2$. 
The expansion in $x$ near $x = \infty$ gives:
\begin{equation}
0 = [x^{-1}] \ \vev{{\x}_{\ib}(x)} = {\ve}_{1}{\ve}_{2}(1 - {\qe}_{\ib}) {\qe}_{\ib} \frac{d}{d{\qe}_{\ib}} {\CalZ}^{\rm inst} + {\qe}_{\ib} {\CalD} {\CalZ}^{\rm inst}
\label{eq:zeqx}
\end{equation}
where ${\CalD}$ is the first order differential operator in ${\qe}_{\jb}$, with ${\jb} \neq \ib$. 
The equation \eqref{eq:zeqx} is the quasilinear partial differential equation of the first order, which can be solved using the method of characteristics. The solution is unique given the initial condition, which can be set at ${\qe}_{\ib}=0$, where the $U(1)_{\ib}$ gauge factor becomes a flavor group. 

\subsubsec{ The linear quiver abelian theories}

As an example of the application of this technique, consider  the Type $I$ theories with the $A_{r}$-type quiver, with $m_{1} = n_{1} = n_{2} = \ldots = n _{r} =  m_{r} =1$, $m_{\jb} = 0, 1< {\jb} < r$. The theory is characterized by the masses ${\mt}_{1}, {\mt}_{r}$ of the fundamental hypermultiplets, the Coulomb moduli ${\ba} = \left( {\ac}_{\ib} \right)_{{\ib} = 1}^{r}$ (which could be traded for the masses of the bi-fundamental hypermultiplets, cf. \cite{Nekrasov:2012xe}) and the couplings 
${\bqt} = ( {\qe}_{\jb})_{{\jb} = 1}^{r}$. Let us introduce the ``momenta'' $p_{i}^{\pm}$, $i = 0, \ldots, r$:
\beq
p_{i}^{+} = {\ve} + {\ac}_{i} - {\ac}_{i+1}, \qquad p_{i}^{-} = {\ac}_{i} - {\ac}_{i+1}, \qquad 
\label{eq:ppm}
\eeq
Using \eqref{eq:chilari},  \eqref{eq:chilar}, we derive:
\begin{multline}
0 = - [x^{-1}] \vev{ {\CalX}_{l}(x)} = \\
 = \sum_{I \subset [ 0 ,  r ] , \,   | I | = l}\, 
z_{I}\, \left(  {\ve}_{1}{\ve}_{2} \sum_{i \in I} {\nabla}_{i}^{z} {\rm log}{\CalZ}^{\rm inst}_{A_{r}}   +  \sum_{i \in I, j \in [0,r] \backslash I, j < i}
 p_{i}^{+} p_{j}^{-} \right) \, .  \label{eq:chileq}
 \end{multline}
 The solution to \eqref{eq:chileq} is given by the simple ``free-field formula'':
\beq
{\CalZ}^{\rm inst}_{A_{r}} ({\ba}, {\bqt}) = \prod_{0 \leq j < i \leq r} ( 1 - z_{i}/z_{j})^{-\frac{p_{i}^{+}p_{j}^{-}}{{\ve}_{1}{\ve}_{2}}}
\label{eq:zr}
\eeq  
Thus, we have derived the formulas conjectured in the sections C.1 and C.2 of \cite{Alday:2009aq}. Our derivation here differs from that in 
\cite{OC:2008}.

For $r=1$ we get:
\[ {\CalZ}_{A_{1}}^{\rm inst} = ( 1 - {\qe} )^{\frac{({\ve} + {\ac}_{0} - {\ac}_{1})({\ac}_{2} - {\ac}_{1})}{{\ve}_{1}{\ve}_{2}}} \]
\subsubsec{ The $D$-type theories}

Let us present now an example of the $D_{4}$-type theory. 
This is the theory with four gauge group factors, which we shall label by $\ib = 0, 1,2,3$,
 with the assignments: $n_{0} =2$, $n_{1}=n_{2}=n_{3}=m_{0} = 1$. 
 The theory is characterized by four couplings ${\bqt} = ({\qe}_{0}, {\qe}_{1}, {\qe}_{2}, {\qe}_{3})$, 
 and six Coulomb and mass parameters: the Coulomb parameters
\[ {\ba} = ( {\ac}_{0,1} = a_{1}, {\ac}_{0,2} = a_{2} , {\ac}_{1,1} = {\mt}_{1}, {\ac}_{2,1} = {\mt}_{2}, {\ac}_{3,1} = {\mt}_{3} ) \] 
two for the $U(2)$ gauge group factor, and three for three $U(1)$ factors, and the mass ${\mt}_{4}$. 

By computing $[x^{-1}] {\x}_{{\ib},0}(x)$ in \eqref{eq:1qqD4} using \eqref{eq:yxki} for ${\ib} = 1,3,4$ we derive three first order differential equations, whose solution give:
\begin{multline}
{\CalZ}_{D_{4}} \left(  {\ba} ; {\mt}_{4} ; {\bqt} \right) =  {\CalZ}_{A_{1}} \left( a_{1}, a_{2}; {\mt}_{1}, {\mt}_{2}, {\mt}_{3}, {\mt}_{4} ; {\qe} \right) \times \\
( 1 - {\qe}_{1} )^{{\mu}_{1}} ( 1 - {\qe}_{2} )^{{\mu}_{2}} ( 1 - {\qe}_{3} )^{{\mu}_{3}} (1 - {\qe}_{0}^{2} {\qe}_{1}{\qe}_{2}{\qe}_{3} )^{{\mu}_{4}}   \times \\
( 1 - {\qe}_{0}{\qe}_{1} )^{{\nu}_{1}} ( 1 - {\qe}_{0}{\qe}_{2} )^{{\nu}_{2}} ( 1 - {\qe}_{0}{\qe}_{3} )^{{\nu}_{3}} (1 - {\qe}_{0} {\qe}_{1}{\qe}_{2}{\qe}_{3} )^{{\nu}_{4}}   \times \\
(1 - {\qe}_{0}{\qe}_{1}{\qe}_{2} )^{{\kappa}_{3}} (1 - {\qe}_{0}{\qe}_{1}{\qe}_{3} )^{{\kappa}_{2}}
(1 - {\qe}_{0}{\qe}_{2}{\qe}_{3} )^{{\kappa}_{1}}  
  \end{multline}
where
\begin{equation}
\begin{aligned}
& {\ve}_{1}{\ve}_{2} {\mu}_{j} = ({\mt}_{j}-a_{1})({\mt}_{j}-a_{2})\ ,  \\
& {\ve}_{1}{\ve}_{2} {\nu}_{j} = (a_{1} + a_{2} + {\ve} )(a_{1} + a_{2} + {\ve} + {\mt}_{j} - {\mt}) - a_{1}a_{2} - {\ve} {\mt}_{4}  +  {\mt}_{j} ({\mt}_{j} - {\mt}) +\sum_{1\leq i < k \leq 4} {\mt}_{i} {\mt}_{k}     \ ,  \\
& {\ve}_{1}{\ve}_{2} {\kappa}_{j} = \left( a_{1} + a_{2} + {\ve} - {\mt}_{j} - {\mt}_{4}\right) \left(a_{1} + a_{2} + {\ve} + {\mt}_{j} -  {\mt} \right)\\
& \qquad\qquad j = 1, \ldots , 4 \, , \quad {\mt} = {\mt}_{1} + {\mt}_{2} + {\mt}_{3} \\
\end{aligned}
\end{equation}
and
\begin{equation}
{\qe} = {\qe}_{0} \frac{(1-{\qe}_{1})(1-{\qe}_{2})(1-{\qe}_{3})(1-{\qe}_{0}^2{\qe}_{1}{\qe}_{2}{\qe}_{3})}{(1-{\qe}_{0}{\qe}_{1})(1-{\qe}_{0}{\qe}_{2})(1-{\qe}_{0}{\qe}_{3})(1-{\qe}_{0}{\qe}_{1}{\qe}_{2}{\qe}_{3})}
\end{equation}

 \subsec{Fractional instantons and quantum differential equations}

The equations of the schematic form:
\beq
{\kappa} \frac{\partial}{{\partial\tau}_a} {\bf\Psi} = {\hat H}_{a} ({\tau}) \cdot {\bf\Psi}
\label{eq:kz}
\eeq
where $a$ label the set of couplings and the operators ${\hat H}_{a}$ on the right hand side are $\kappa$-independent, show up in mathematical physics on  several occasions ( Knizhnik-Zamolodchikov connection \cite{Knizhnik:1984}, \cite{Varchenko:1991}, $t$-part of $tt^*$-connection \cite{Cecotti:1991me}, Gauss-Manin connection for exponential periods  \cite{lone}, \cite{loma}, ${\lambda}$-connection associated to the solution of the WDVV equations \cite{Kontsevich:1994qz}, \cite{Givental:1996}, and more recently, e.g. \cite{BMO:2010}). The consistency of \eqref{eq:kz}, i.e. the flatness of the corresponding connection for any value of $\kappa$, is equivalent to two sets of equations:
\beq
\begin{aligned}
& \qquad  [ {\hat H}_{a}, {\hat H}_{b} ] = 0, \\
&  \frac{\partial}{{\partial\tau}_a} {\hat H}_{b} -  
\frac{\partial}{{\partial\tau}_b} {\hat H}_{a} = 0 \end{aligned}
\label{eq:flatkz}
\eeq
The first set of equations imply that at each value of ${\tau}$ one has a quantum integrable system (if the number of the operators ${\hat H}_{a}$ is maximal in the appropriate sense). 

In the present case the meaning of these equations is the following. We have a quantum field theory with some set of couplings ${\tau}_{a}$ in which we study a codimension two defect, which has its own couplings ${\tilde\tau}_{a, \omega}$. Differentiating the partition function of the theory with defect brings down the corresponding observable ${\CalO}_{a}$, deforming the Lagrangian. Integration over the positions of ${\CalO}_{a}$'s has a contribution of the region where ${\CalO}_{a}$ approaches the defect. When ${\CalO}_{a}$ hits the defect, it fractionalizes, and splits into the observables of the defect theory:
\beq
{\CalO}_{a} \sim \sum_{\omega} f^{(1)}_{a, {\omega}} ({\tau}, {\tilde\tau} ) {\tilde\CalO}_{a, \omega} +
\sum_{\omega' , \omega''} f^{(2)}_{a, {\omega}',{\omega}''} ({\tau}, {\tilde\tau} ) {\tilde\CalO}_{a, \omega'} {\tilde\CalO}_{a, \omega''} + \ldots
\eeq  
The equation we derive in \cite{Nekrasov:2015iii} is an example of such a relation, where the bulk operator
${\CalO}_{a}$ is, in fact, the familiar ${\Tr}{\bf\Phi}^{2}$, and its supersymmetric descendents (which are all equal up to the powers of $\ve_{2}$ in cohomology of the $\Omega$-deformed supersymmetry). What about other operators, such as
${\Tr}{\Phi}^{k}$ for $k > 2$?

The operators deforming
the gauge Lagrangian by 
\beq
{\delta}_{\tau}{\mathbb{L}} = \sum_{k > 2} \frac{{\tau}_{k}}{k!} \int d^{4}xd^{4} {\vartheta} {\Tr} {\Phi}^{k} \sim \frac{1}{(k-2)!} \int {\Tr} {\Phi}^{k-2} F^2 + \ldots 
\eeq
are irrelevant and lead to non-renormalizable theories. We can, nevertheless, study them by treating ${\tau}_{k}$ as formal variables (i.e. assuming some power ${\tau}_{k}^{n_{k}}$ of $\tau_k$ to vanish). 
The $qq$-characters are modified by the introduction of the higher times. 
 In the $A_1$ case, for example, the $qq$-character modifies to
 \beq
 {\y}(x+{\ve}) + {\y}(x)^{-1} {\qe} P(x)\, {\exp} \sum_{l=1}^{\infty} \frac{1}{l!} {\tau}_{l} x^{l}
 \eeq
In this
way we get the realization of the $W$-algebra and its $qq$-deformation in gauge theory. We also get a new perspective
on the r\^ole of Whitham hierarchies \cite{Krichever:1992qe}, \cite{Gorsky:1995zq} and their quantum and $qq$-deformations 
in gauge theory. 

See also \cite{Borodin:2015} for more applications of $qq$-characters in the $U(1)$ case.

\secc{ Discussion\ and\ open\ questions}

Of course, the most interesting question is to extend our formalism of non-perturbative
Dyson-Schwinger equations beyond the BPS limit, even beyond the realm of supersymmetric theories. 

However, even in the world of moderately supersymmetric theories our approach seems to be useful. It appears
that the exact computations of \cite{Dijkgraaf:2002dh}, \cite{Dijkgraaf:2002vw} of effective
superpotentials of ${\CalN}=1$ theories and their gravitational descendands can be cast in the form
of the non-perturbative Dyson-Schwinger identities. The precise definition of the $qq$-characters
in ${\CalN}=1$ theories will be discussed elsewhere.

Another exciting problem is to find the string theory analogue of our $qq$-characters and the stringy version of the large field redefinitions (see \cite{Gerasimov:2000ga} for a discussion of stringy symmetries). 

The considerations of this paper and its companions are local, they describe gauge theories in the vicinity of
a fixed point of rotational symmetry. 
In \cite{Nekrasov:2003vi} four dimensional ${\CalN}=2$ gauge theories on the smooth toric surfaces were studied. It was found that the partition function of the gauge theory on a toric surface $S$ has the \emph{topological vertex} structure:
\beq
{\CalZ}_{S} \sim \sum_{\rm lattice} \, \prod_{v \in S} {\CalZ}_{\BR^4} ({\rm local\ Coulomb\ parameters}, {\rm local}\ {\Omega}- \ {\rm parameters} )
\label{eq:topvert}
\eeq
where the sum goes over the lattice of magnetic fluxes
$H^{2}(S, {\Lambda}_{\rm w})$, the product is over the fixed points of the two-torus action on $S$ (see \cite{Bershtein:2015xfa} for the recent progress in this direction). 
Our generalized gauge theories involving intersecting four dimensional space-times naturally live on Calabi-Yau fourfolds.
They describe generalized complex surfaces which may have several components with different multiplicities. 
It would be interesting to apply these ideas to topological strings and to topological gravity. 

On a more mathematical note, let us discuss the relation of our $qq$-characters to the
 $t$-deformation of $q$-characters of \cite{Frenkel:1998}, introduced by H.~Nakajima in 
 \cite{Nakajima:2001, Nakajima:2003, Nakajima:2004, Nakajima:2010}. His definition is basically 
 the weighted sum of the Poincare polynomials of the $H_{{\bw}, \gamma}$-fixed loci on ${\qM}({\bw}, {\bv})$. 
Let us observe that if in the formula \eqref{eq:cxwnqq} we pull the 
${\y}_{\ib}$'s and $P_{\ib}$'s out of the integral, with some clever choice of the arguments replacing those in \eqref{eq:xifi}, 
the remaining integral, for each $\bv$ would compute 
\beq
\sum_{j} (-q_{2})^{-j} {\chi} ( {\qM}({\bw}, {\bv}) , {\Omega}^{j}_{{\qM}({\bw}, {\bv}) } )
\label{eq:holpol}
\eeq
i.e. the holomorphic Poincare polynomial. In other words,
if the $qq$-operator is viewed as the difference-differential operator on the functions ${\y}_{\ib}$, then the $t$-deformed $q$-character looks like its symbol. 
It would be interesting to develop some kind of deformation quantization scheme, allowing to compute our $qq$-characters using the knowledge of the $t$-deformed $q$-characters \cite{Nakajima:2010}, and to apply them to rederive the results of \cite{ONC}. The paper \cite{Pestun:2015} is a step in this direction.

\appendix

\bibliographystyle{acm}
\bibliography{Nekrasov-BPSCFT}

\def\cprime{$'$}
\begin{thebibliography}{100}

\bibitem{Alday:2009fs}
{\sc Alday, L.~F., Gaiotto, D., Gukov, S., Tachikawa, Y., and Verlinde, H.}
\newblock {Loop and surface operators in N=2 gauge theory and Liouville modular
  geometry}.

\bibitem{Alday:2009aq}
{\sc Alday, L.~F., Gaiotto, D., and Tachikawa, Y.}
\newblock {Liouville Correlation Functions from Four-dimensional Gauge
  Theories}.
\newblock {\em Lett. Math. Phys. 91\/} (2010), 167--197.

\bibitem{Alday:2010vg}
{\sc Alday, L.~F., and Tachikawa, Y.}
\newblock {Affine SL(2) conformal blocks from 4d gauge theories}.
\newblock {\em Lett.Math.Phys. 94\/} (2010), 87--114.

\bibitem{alekseev:1988}
{\sc Alekseev, A., Faddeev, L., and Shatashvili, S.}
\newblock Quantization of symplectic orbits of compact {L}ie groups by means of
  the functional integral.
\newblock {\em Journal of Geometry and Physics 5}, 3 (1988), 391--406.

\bibitem{alekseev:1989}
{\sc Alekseev, A., and Shatashvili, S.}
\newblock Path integral quantization of the coadjoint orbits of the {V}irasoro
  group and 2-d gravity.
\newblock {\em Nuclear Physics B 323}, 3 (1989), 719--733.

\bibitem{AlvarezGaume:1983at}
{\sc Alvarez-Gaume, L.}
\newblock {Supersymmetry and the Atiyah-Singer Index Theorem}.
\newblock {\em Commun.Math.Phys. 90\/} (1983), 161.

\bibitem{Arnold:1985}
{\sc Arnold, V., Gusein-Zade, S., and Varchenko, A.}
\newblock {Singularities of Differentiable Maps}.

\bibitem{Atiyah:1978ri}
{\sc Atiyah, M., Hitchin, N.~J., Drinfeld, V., and Manin, Y.}
\newblock {Construction of Instantons}.
\newblock {\em Phys.Lett. A65\/} (1978), 185--187.

\bibitem{BFSS}
{\sc Banks, T., Fischler, W., Shenker, S., and Susskind, L.}
\newblock {M theory as a matrix model: A Conjecture}.
\newblock {\em Phys.Rev. D55\/} (1997), 5112--5128.

\bibitem{Baxter:1985}
{\sc Baxter, R.~J.}
\newblock {\em Exactly solved models in statistical mechanics}, vol.~1 of {\em
  Ser. Adv. Statist. Mech.}
\newblock World Sci. Publishing, Singapore, 1985.

\bibitem{obs:1996}
{\sc {Behrend}, K., and {Fantechi}, B.}
\newblock {The intrinsic normal cone}.
\newblock {\em {arXiv:alg-geom/9601010}\/}.

\bibitem{Bershtein:2015xfa}
{\sc Bershtein, M., Bonelli, G., Ronzani, M., and Tanzini, A.}
\newblock {Exact results for ${\CalN}=2$ supersymmetric gauge theories on
  compact toric manifolds and equivariant Donaldson invariants}.
\newblock {\em 1509.00267\/} (2015).

\bibitem{Borodin:2015}
{\sc {Borodin}, A., {Gorin}, V., and {Guionnet}, A.}
\newblock {Gaussian asymptotics of discrete $\beta$-ensembles}.
\newblock {\em math.PR:1505.03760\/}.

\bibitem{Bourgine:2015szm}
{\sc Bourgine, J.-E., Mastuo, Y., and Zhang, H.}
\newblock {Holomorphic field realization of SH$^c$ and quantum geometry of
  quiver gauge theories}.
\newblock {\em hep-th:1512.02492\/}.

\bibitem{BMO:2010}
{\sc {Braverman}, A., {Maulik}, D., and {Okounkov}, A.}
\newblock {Quantum cohomology of the Springer resolution}.
\newblock {\em ArXiv:math.AG/1001.0056\/} (Dec. 2010).

\bibitem{Bullimore:2014awa}
{\sc Bullimore, M., Kim, H.-C., and Koroteev, P.}
\newblock {Defects and Quantum Seiberg-Witten Geometry}.
\newblock {\em arXiv:hep-th/1412.6081\/} (2014).

\bibitem{ONC}
{\sc {Carlsson}, E., {Nekrasov}, N., and {Okounkov}, A.}
\newblock {Five dimensional gauge theories and vertex operators}.
\newblock {\em {math.RT/1308.2465}\/} ({2013}).

\bibitem{OC:2008}
{\sc {Carlsson}, E., and {Okounkov}, A.}
\newblock {Exts and Vertex Operators}.
\newblock {\em math.AG/0801.2565\/}.

\bibitem{Cecotti:1991me}
{\sc Cecotti, S., and Vafa, C.}
\newblock Topological-antitopological fusion.
\newblock {\em Nucl. Phys. B367\/} (1991), 359--461.

\bibitem{Chari:2004}
{\sc Chari, V., and Moura, A.~A.}
\newblock Characters and blocks for finite-dimensional representations of
  quantum affine algebras.
\newblock {\em Int. Math. Res. Not.}, 5 (2005), 257--298.

\bibitem{Chari:1991}
{\sc Chari, V., and Pressley, A.}
\newblock Quantum affine algebras.
\newblock {\em Comm. Math. Phys. 142}, 2 (1991), 261--283.

\bibitem{Chari:1994}
{\sc Chari, V., and Pressley, A.}
\newblock Quantum affine algebras and their representations.
\newblock {\em hep-th/9411145 16\/} (1995), 59--78.

\bibitem{Chari:1996}
{\sc Chari, V., and Pressley, A.}
\newblock Yangians: their representations and characters.
\newblock {\em Acta Appl. Math. 44}, 1-2 (1996), 39--58.
\newblock Representations of Lie groups, Lie algebras and their quantum
  analogues.

\bibitem{Dijkgraaf:2002dh}
{\sc Dijkgraaf, R., and Vafa, C.}
\newblock A perturbative window into non-perturbative physics.
\newblock {\em hep-th/0208048\/}.

\bibitem{Dijkgraaf:2002vw}
{\sc Dijkgraaf, R., and Vafa, C.}
\newblock On geometry and matrix models.
\newblock {\em Nucl. Phys. B644\/} (2002), 21--39.

\bibitem{Verlinde:1995}
{\sc {Dijkgraaf, R. and Verlinde, E. and Verlinde, H.}}
\newblock Matrix string theory.
\newblock {\em Nucl. Phys. B 500}, 1 (1997), 43--61.

\bibitem{Donagi:1995cf}
{\sc Donagi, R., and Witten, E.}
\newblock {Supersymmetric Yang-Mills theory and integrable systems}.
\newblock {\em Nucl.Phys. B460\/} (1996), 299--334.

\bibitem{Douglas:1995bn}
{\sc Douglas, M.~R.}
\newblock {Branes within branes}.
\newblock {\em hep-th/9512077\/}.

\bibitem{Douglas:1996sw}
{\sc Douglas, M.~R., and Moore, G.~W.}
\newblock {D-branes, quivers, and ALE instantons}.
\newblock {\em hep-th/9603167\/} (1996).

\bibitem{Drinfeld:1985}
{\sc Drinfeld, V.~G.}
\newblock Hopf algebras and the quantum {Y}ang-{B}axter equation.
\newblock {\em Dokl. Akad. Nauk SSSR 283}, 5 (1985), 1060--1064.

\bibitem{Drinfeld:1987}
{\sc Drinfeld, V.~G.}
\newblock A new realization of {Y}angians and of quantum affine algebras.
\newblock {\em Dokl. Akad. Nauk SSSR 296}, 1 (1987), 13--17.

\bibitem{Drinfeld:1986}
{\sc Drinfeld, V.~G.}
\newblock Quantum groups.
\newblock {\em Proceedings of the {I}nternational {C}ongress of
  {M}athematicians, {V}ol. 1, 2 ({B}erkeley, {C}alif., 1986)\/} (1987),
  798--820.

\bibitem{Faddeev:1987ih}
{\sc Faddeev, L., Reshetikhin, N.~Y., and Takhtajan, L.}
\newblock {Quantization of Lie Groups and Lie Algebras}.
\newblock {\em Leningrad Math.J. 1\/} (1990), 193--225.

\bibitem{Freed:1999}
{\sc Freed, D.~S.}
\newblock Special {K}\"ahler manifolds.
\newblock {\em Comm. Math. Phys. 203}, 1 (1999), 31--52.

\bibitem{Frenkel:2013dh}
{\sc {Frenkel}, E., and {Hernandez}, D.}
\newblock {Baxter's Relations and Spectra of Quantum Integrable Models}.
\newblock {\em math.QA/1308.3444\/}.

\bibitem{Frenkel:2001}
{\sc Frenkel, E., and Mukhin, E.}
\newblock Combinatorics of {$q$}-characters of finite-dimensional
  representations of quantum affine algebras.
\newblock {\em Comm. Math. Phys. 216}, 1 (2001), 23--57.

\bibitem{Frenkel:1996}
{\sc Frenkel, E., and Reshetikhin, N.}
\newblock Quantum affine algebras and deformations of the {V}irasoro and
  {$\mathcal{W}$}-algebras.
\newblock {\em Comm. Math. Phys. 178}, 1 (1996), 237--264.

\bibitem{Frenkel:1998}
{\sc Frenkel, E., and Reshetikhin, N.}
\newblock The {$q$}-characters of representations of quantum affine algebras
  and deformations of {$\mathscr W$}-algebras.
\newblock {\em Contemp. Math. 248\/} (1999), 163--205.

\bibitem{Frenkel:1992}
{\sc Frenkel, I.~B., and Reshetikhin, N.~Y.}
\newblock Quantum affine algebras and holonomic difference equations.
\newblock {\em Comm. Math. Phys. 146}, 1 (1992), 1--60.

\bibitem{Furuuchi:1999kv}
{\sc Furuuchi, K.}
\newblock {Instantons on noncommutative ${\BR}^4$ and projection operators}.
\newblock {\em Prog. Theor. Phys. 103\/} (2000), 1043--1068.

\bibitem{Gaiotto:2014ina}
{\sc Gaiotto, D., and Kim, H.-C.}
\newblock {Surface defects and instanton partition functions}.
\newblock {\em arXiv:hep-th/1412.2781\/}.

\bibitem{Gerasimov:2000ga}
{\sc Gerasimov, A.~A., and Shatashvili, S.~L.}
\newblock {Stringy Higgs mechanism and the fate of open strings}.
\newblock {\em JHEP 01\/} (2001), 019.

\bibitem{Gerasimov:2007ap}
{\sc Gerasimov, A.~A., and Shatashvili, S.~L.}
\newblock {Two-dimensional Gauge Theories and Quantum Integrable Systems}.
\newblock {\em arXiv:hep-th/0711.1472\/} (2007).

\bibitem{Gerasimov:2006zt}
{\sc Gerasimov, A.~A., and Shatashvili, S.~L.}
\newblock {Higgs Bundles, Gauge Theories and Quantum Groups}.
\newblock {\em Commun.Math.Phys. 277\/} (2008), 323--367.

\bibitem{Givental:1996}
{\sc {Givental}, A.~B.}
\newblock {Equivariant Gromov - Witten Invariants}.
\newblock {\em arXiv:alg-geom/9603021\/}.

\bibitem{Gorsky:1995zq}
{\sc Gorsky, A., Krichever, I., Marshakov, A., Mironov, A., and Morozov, A.}
\newblock {Integrability and Seiberg-Witten exact solution}.
\newblock {\em Phys.Lett. B355\/} (1995), 466--474.

\bibitem{Gorsky:2014mva}
{\sc Gorsky, A., and Vysotsky, M.}
\newblock {Proceedings, 100th anniversary of the birth of I.Ya. Pomeranchuk
  (Pomeranchuk 100)}.

\bibitem{Gutperle:2002ai}
{\sc Gutperle, M., and Strominger, A.}
\newblock {Space - like branes}.
\newblock {\em JHEP 0204\/} (2002), 018.

\bibitem{Hernandez:2008}
{\sc Hernandez, D.}
\newblock Quantum toroidal algebras and their representations.
\newblock {\em Selecta Math. (N.S.) 14}, 3-4 (2009), 701--725.

\bibitem{Hietamaki:1991qp}
{\sc Hietamaki, A., Morozov, A.~Y., Niemi, A.~J., and Palo, K.}
\newblock {Geometry of N=1/2 supersymmetry and the Atiyah-Singer index
  theorem}.
\newblock {\em Phys.Lett. B263\/} (1991), 417--424.

\bibitem{Howe:1983wj}
{\sc Howe, P.~S., Stelle, K., and West, P.~C.}
\newblock {A Class of Finite Four-Dimensional Supersymmetric Field Theories}.
\newblock {\em Phys.Lett. B124\/} (1983), 55.

\bibitem{IKKT}
{\sc Ishibashi, N., Kawai, H., Kitazawa, Y., and Tsuchiya, A.}
\newblock {A Large {\it N} reduced model as superstring}.
\newblock {\em Nucl.Phys. B498\/} (1997), 467--491.

\bibitem{Jimbo:1985}
{\sc Jimbo, M.}
\newblock A {$q$}-difference analogue of {$U({\mathfrak g})$} and the
  {Y}ang-{B}axter equation.
\newblock {\em Lett. Math. Phys. 10}, 1 (1985), 63--69.

\bibitem{Johnson:1996py}
{\sc Johnson, C.~V., and Myers, R.~C.}
\newblock {Aspects of type IIB theory on ALE spaces}.
\newblock {\em Phys.Rev. D55\/} (1997), 6382--6393.

\bibitem{Kac:1990}
{\sc Kac, V.~G.}
\newblock {\em Infinite-dimensional {L}ie algebras}, third~ed.
\newblock Cambridge University Press, Cambridge, 1990.

\bibitem{Kanno:2013aha}
{\sc Kanno, S., Matsuo, Y., and Zhang, H.}
\newblock {Extended Conformal Symmetry and Recursion Formulae for Nekrasov
  Partition Function}.
\newblock {\em JHEP 1308\/} (2013), 028.

\bibitem{Katz:1997eq}
{\sc Katz, S., Mayr, P., and Vafa, C.}
\newblock {Mirror symmetry and exact solution of 4-D ${\CalN}=2$ gauge
  theories: 1.}
\newblock {\em Adv.Theor.Math.Phys. 1\/} (1998), 53--114.

\bibitem{Kimura:2015rgi}
{\sc Kimura, T., and Pestun, V.}
\newblock {Quiver W-algebras}.
\newblock {\em arXiv:1512.08533\/} (2015).

\bibitem{Kirillov:1999}
{\sc Kirillov, A.~A.}
\newblock {Merits and demerits of the orbit method}.
\newblock {\em Bull. Amer. Math. Soc. 36\/} (1999), 433--488.

\bibitem{Knight:1995}
{\sc Knight, H.}
\newblock Spectra of tensor products of finite-dimensional representations of
  {Y}angians.
\newblock {\em J. Algebra 174}, 1 (1995), 187--196.

\bibitem{Knizhnik:1984}
{\sc {Knizhnik, V.G. and Zamolodchikov, A.B.}}
\newblock {Current algebra and Wess-Zumino model in two dimensions}.
\newblock {\em Nucl. Phys. B247}, 1 (1984), 83--103.

\bibitem{Kontsevich:1994qz}
{\sc Kontsevich, M., and Manin, Y.}
\newblock {Gromov-Witten classes, quantum cohomology, and enumerative
  geometry}.
\newblock {\em Commun.Math.Phys. 164\/} (1994), 525--562.

\bibitem{Krichever:1992qe}
{\sc Krichever, I.~M.}
\newblock {The tau function of the universal Whitham hierarchy, matrix models
  and topological field theories}.
\newblock {\em Commun. Pure Appl. Math. 47\/} (1994), 437.

\bibitem{Kronheimer:1990nk}
{\sc {Kronheimer}, P., and {Nakajima}, H.}
\newblock {Y}ang-{M}ills instantons on {ALE} gravitational instantons.
\newblock {\em Math. Ann. 288\/} (1990), 263--307.

\bibitem{Lawrence:1998ja}
{\sc Lawrence, A.~E., Nekrasov, N., and Vafa, C.}
\newblock {On conformal field theories in four-dimensions}.
\newblock {\em Nucl.Phys. B533\/} (1998), 199--209.

\bibitem{loma}
{\sc {Losev}, A., and {Manin}, Y.}
\newblock {New moduli spaces of pointed curves and pencils of flat
  connections}.
\newblock {\em math/0001003, and the refs [Lo1], [Lo2] there\/}.

\bibitem{Losev:2003py}
{\sc Losev, A., Marshakov, A., and Nekrasov, N.}
\newblock {Small instantons, little strings and free fermions}.
\newblock {\em hep-th/0302191\/} (2003).

\bibitem{lone}
{\sc {Losev}, A., and {Nekrasov}, N.}
\newblock Discussions on {K}.~{S}aito's pairings and contact terms in
  {L}andau-{G}inzburg theory.
\newblock {\em Mathematical physics conference, Rakhov, Ukraine\/} (Oct. 1992).

\bibitem{MakMig}
{\sc Makeenko, Y., and Migdal, A.}
\newblock {Quantum chromodynamics as dynamics of loops}.
\newblock {\em Nucl.Phys. B 188\/} (1981), 269--316.

\bibitem{Marshakov:2006ii}
{\sc Marshakov, A., and Nekrasov, N.}
\newblock {Extended Seiberg-Witten theory and integrable hierarchy}.
\newblock {\em JHEP 01\/} (2007), 104.

\bibitem{McKay:1980}
{\sc McKay, J.}
\newblock Graphs, singularities, and finite groups.
\newblock {\em The {S}anta {C}ruz {C}onference on {F}inite {G}roups ({U}niv.
  {C}alifornia, {S}anta {C}ruz, {C}alif., 1979) 37\/} (1980), 183--186.

\bibitem{Migdal:1984gj}
{\sc Migdal, A.~A.}
\newblock {Loop Equations and 1/N Expansion}.
\newblock {\em Phys. Rept. 102\/} (1983), 199--290.

\bibitem{Moore:1998et}
{\sc Moore, G., Nekrasov, N., and Shatashvili, S.}
\newblock {D-particle bound states and generalized instantons}.
\newblock {\em Commun.Math.Phys. 209\/} (2000), 77--95.

\bibitem{Moore:1997dj}
{\sc Moore, G.~W., Nekrasov, N., and Shatashvili, S.}
\newblock {Integrating over Higgs branches}.
\newblock {\em Commun. Math. Phys. 209\/} (2000), 97--121.

\bibitem{Nakajima:2001}
{\sc Nakajima, H.}
\newblock {$t$}-analogue of the {$q$}-characters of finite dimensional
  representations of quantum affine algebras.
\newblock {\em ``Physics and combinatorics'', 196--219, World Sci. Publ., River
  Edge, NJ 2001\/}.

\bibitem{Nakajima:1994r}
{\sc Nakajima, H.}
\newblock Gauge theory on resolutions of simple singularities and simple {L}ie
  algebras.
\newblock {\em Internat. Math. Res. Notices}, 2 (1994), 61--74.

\bibitem{Nakajima:1994}
{\sc Nakajima, H.}
\newblock Instantons on {ALE} spaces, quiver varieties, and {K}ac-{M}oody
  algebras.
\newblock {\em Duke Math. J. 76}, 2 (1994), 365--416.

\bibitem{Nakajima:2003}
{\sc Nakajima, H.}
\newblock {$t$}-analogs of {$q$}-characters of quantum affine algebras of type
  {$A_n,D_n$}.
\newblock {\em Contemp. Math. 325\/} (2003), 141--160.

\bibitem{Nakajima:2004}
{\sc Nakajima, H.}
\newblock Quiver varieties and {$t$}-analogs of {$q$}-characters of quantum
  affine algebras.
\newblock {\em Ann. of Math. (2) 160}, 3 (2004), 1057--1097.

\bibitem{Nakajima:2010}
{\sc Nakajima, H.}
\newblock {$t$}-analogs of {$q$}-characters of quantum affine algebras of type
  {$E_6,E_7,E_8$}.
\newblock {\em Progr. Math. 284\/} (2010), 257--272.

\bibitem{Nekrasov:2009st}
{\sc {Nekrasov}, N.}
\newblock Supersymmetric gauge theories and quantization of integrable systems.
\newblock {\em Lecture at Strings'2009\/}.

\bibitem{Nekrasov:1996cz}
{\sc Nekrasov, N.}
\newblock {Five dimensional gauge theories and relativistic integrable
  systems}.
\newblock {\em Nucl. Phys. B531\/} (1998), 323--344.

\bibitem{Nekrasov:2004sem}
{\sc {Nekrasov}, N.}
\newblock {{\sl On the BPS/CFT correspondence},}.
\newblock {\em {L}ecture at the {U}niversity of {A}msterdam string theory group
  seminar\/} (Feb. 3, 2004).

\bibitem{Nekrasov:2003rj}
{\sc {Nekrasov}, N., and {Okounkov}, A.}
\newblock {Seiberg-Witten theory and random partitions}.
\newblock {\em The Unity of Mathematics, In Honor of the Ninetieth Birthday of
  I.M. Gelfand, Progress in Mathematics, Vol. 244 P.~Etingof, V.~Retakh,
  I.M.~Singer (Eds.) 2006, XXII, Birkh\"auser Basel\/} (2003).

\bibitem{Nekrasov:2012xe}
{\sc Nekrasov, N., and Pestun, V.}
\newblock {Seiberg-Witten geometry of four dimensional N=2 quiver gauge
  theories}.
\newblock {\em arXiv:1211.2240 [hep-th]\/} ({2012}).

\bibitem{Nekrasov:2013xda}
{\sc Nekrasov, N., Pestun, V., and Shatashvili, S.}
\newblock {Quantum geometry and quiver gauge theories}.
\newblock {\em arXiv:1312.6689 [hep-th]\/} (2013).

\bibitem{Nekrasov:2011bc}
{\sc Nekrasov, N., Rosly, A., and Shatashvili, S.}
\newblock {Darboux coordinates, Yang-Yang functional, and gauge theory}.
\newblock {\em Nucl.Phys.Proc.Suppl. 216\/} (2011), 69--93.

\bibitem{Nekrasov:1998ss}
{\sc Nekrasov, N., and Schwarz, A.~S.}
\newblock {Instantons on noncommutative ${\BR}^{4}$ and $(2,0)$-superconformal
  six dimensional theory}.
\newblock {\em Commun. Math. Phys. 198\/} (1998), 689--703.

\bibitem{Nekrasov:2009zz}
{\sc Nekrasov, N., and Shatashvili, S.}
\newblock {Bethe Ansatz and supersymmetric vacua}.
\newblock {\em AIP Conf. Proc. 1134\/} (2009), 154--169.

\bibitem{Nekrasov:2010ka}
{\sc Nekrasov, N., and Witten, E.}
\newblock {The Omega Deformation, Branes, Integrability, and Liouville Theory}.
\newblock {\em JHEP 1009\/} (2010), 092.

\bibitem{Nekrasov:2002kc}
{\sc Nekrasov, N.~A.}
\newblock {Lectures on open strings, and noncommutative gauge fields}.
\newblock {\em arXiv:hep-th/0203109\/}.

\bibitem{Nekrasov:2003vi}
{\sc Nekrasov, N.~A.}
\newblock {Localizing gauge theories}.
\newblock Prepared for 14th International Congress on Mathematical Physics
  (ICMP 2003), Lisbon, Portugal, 28 Jul - 2 Aug 2003.

\bibitem{Nekrasov:2000zz}
{\sc Nekrasov, N.~A.}
\newblock {Noncommutative instantons revisited}.
\newblock {\em Commun.Math.Phys. 241\/} (2003), 143--160.

\bibitem{Nekrasov:2002qd}
{\sc Nekrasov, N.~A.}
\newblock {Seiberg-Witten prepotential from instanton counting}.
\newblock {\em Adv.Theor.Math.Phys. 7\/} (2004), 831--864.

\bibitem{Nekrasov:2005wp}
{\sc Nekrasov, N.~A.}
\newblock {Lectures on nonperturbative aspects of supersymmetric gauge
  theories}.
\newblock {\em Class. Quant. Grav. 22\/} (2005), S77--S105.

\bibitem{Nekrasov:2009rc}
{\sc Nekrasov, N.~A., and Shatashvili, S.~L.}
\newblock {Quantization of Integrable Systems and Four Dimensional Gauge
  Theories}.
\newblock {\em arXiv:0908.4052\/} (2009).

\bibitem{Nekrasov:2009ui}
{\sc Nekrasov, N.~A., and Shatashvili, S.~L.}
\newblock {Quantum integrability and supersymmetric vacua}.
\newblock {\em Prog. Theor. Phys. Suppl. 177\/} (2009), 105--119.

\bibitem{Nekrasov:2009uh}
{\sc Nekrasov, N.~A., and Shatashvili, S.~L.}
\newblock {Supersymmetric vacua and Bethe ansatz}.
\newblock {\em Nucl. Phys. Proc. Suppl. 192-193\/} (2009), 91--112.

\bibitem{Nekrasov:2000ih}
{\sc {Nekrasov, Nikita}}.
\newblock {Trieste lectures on solitons in noncommutative gauge theories}.
\newblock {\em hep-th/0011095\/} (2000).

\bibitem{Nekrasov:2015ii}
{\sc {Nekrasov, Nikita}}.
\newblock {BPS/CFT correspondence: Gauge origami and qq-characters}.

\bibitem{Nekrasov:2015iim}
{\sc {Nekrasov, Nikita}}.
\newblock {BPS/CFT correspondence: Instantons at crossroads and Gauge origami}.

\bibitem{Nekrasov:2015iii}
{\sc {Nekrasov, Nikita}}.
\newblock {BPS/CFT correspondence: KZ and BPZ equations from Non-perturbative
  Dyson-Schwinger equations}.

\bibitem{Nekrasov:2015is}
{\sc {Nekrasov, Nikita}}.
\newblock {BPS/CFT correspondence: Non-perturbative Dyson-Schwinger equations
  and surface operators}.

\bibitem{Pressley:1986}
{\sc Pressley, A., and Segal, G.}
\newblock {\em {L}oop groups}.
\newblock Oxford University Press, 1986.

\bibitem{Varchenko:1991}
{\sc Schechtman, V., and Varchenko, A.}
\newblock {\em Invent. Math. 106\/} (1991), 139.

\bibitem{Seiberg:1994rs}
{\sc Seiberg, N., and Witten, E.}
\newblock Electric - magnetic duality, monopole condensation, and confinement
  in {N=2} supersymmetric {Y}ang-{Mi}lls theory.
\newblock {\em Nucl. Phys. B426\/} (1994), 19--52.

\bibitem{Seiberg:1994aj}
{\sc Seiberg, N., and Witten, E.}
\newblock Monopoles, duality and chiral symmetry breaking in {N=2}
  supersymmetric {QCD}.
\newblock {\em Nucl. Phys. B431\/} (1994), 484--550.

\bibitem{Seiberg:1996nz}
{\sc Seiberg, N., and Witten, E.}
\newblock {Gauge dynamics and compactification to three-dimensions}.
\newblock {\em {hep-th/9607163}\/} ({1996}).

\bibitem{Sklyanin:1980ij}
{\sc Sklyanin, E.}
\newblock {Quantum version of the method of inverse scattering problem}.
\newblock {\em J. Sov. Math. 19\/} (1982), 1546--1596.

\bibitem{Sklyanin:1982tf}
{\sc Sklyanin, E.}
\newblock {Some algebraic structures connected with the Yang-Baxter equation}.
\newblock {\em Funct.Anal.Appl. 16\/} (1982), 263--270.

\bibitem{Sklyanin:1978pj}
{\sc Sklyanin, E., and Faddeev, L.}
\newblock {Quantum Mechanical Approach to Completely Integrable Field Theory
  Models}.
\newblock {\em Sov. Phys. Dokl. 23\/} (1978), 902--904.

\bibitem{Sklyanin:1979}
{\sc Sklyanin, E.~K., Takhtajan, L.~A., and Faddeev, L.~D.}
\newblock Quantum inverse problem method. {I}.
\newblock {\em Teoret. Mat. Fiz. 40}, 2 (1979), 194--220.

\bibitem{Smirnov:1992vz}
{\sc Smirnov, F.}
\newblock {Form-factors in completely integrable models of quantum field
  theory}.
\newblock {\em Adv.Ser.Math.Phys. 14\/} (1992), 1--208.

\bibitem{Smirnov:1991me}
{\sc Smirnov, F.~A.}
\newblock {Dynamical symmetries of massive integrable models, 1. Form-factor
  bootstrap equations as a special case of deformed Knizhnik- Zamolodchikov
  equations}.
\newblock {\em Int.J.Mod.Phys. A71B\/} (1992), 813--837.

\bibitem{Takhtajan:1979iv}
{\sc Takhtajan, L., and Faddeev, L.}
\newblock {The Quantum method of the inverse problem and the Heisenberg XYZ
  model}.
\newblock {\em Russ.Math.Surveys 34\/} (1979), 11--68.

\bibitem{Tong:2014yna}
{\sc Tong, D.}
\newblock {The holographic dual of $AdS_{3} \times S^{3} \times S^{3} \times
  S^{1}$}.
\newblock {\em JHEP 1404\/} (2014), 193.

\bibitem{Tong:2014cha}
{\sc Tong, D., and Wong, K.}
\newblock {ADHM Revisited: Instantons and Wilson Lines}.
\newblock {\em arXiv:1410.8523\/} (2014).

\bibitem{Vafa:1994tf}
{\sc {Vafa}, C., and {Witten}, E.}
\newblock {A Strong coupling test of S-duality}.
\newblock {\em Nucl.Phys. B431\/} (1994), 3--77.

\bibitem{Witten:1995zh}
{\sc Witten, E.}
\newblock {Some comments on string dynamics}.
\newblock {\em hep-th/9507121\/} ({1995}).

\bibitem{Wyllard:2009hg}
{\sc Wyllard, N.}
\newblock {$A_{N-1}$ conformal Toda field theory correlation functions from
  conformal ${\CalN} = 2$ $SU(N)$ quiver gauge theories}.
\newblock {\em JHEP 0911\/} (2009), 002.

\end{thebibliography}

\end{document}